\documentclass[3p,12pt]{elsarticle}
\usepackage{epsf,epsfig,float,amssymb,latexsym,amsmath,amsthm,fancyhdr}
\usepackage{graphics,psfrag,longtable}
\usepackage{slashed}
\usepackage{colortbl}
\usepackage[pagewise]{lineno}
\newcommand{\vp}{\mathbf{p}}
\newcommand{\bl}{\begin{aligned}}
\newcommand{\el}{\end{aligned}}

\newcommand{\la}{\langle}
\newcommand{\ra}{\rangle}

\newcommand{\Op}{{\mathcal{O}}(p)}
\newcommand{\Opd}{{\mathcal{O}}(p^2)}
\newcommand{\Opt}{{\mathcal{O}}(p^3)}
\newcommand{\Opc}{{\mathcal{O}}(p^4)}

\newcommand{\be}{\begin{equation}}
\newcommand{\ee}{\end{equation}}
\newcommand{\ba}{\begin{eqnarray}}
\newcommand{\ea}{\end{eqnarray}}
\newcommand{\bg}{\begin{align}}
\newcommand{\egg}{\end{align}}

\newcommand{\nn}{\nonumber}

\newcommand{\barr}[1]{\not\mathrel  #1}

\newcommand{\logMm}{\log \left( \frac{M^2}{m^2} \right)}

\newcommand{\logMpimN}{\log \left( \frac{M_\pi^2}{m_N^2} \right)}
\newcommand{\logMmu}{\log \left( \frac{M^2}{\mu^2} \right)}
\newcommand{\logmmu}{\log \left( \frac{m^2}{\mu^2} \right)}

\newcommand{\dkd}{\frac{d^d k}{(2 \pi)^d}}
\newcommand{\HH}{\mathcal{H}}
\newcommand{\RR}{\mathcal{R}}
\newcommand{\sigmaterm}{\sigma_{\pi N}}

\newcommand{\Deltaless}{\slashed{\Delta}}

\def\slashp{p \!\!\! \slash}

\begin{document}
\begin{frontmatter}
\title{Improved description of the $\pi N$-scattering phenomenology in covariant baryon chiral perturbation theory}

\author[kph]{J. M. Alarc\'on}
\author[uos]{J. Martin Camalich}
\author[uom]{J.A. Oller}
\address[kph]{Institut f\"ur Kernphysik, Johannes Gutenberg Universit\"at, Mainz D-55099, Germany}
\address[uos]{Department of Physics and Astronomy, University of Sussex, BN1 9QH, Brighton, UK}
\address[uom]{ Departamento de F\'{\i}sica. Universidad de Murcia. E-30071,
Murcia, Spain}

\begin{abstract}
We present a novel analysis of the $\pi N$ scattering amplitude in covariant baryon chiral perturbation theory up to ${\cal O}(p^3)$ within the extended-on-mass-shell renormalization scheme and including the $\Delta(1232)$ explicitly in the $\delta$-counting. We take the hadronic phase shifts provided by partial wave analyses as basic experimental information to fix the low-energy constants. Subsequently, we study in detail the various observables and low-energy theorems related to the $\pi N$ scattering amplitude. In particular, we discuss the results and chiral expansion of the phase shifts, the threshold coefficients, the Goldberger-Treiman relation, the pion-nucleon sigma term and the extrapolation onto the subthreshold region. The chiral representation of the amplitude in the theory with the $\Delta$ presents a good convergence from very low energies in the subthreshold region up to energies well above threshold, leading also to a phenomenological description perfectly consistent with the one 
reported by the respective partial wave analyses and independent determinations. We conclude that a model-independent and systematic framework to analyze $\pi N$-scattering observables using directly experimental data shall be possible in covariant baryon chiral perturbation theory.
\end{abstract}

\end{frontmatter}


\section{Introduction}

 Pion-nucleon ($\pi N$) scattering is a hadronic reaction that gives access to some of the most prominent and fundamental questions related to the strong interactions~\cite{hoehler}. 
At low energies, it allows to test the dynamical constraints imposed by the chiral symmetry of QCD in one of the simplest processes involving a nucleon~\cite{gasser2}. 
Also, understanding $\pi N$ scattering is essential for a first-principle approach to the nuclear structure, given that the long-range part of the $NN$ interactions is given by the exchange of pions~\cite{Epelbaum:2012vx}.
Experimental data on differential cross sections and polarization observables have been collected in the last 50 years, and more intensively in the last decade thanks to fully dedicated experiments run in meson factories. 
The usual way to organize the experimental information is by means of energy-dependent parameterizations of the scattering amplitude projected in partial waves, fitted to the data and supplemented with unitarity and analyticity constraints. 
These partial wave analyses (PWAs) provide an accurate representation of the data included in the parameterizations, which can be used to extract values of scattering parameters and strong coupling constants or to identify the effect of resonances in the different isospin-angular momentum channels.\footnote{See the classical treatise of H\"ohler~\cite{hoehler} for an exhaustive review on the $\pi N$ scattering amplitude. For updated descriptions of the current experimental situation and data see Refs.~\cite{WI08,EM06}.} 

In spite of the very long and sustained effort in studying the $\pi N$ scattering amplitude, there are fundamental questions concerning this process that have not been satisfactorily answered yet. For instance, it remains unclear what is the exact value of the pion-nucleon coupling constant $g_{\pi N}$ or of the pion-nucleon sigma term, $\sigma_{\pi N}$. Besides that, it is important to determine accurately the threshold parameters. The scattering lengths ought to be compared with those obtained from the analysis of the accurately measured $1S$ level shift in pionic hydrogen and deuterium~\cite{baru}. An alternative source of phenomenological information on the scattering parameters is given by the analyses of the $NN$ interaction~\cite{Epelbaum:2012vx,NNRefNuevas}. On the other hand, it remains a challenge for the theory to understand hadronic processes directly from the parameters and dynamics of the underlying QCD. Important progress in the computation of hadron-hadron scattering lengths in unquenched LQCD has been 
reported~\cite{Beane:2010em}, although those of the $\pi N$ system are still computationally prohibitive. 
Nevertheless, results were obtained in a pioneering quenched calculation~\cite{Fukugita:1994ve} and, more recently, in an unquenched one~\cite{Beane:2009kya}. The LQCD simulations are often run with quark masses heavier than the physical ones and their results require a careful chiral extrapolation to the physical point.          

Chiral perturbation theory (ChPT), as the effective field theory of QCD at low energies~\cite{weinberg,gasser1,GasserLeutwylerFound,LeutwylerFound}, is a suitable framework to build a model-independent representation of the $\pi N$ scattering amplitude and to tackle some of these problems (see Ref.~\cite{ChPTReviews} for comprehensive reviews). An interesting feature of ChPT is that, regardless of the specific values of its parameters or low-energy constants (LECs), it inherits the chiral Ward-Takashi identities of QCD among different Green functions~\cite{LeutwylerFound}. 
Some of these identities at leading order in the chiral expansion were obtained using PCAC and Current-Algebra methods in the fifties and sixties, conforming what has been know as low-energy theorems since then~\cite{DashenAdlerBook}. Remarkable examples of these theorems in the $\pi N$ system are the Goldberger-Treiman (GT) relation between $g_{\pi N}$ and $g_A$~\cite{Goldberger:1958tr} and the one at the Cheng-Dashen (CD) point between $\sigma_{\pi N}$ and the scattering amplitude~\cite{cheng-dashen-theorem}. Other interesting examples are the Weinberg predictions for the scattering lengths~\cite{Weinberg:1966fm} or the Adler condition for the $\pi N$ scattering amplitude~\cite{adler}.

ChPT allows one to investigate up to which extent the low-energy theorems apply since it provides a method to compute systematically the higher-orders in the chiral expansions of the correlation functions entering corresponding Ward-Takahashi identities. A first attempt to apply baryon ChPT (BChPT) to elastic $\pi N$ scattering was undertaken by Gasser {\it et al.} in Ref.~\cite{gasser2}, where an {\it off-shell} amplitude was obtained up to $\mathcal{O}(p^3)$ in a manifestly Lorentz covariant formalism (for reviews on BChPT see Refs.~\cite{review,BernardII}). However, it was shown that the presence of the nucleon mass as a new large scale in the chiral limit invalidated na\"ive power counting arguments in the baryon sector. This problem can be solved in the heavy-baryon formalism (HBChPT)~\cite{jenkins}, in which one recovers a neat power-counting scheme at the cost of manifestly Lorentz covariance. 
Calculations of the $\pi N$ scattering amplitude up to $\mathcal{O}(p^3)$~\cite{fettes3} and $\mathcal{O}(p^4)$~\cite{fettes4} accuracies have been performed in the HB formalism by Fettes {\it et al.}, showing a good description of the $S$- and $P$-wave phase shifts of different PWAs at low energies.

It was latter shown that the HB approach is not well suited for studying some of the low-energy theorems involving the $\pi N$ scattering amplitude~\cite{becher}. The problem is that the non-relativistic expansion implemented in HBChPT alters the analytical structure of the baryon propagator such that the chiral expansion of some Green functions does not converge in certain parts of the low-energy region~\cite{becher,scherer1}. This problem shows up in the analytic continuation of the $\pi N$ scattering amplitude onto the subthreshold region~\cite{fettes4} or in the behavior of the form factors close to the two-pion threshold, $t=4M_\pi^2$~\cite{review}. 
Besides that, it has been shown that the non-relativistic expansion may have a problematic convergence in some other cases~\cite{Holstein:2005db,gorgorito,MartinCamalich:2010fp}. 

Based on the ideas previously discussed in Ref.~\cite{elli}, Becher {\it et al.} proposed the infrared scheme (IRChPT)~\cite{becher} as a solution to the problems of the HB formalism. The IR scheme is a manifestly Lorentz covariant approach to BChPT that preserves the HB power counting at the same as it resumms the kinetic terms of the positive-energy part of the baryon propagators, curing the analyticity problems of the HB approach. The $\pi N$ scattering amplitude has been also calculated at $\mathcal{O}(p^3)$~\cite{elli2,nuestroIR} and $\mathcal{O}(p^4)$~\cite{beche2} accuracies in the IR scheme. 
At $\mathcal{O}(p^4)$ the amplitude rapidly converges in the proximity of the CD point so one can investigate meticulously the corresponding low-energy theorems. However, and similarly to HBChPT~\cite{fettes4}, the convergence at $\mathcal{O}(p^4)$ in IR is such that one fails to connect the subthreshold and threshold regions.

A serious drawback of the IR method is that, in curing the problems of the HB expansion, it runs into its own ones with the analytic properties of the loop integrals~\cite{Ledwig:2011cx}. This is related to the fact that the IR resummation of kinetic terms performed on the HB propagators completely omits the inclusion of negative-energy pole or anti-nucleon contribution, violating charge conjugation symmetry and, therefore, causality~\cite{causalidad}. The most striking consequence of this is the appearance of unphysical cuts. Despite lying outside the range of applicability of ChPT, these cuts can have sizable contributions to the Green functions at low-energies, disrupting the convergence of the respective chiral expansions. This has been indeed shown in different applications of ChPT in the baryon sector such as the chiral extrapolation of the nucleon magnetic moment~\cite{Holstein:2005db}, the $SU(3)_F$ breaking of the baryon magnetic moments~\cite{gorgorito} or the unitarized description of the $\pi N$ 
scattering amplitude~\cite{nuestroIR}. 
        
A completely different difficulty in the baryon sector of ChPT is related to the $\Delta(1232)$-resonance. Its contributions can be important at very low energies since this resonance is very close in mass to the nucleon. 
In the conventional chiral expansion, these effects are accounted for by the LECs but the radius of convergence, in this case, becomes drastically reduced. This problem is prominent in $\pi N$ scattering as the threshold for this process is at a center-of-mass (CM) energy $\delta W= M_\pi$ away from the point around which the chiral expansion is performed. An improved convergence of the chiral series can be obtained including the $\Delta$ resonance as an explicit degree of freedom~\cite{Jenkins:1991es,Hemmert:1997ye,Pascalutsa:2006up}. In this case, one introduces a power counting for the new scale $\epsilon=m_\Delta-m_N$ and computes the $\Delta$ contributions accordingly. The $\Delta$ corrections to the $\pi N$ scattering amplitude have been calculated in the HB~\cite{fettes_ep} and IR~\cite{elli2} schemes up to $\mathcal{O}(\epsilon^3)$ within the small-scale expansion (SSE), which considers $\epsilon\sim\mathcal{O}(p)$~\cite{Hemmert:1997ye}. In case of the HB calculation, the 
inclusion of the $\Delta$ increases the range of energies described as compared with the $\Delta$-less case at $\mathcal{O}(p^3)$, although the values of the LECs strongly depend on the fitted data, precluding a clear discussion on the extracted values of the observables related to $\pi N$ scattering (e.g. $\sigma_{\pi N}$). In the IR calculation, the $\Delta$ corrections worsen severely the description of the different PWA phase-shifts~\cite{elli2}.

In this paper we present a novel chiral representation of the $\pi N$ scattering amplitude with two main differences as compared with previous work. In the first place, we use Lorentz covariant BChPT with a consistent power counting obtained via the extended-on-mass-shell (EOMS) renormalization scheme~\cite{eoms1,eoms2}. The main advantage of EOMS over the HB and IR schemes is that it preserves the proper analytic structure of the Green functions~\cite{Ledwig:2011cx}. Secondly, we explicitly include the $\Delta(1232)$ taking into account that, below the resonance region, the diagrams with the $\Delta$ are suppressed in comparison with those with the nucleon~\cite{pascalutsa-phillips,Pascalutsa:2006up}. 

The paper is organized as follows: We first present in Sec.~\ref{Sec:Formalism} a detailed description of the formalism employed, which comprises the representations of the $\pi N$ scattering amplitude and the ChPT formalism. Special emphasis is put on explaining the EOMS scheme (Sec~\ref{Sec:eomsrenormalization}) and the inclusion of $\Delta(1232)$ degrees of freedom (Sec.~\ref{Sec:DeltaIntrosection}). A brief discussion of isospin breaking effects is also presented here (Sec.~\ref{Sec:isospinbreakingcorrections}). In Sec.~\ref{Sec:ScatteringAmp}, we explain in some detail the calculation of the scattering amplitude and the application of the EOMS renormalization scheme. The reader is addressed to the Appendices for the complete results and some technical details.  Sec.~\ref{Sec:phaseshifts} focuses on the determination of the LECs of BChPT without ($\Deltaless$-ChPT) and with ($\Delta$-ChPT) the contributions of the $\Delta(1232)$ as an explicit degree of freedom and using various phase-shifts sets provided by 
different PWAs. The convergence of the chiral expansion of the amplitude in either case is studied in Secs.~\ref{Sec:convergenceofthechiralseries1} and \ref{Sec:convergenceofthechiralseries2}. Once the LECs are determined, Sec.~\ref{Sec:piNpheno} is dedicated to discuss the results and the chiral structure of the different $\pi N$-scattering observables. The analysis of the threshold parameters, including a comparison with LQCD results, the GT relation and the pion-nucleon sigma term are then presented in Secs.~\ref{Sec:thresholdparameters}, \ref{Sec:GT} and \ref{Sec:sigmaterm} respectively. Finally, we study the important issue of the connection between the threshold and subthreshold regions of the $\pi N$ scattering amplitude in Sec.~\ref{Sec:subthresholdregion}. We close the paper with a summary of our findings and results together with a brief outlook.                 

\section{Formalism}
\label{Sec:Formalism}
\subsection{The scattering amplitude}
\label{Sec:Form-ScatterinAmp}
\begin{figure}
\begin{center}
\epsfig{file=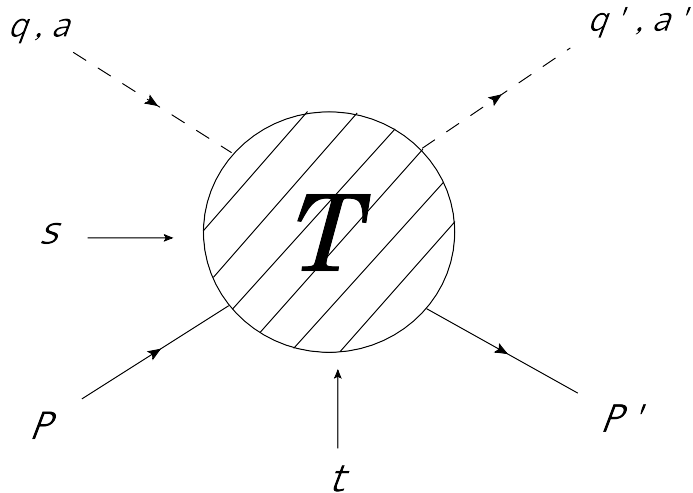,width=0.55\textwidth,angle=0}
\caption[pilf]{\protect \small Diagrammatic representation of the Lorentz- and isospin-invariant amplitudes. In this figure, $s$ and $t$ correspond to the Mandelstam variables, while $P$ and $P'$ correspond to the momentum of the incoming and outgoing nucleon, respectively. 
On the other hand, $q$ ($q'$) corresponds to the momentum of the incoming (outgoing) pion with Cartesian isospin index $a$ ($a'$). \label{TmatrixFig}}
\end{center}
\end{figure}

We consider the process $\pi^a(q) N(p,\sigma;\alpha)\to \pi^{a'}(q') N(p',\sigma';\alpha')$. 
Here $a$ and $a'$ denote the Cartesian coordinates in the isospin space of the initial and final pions with four-momentum $q$ and $q'$, respectively. 
Regarding the nucleons, $\sigma$($\sigma'$) and $ \alpha(\alpha')$ correspond to the third-components of spin and isospin of the initial (final) states, in order.
Since we calculate the on-shell amplitude, the usual Mandelstam variables $s=(p+q)^2=(p'+q')^2$, $t=(q-q')^2=(p-p')^2$ and $u=(p-q')^2=(p'-q)^2$, fulfill $s+t+u=2M_\pi^2+2 m_N^2$, with $m_N$ and $M_\pi$ the nucleon and pion mass, respectively. 
Exact isospin symmetry is assumed in the following, so it is convenient to consider Lorentz- and isospin-invariant amplitudes.  
We then decompose the scattering amplitude as~\cite{hoehler}

\begin{align}
T_{ a a' }&=\delta_{a'a} T^++\frac{1}{2}[\tau_a,\tau_{a'}]T^{-}\nn,\\
T^{\pm}&=\bar{u}(p',\sigma')\left[A^{\pm}+\frac{1}{2}(\barr{q}+{\barr{q}}\,')B^{\pm}\right]u(p,\sigma),
\label{apmbpmdef}
\end{align}
where the Pauli matrices are indicated by $\tau_c$  and the definitions of the indices and momenta are shown in Fig. \ref{TmatrixFig}. 

However, the decomposition \eqref{apmbpmdef} is not transparent from the power counting point of view, since the leading order contributions of $A^\pm$ and $B^\pm$ cancel. 
Therefore, it is convenient to write the scattering amplitude in terms of the energy variable $\nu=\frac{s-u}{4m_N}$,\footnote{In fact, the chiral expansion can be understood as an expansion around the {\it soft point}, in which $\nu=0,\,t=0$ and  $M_\pi=0$.} that scales as $\nu\sim \Op$, and the amplitudes $D \equiv A+ \nu B$ and $B$.
In this representation, the scattering amplitude takes the form
\begin{align}\label{DyB}
T^{\pm}&=\bar{u}(p',\sigma')\left[D^{\pm}-\frac{1}{4m_N}[\barr{q}\,',\barr{q}]B^{\pm}\right]u(p,\sigma).
\end{align}
Furthermore, one can construct the representation in partial waves of the amplitude, which is obtained by projecting Eqs.~(\ref{apmbpmdef}) onto states with definite total angular momentum $J$ and orbital angular momentum $\ell$,
\begin{align}
T_{J\ell}(a',\alpha';a,\alpha)&=\frac{1}{\sqrt{4\pi(2\ell+1)}(0\sigma\sigma|\ell\frac{1}{2}J)} \nonumber \\
\times\sum_{m,\sigma'} & \int d\hat{\vp}'\,\la \pi(-\vp';a')N(\vp',\sigma';\alpha')|T|\pi(-\vp;a)N(\vp,\sigma;\alpha)\ra  (m\sigma'\sigma|\ell\frac{1}{2}L) Y_\ell^m(\hat{\vp}')^*.
\end{align}   
This representation is often used in $\pi N$  scattering phenomenology and, in fact, global analyses of the experimental database are often presented in terms of partial-wave phase shifts~\cite{hoehler,KA85,WI08,EM06}. In the \ref{Sec:partialwavedecomposition} we specify the conventions used in this paper for the partial-wave representation of the scattering amplitude and the formulas used to extract the different phase shifts from the chiral representation.   

\subsection{Chiral perturbation theory and chiral Lagrangians}
\label{Sec:ChPTandLags}

In ChPT one develops a power counting  to express the Green functions of QCD as an expansion in the light quark mass and soft external momentum (generically denoted as $p$), which are small compared with the chiral symmetry breaking scale, $\Lambda_{\chi}\simeq4\pi f_\pi\simeq1$ GeV. These are called chiral expansions and they contain analytic and non-analytic pieces. 
The latter ones are genuine consequences of the spontaneous breaking of chiral symmetry, as they arise from quantum fluctuations or loops of the corresponding pseudo-Goldstone bosons (pions in the $SU(2)_F$ QCD case). They provide the complex contributions required by unitarity and come accompanied by UV divergences. Renormalization is ensured by the coefficients of the local terms in the expansion, the LECs. The values of these parameters are not determined by chiral symmetry and they enclose information on the dynamical content of the underlying theory, i.e. QCD. A consistent chiral power counting including baryons can be established considering the latter as non-relativistic fields~\cite{wein}; for processes with a single baryon, a Feynman diagram with $L$ loops, $N_M$ meson propagators, $N_B$ baryon propagators and $V_k$ vertices of $k$th order Lagrangian scales as $\mathcal{O}(p^n)$ where
\begin{equation}
\label{n.def}
n=4L-2N_M-N_B+\sum_k k V_k. 
\end{equation}
The chiral Lagrangian in the nucleon sector up to $\Opt$ can be expressed then as~\cite{gasser1,fettes3}
\begin{align}
{\cal L}_{\rm ChPT}&={\cal L}_{\pi\pi}^{(2)}+
{\cal L}_{\pi\pi}^{(4)}+
{\cal L}_{\pi N}^{(1)}+
{\cal L}_{\pi N}^{(2)}+
{\cal L}_{\pi N}^{(3)}~,
\end{align}
where the superscript is the chiral order $n$. Here, ${\cal L}_{\pi\pi}^{(n)}$ refers to the purely mesonic Lagrangian without baryons and ${\cal L}_{\pi N}^{(n)}$ corresponds to bilinears in the baryon fields. 
The explicit form for the mesonic Lagrangian is
 \begin{align}
{\cal L}_{\pi\pi}^{(2)}&=\frac{f^2}{4}\la u_\mu u^\mu+\chi_+\ra, ~\nn~\\ 
{\cal L}_{\pi\pi}^{(4)}&=\frac{1}{16}\ell_4 \left(2 \la u_\mu u^\mu\ra \la \chi_+ \ra
+\la \chi_+\ra^2\right)+\ldots,
\label{lagpi}
\end{align} 
where the ellipsis indicates terms that are not needed in the calculations given here and $\la \cdots \ra$ denote the trace of the resulting $2\times 2$ matrix in the flavor space. 
For the different symbols, $f$ is the pion weak decay constant in the chiral limit and
\begin{align}
u^2&=U~,~u_\mu=i u^\dagger \partial_\mu U\, u^\dagger~,~\chi_{\pm}=u^\dagger \chi u^\dagger\pm u \chi^\dagger u.~
\end{align}
The explicit chiral symmetry breaking due to the non-vanishing quark masses (in the isospin limit $m_u=m_d=\hat{m}$) is introduced through $\chi=2 B_0 \hat{m}$. 
The constant $B_0$ is proportional to the quark condensate in  the chiral limit ($m_u=m_d=0$), $\la 0|\bar{q}^j q^i|0\ra=-B_0 f^2 \delta^{ij}$.
In the following we employ the so-called sigma-parameterization where

\begin{align}
U(x)&=\sqrt{1-\frac{\vec{\pi}(x)^2}{f^2}}+i\frac{\vec{\pi}(x)\cdot \vec{\tau}}{f}.
\end{align}

For the pion-nucleon Lagrangian we have~\cite{fettes3}

\begin{align}
{\cal L}_{\pi N}^{(1)}&=\bar{\psi}(i\barr{D}-m)\psi+\frac{g}{2}\bar{\psi}\barr{u}\gamma_5 \psi~,\nn\\
{\cal L}_{\pi N}^{(2)}&=c_1 \la \chi_+\ra \bar{\psi}\psi-\frac{c_2}{4m^2}\la u_\mu u_\nu\ra(\bar{\psi}D^\mu D^\nu \psi+\hbox{h.c.})+\frac{c_3}{2}\la u_\mu u^\mu\ra \bar{\psi}\psi-\frac{c_4}{4}\bar{\psi}\gamma^\mu\gamma^\nu[u_\mu,u_\nu]\psi+\ldots~,\nn\\
{\cal L}_{\pi N}^{(3)}&=\bar{\psi}\Biggl(-\frac{d_1+d_2}{4m}([u_\mu,[D_\nu,u^\mu]+[D^\mu,u_\nu]]D^\nu 
+\hbox{h.c.})\nn\\
&+\frac{d_3}{12 m^3}([u_\mu,[D_\nu,u_\lambda]](D^\mu D^\nu D^\lambda+\hbox{sym.})+\hbox{h.c.})
+i\frac{d_5}{2 m}([\chi_-,u_\mu]D^\mu+\hbox{h.c.})\nn\\
&+i\frac{d_{14}-d_{15}}{8 m}\left(\sigma^{\mu \nu}\la [D_\lambda,u_\mu]u_\nu-u_\mu [D_\nu,u_\lambda]\ra 
D^\lambda+\hbox{h.c.}\right)\nn\\
&+\frac{d_{16}}{2}\gamma^\mu\gamma_5\la\chi_+\ra u_\mu+\frac{id_{18}}{2}\gamma^\mu \gamma_5 [D_\mu,\chi_-]\Biggr) \psi
+\ldots
\label{lagN}
\end{align}
In the previous equation $m$ is the nucleon mass in the chiral limit and the covariant derivative, $D_\mu$, acting on the baryon fields is given by $D_\mu=\partial_\mu+\Gamma_\mu$ with $\Gamma_\mu= [u^\dagger,\partial_\mu u]/2$. 
On the other hand, the LECs ($c_i$ and $d_i$) are determined by fitting them to $\pi N$ scattering data.  
Again, only the terms needed for the present study are shown in Eq.~\eqref{lagN}. 
For further details on the definition and derivation of the different monomials  we refer to Refs.~\cite{fettes3,opv}.

\subsection{The Extended-On-Mass-Shell renormalization scheme}
\label{Sec:eomsrenormalization}

The chiral power counting in the presence of baryons, Eq.~\eqref{n.def}, is blurred due to the presence of a new large scale, the nucleon mass $m_{N}$, that is finite (does not vanish) in the chiral limit~\cite{gasser2} and breaks the homogeneity of the amplitudes in the small scale $p/\Lambda_\chi$. 
More precisely, in a Lorentz covariant framework, the loop contributions produce divergent pieces that violate the power counting formula requiring the renormalization of the lower-order LECs. Any of these pieces will be denoted in the following as a power-counting breaking term (PCBT).

The non-relativistic treatment of the baryon fields leading to Eq.~(\ref{n.def}) is implemented from the outset and in a systematic fashion within the HBChPT formalism~\cite{jenkins}. 
As in the theory for mesons, the renormalization of the LECs in HBChPT can be completed order by order according to Eq.~(\ref{n.def}), although at the price of loosing manifest Lorentz covariance. 
Moreover, as it was explained above, the non-relativistic expansion of the baryon propagators performed in the HBChPT scheme alters the analytic properties that a theory with dynamical nucleons should have, causing problems of convergence in some parts of the low-energy region.

These problems can be overcome using modern manifestly covariant approaches. 
A consistent organization of the chiral power counting in covariant BChPT arises from the following crucial observation: The leading non-analytical behavior of the baryonic loop graphs obeys the power-counting formula (\ref{n.def}) and agrees with the one given by HB~\cite{becher,eoms1}. 
This means that all the terms breaking the counting are analytical in quark masses and momenta and, consequently, of the same type as those given at tree-level by the most general chiral Lagrangian. A corollary is that one can trade the power-counting problem of the covariant approach for a renormalization prescription issue.

The EOMS approach is a dimensional regularization scheme in which the bare LECs are adjusted to cancel the PCBTs present in the loop contributions. Notice that in this scheme one preserves exactly the {\it right} \footnote{By {\it right} analytic properties we mean those derived from $S$-matrix theory and implemented automatically in a (Lorentz covariant) quantum field theory of dynamical pions and nucleons.}  analytic properties of the theory, in contrast with the also covariant IR renormalization scheme, in which a resummation of {\it only} the recoil corrections of the positive-energy part (particle) of the propagator is performed. To apply the EOMS scheme it is necessary to calculate analytically the terms coming from the loop integrals that can generate PCBT in the full amplitude. 
The technique we use is explained in detail in \ref{Sec:PCBTs}, and the EOMS renormalization of $m$, $g$ and the $\mathcal{O}(p^2)$ LECs is shown in \ref{Sec:lecsrenormalization}.

\subsection{The $\Delta(1232)$ resonance}
\label{Sec:DeltaIntrosection}

The resonances are an important feature of the low-energy hadronic spectrum and strong-interaction phenomenology. They appear as poles in the complex plane of the scattering amplitude or, more generically, of the QCD correlators. At low energies, the contribution of these poles can be expanded in Taylor series of $p/\delta$ around $p=0$, where $\delta$ is the scale of the mass gap between the ground state and the mass of the resonance. The role of the resonances in ChPT can be understood by means of a chiral effective field theory which includes not only the nucleons but {\it all} the resonances as dynamical degrees of freedom~\cite{ecker}. The conventional chiral Lagrangian is recovered by integrating out these resonances one-by-one and storing the information of their effects into the LECs. In fact, one can set the so-called Resonance Saturation Hypothesis (RSH) stating that all the short-range nature of the QCD interactions at low-energies is 
mediated by the resonances~\cite{ecker,aspects}. On the other hand, in cases where the mass gap $\delta$ is not large enough, the $p/\delta$ expansion has a small radius of convergence, ruining the behavior of the chiral series. In this case, integrating out the resonance fields is not justified, so they ought to remain as genuine dynamical degrees of freedom in the theory. A remarkable example of this problem is the effect of the $\Delta(1232)$ in the $\pi N$ system for which $\delta\simeq m_\Delta-m_N\simeq300$ MeV, that is well within the expected region of validity of the chiral expansion.

Treating the $\Delta(1232)$ as an explicit degree of freedom in a ChPT setup introduces two types of difficulties. The first one is related to the appearance of the new scale $\delta$ and, thus, to the power counting associated to the Feynman diagrams including resonance lines. A power-counting method that takes into account the fact that $M_\pi<\delta<\Lambda_\chi$ is the $\delta$-counting~\cite{pascalutsa-phillips}. In this method, one employs the power-counting formula Eq.~(\ref{n.def}) for the resonant contributions but using the assignment $\delta\sim p^{1/2}$. This means that the $\Delta$ propagators, which count as $\sim 1/\delta$, receive a suppression of $\mathcal{O}(p^{1/2})$ with respect to the nucleon propagators. Although in the $\delta$-counting the expansion parameter is $\delta/\Lambda_\chi\simeq0.3$, the convergence of the chiral series is expected to be faster than in the theory without the $\Delta$. Notice also that this counting is only 
valid for the description of energies and pion masses below the scale $\delta$, so it is not well suited for LQCD extrapolations or for describing the resonance region. In the latter case, the $\delta$-counting changes to take into account the prominence of the $\Delta$-pole~\cite{pascalutsa-phillips,Pascalutsa:2006up}.
 
The second issue is related to the representation of the resonances and the construction of suitable chiral Lagrangians. The $\Delta(1232)$ is a spin-3/2 resonance that can be described in terms of an isospin multiplet of Rarita-Schwinger (RS) fields $\Delta_\mu=(\Delta^{++},\Delta^{+},\Delta^0,\Delta^{-})_\mu^{T}$, where $\mu$ is the Lorentz index. The free RS Lagrangian is
\begin{equation}
\mathcal{L}_{3/2}=\bar{\Delta}_{\mu}\left(i\gamma^{\mu\nu\rho}\partial_\rho-m_\Delta\gamma^{\mu\nu}\right)\Delta_\nu,\label{Eq:DeltaFreeLag}
\end{equation}
where $\gamma^{\mu\nu}$ and $\gamma^{\mu\nu\rho}$ are the anti-symmetric combinations of Dirac matrices. A very well known difficulty of a quantum field theory of high-spin particles concerns the $consistency$ problem, which is due to the fact that the fields we use to represent these particles contain more components than physical degrees of freedom. For instance, the RS field is an object with 16 components (4 of the spinor times 4 of the Lorentz index) of which only 8 (4 in case of a massless field) correspond to the physical spin-3/2 particle and antiparticle. The Euler-Lagrange equations derived from Eq.~(\ref{Eq:DeltaFreeLag}) provide the necessary constraints to guarantee, in the free case, that the independent components of the RS field are those corresponding to the physical degrees of freedom~\cite{Pascalutsa:1998pw}. 

Introducing interactions that do not fulfill the right constraints upon the unphysical components of the RS field can lead to well known pathologies (see e.g. Refs.~\cite{Pascalutsa:1998pw,Pascalutsa:2006up} and references therein). In order to tackle this problem, we adopt the {\it consistent} interactions, which are invariant under the transformation $\Delta_\mu(x)\rightarrow\Delta_\mu(x)+\partial_\mu\epsilon(x)$. A remarkable property of these interactions is that they fulfill the same constraints as the free Lagrangian so they only account for the dynamical effects of the physical degrees of freedom~\cite{Pascalutsa:1998pw}. A systematic procedure can be set to construct chiral-invariant Lagrangians describing interactions among pions, nucleons and $\Delta$'s that are also consistent~\cite{Pascalutsa:2006up,Krebs:2008zb}. For instance, the leading $\pi N \Delta$ $chiral$ coupling is given by~\cite{Pascalutsa:1998pw,Pascalutsa:2006up}  
\begin{eqnarray}
&&\mathcal{L}^{(1)}_{\pi N\Delta}= \frac{i\, h_A}{2 f_\pi m_{\Delta}}\;\left(\partial_\rho\bar{\Delta}_\mu\right)T^{\dagger a}\gamma^{\rho\mu\nu}N\partial_\nu \pi^a+{\rm h.c.},\label{Eq:CnsBTLag}
\end{eqnarray}
where $T^a$ are the spin-3/2 $\longrightarrow$ spin-1/2 projectors, which verify $T^a T^{\dagger b}=\delta^{ab}-\tau^a\tau^b/3$. The parameter $h_A$ is the $N\Delta$ axial coupling, which is poorly determined but related with the $\pi N \Delta$ coupling through the off-diagonal Goldberger-Treiman relation \cite{Geng:2008bm}. 
Using the $\Delta(1232)$ decay rate, corresponding to its Breit-Wigner width, $\Gamma_\Delta=118(2)$~MeV~\cite{PDG}, one obtains the value $h_A=2.90(2)$~\cite{Pascalutsa:2006up}. 

At the order we work, higher-order $\pi N \Delta$ couplings can also contribute. In particular, the following $\mathcal{O}(p^2)$ chiral Lagrangians can be constructed~\cite{Geng:2008bm},  
\begin{eqnarray}
 &&\mathcal{L}^{(2)}_{N\Delta}=\frac{d_3^\Delta}{m_{\Delta}}\;\bar{N}T^a\omega^a_{\mu\nu}\gamma^\mu\gamma^{\nu\rho\sigma}D_\rho\Delta_\sigma-\frac{id_4^\Delta}{m_{\Delta}^2}\;\bar{N}T^a\omega^a_{\mu\nu}\gamma^{\nu\rho\sigma}D^\mu D_\rho\Delta_\sigma+{\rm h.c.},\label{Eq:CnsBTLag2}
\end{eqnarray} 
where $\omega^a_{\mu\nu}=\langle \tau^a [\partial_\mu,u_\nu]\rangle/2$. These Lagrangians are consistent and the on-shell equivalent of those accompanying the LECs $b_3$ and $b_8$ in Ref.~\cite{fettes_ep}.

For the sake of completeness, we also display the propagator of the $\Delta$ field, which is the $d$-dimensional inverse operator of Eq.~(\ref{Eq:DeltaFreeLag}),
\begin{eqnarray}
S_{\mu\nu}(p)&=&-\frac{\slashp+m_{\Delta}}{p^2-m_{\Delta}^2+i\epsilon}\left[
g_{\mu\nu}-\frac{1}{d-1}\gamma_\mu\gamma_\nu\right.\nonumber\\
&&\left.-\frac{1}{(d-1)m_{\Delta}}\left(\gamma_\mu\, p_\nu-\gamma_\nu\, p_\mu\right)
-\frac{d-2}{(d-1)m_{\Delta}^2} p_\mu p_\nu \right].
\label{Eq:RSpropagator}
\end{eqnarray}
Notice that this propagator does not include the width of the resonance, which should be incorporated consistently order by order in the calculation of the specific Green functions.  

\subsection{Isospin breaking}
\label{Sec:isospinbreakingcorrections}

Isospin symmetry implies a degenerate mass in the nucleon isospin doublet.
However, the mass difference between the $u$ and $d$ quarks introduces a difference of $2.5$~MeV in the neutron mass respect to the proton one. This splitting is compensated by electromagnetic interactions that rises the proton mass in $1.2$~MeV. 
Thus, the overall correction to the neutron mass over the proton one due to the isospin breaking effects is of about $1.3$~MeV \cite{WeinbergVol2}. Regarding the pions, the situation is reversed. In that case, the electromagnetic interaction produces the dominant isospin breaking
corrections, while the quark mass difference starts to contribute at order $(m_u-m_d)^2$ \cite{fettesIsospin}. This gives a total splitting of the charged pions respect to the neutral one of approximately 5 MeV. These observations leads to the question about the impact of isospin violation in strong interaction processes like the one we are considering here. In fact, $\pi N$ scattering is an excellent test ground for studying this question. High-quality data opened the possibility of studying the isospin breaking corrections from the experimentally accessible reactions $\pi^+ p$ and $\pi^- p$ scattering and the single charge exchange (SCX) reaction $\pi^-p \rightarrow \pi^0 n$. A study of the isospin and electromagnetic corrections within the framework of ChPT has been performed in Ref.~\cite{fettesIsospin}. In that paper, it is concluded that the maximal effects of isospin breaking in the $\pi N$ scattering amplitude are approximately of the order of $\approx-0.7\%$ for the $S$-waves and of $\sim-4\%$ for the the $P$-waves, 
when CM energies below $\approx 1.11$~GeV are considered. In the present paper we work in the isospin limit and we will assume possible isospin breaking effects of this size when assigning errors to the PWAs that we use in Sec.~\ref{Sec:phaseshifts}.

\section{Calculation of the Scattering Amplitude}
\label{Sec:ScatteringAmp}

In the following we display the details of the calculation of the $\pi N$ scattering amplitude used in this paper. As it has been explained in the introduction, chiral representations of this amplitude have been obtained before in many different approaches and up to different degrees of accuracy. We present a ChPT analysis of the $\pi N$-scattering amplitude up to $\mathcal{O}(p^3)$ accuracy that includes two main improvements over previous work. In the first place, we use Lorentz covariant B$\chi$PT in the EOMS scheme and, secondly, we explicitly include the $\Delta$ in the $\delta$-counting and filtering the unphysical components of the RS spinors via the consistent couplings. We show in the next sections that this approach solves many of the problems reported by previous works. More precisely, it presents a natural convergence of the chiral expansion from the subthreshold region up to energies well above threshold, providing a suitable framework to undertake a model-independent analysis of all the phenomenology related to $\pi N$ 
elastic scattering. We first revisit the chiral expansions up to $\mathcal{O}(p^3)$ of the nucleon mass $m_N$, the nucleon wave-function renormalization $Z_N$ and the axial coupling $g_A$, which are essential ingredients for the respective calculation of the $\pi N$ scattering amplitude. 

\subsection{The nucleon mass, wave-function and axial coupling}      
\label{Sec:MassandgA}

\begin{figure}[ht]
\begin{center}
\epsfig{file=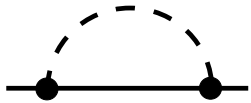,width=.20\textwidth,angle=0}
\caption[pilf]{\protect One-loop self-energy of the nucleon at $\mathcal{O}(p^3)$. \label{selfenergydiagram}}
\end{center}
\end{figure}

The chiral expansions of the nucleon mass and axial coupling are ubiquitous in baryon ChPT literature~\cite{ChPTReviews}. Expressions for $m_N$ and $g_A$ in the EOMS scheme and up to $\mathcal{O}(p^4)$ accuracy can be found in Ref.~\cite{eoms2,Pascalutsa:2005nd} and~\cite{Ando:2006xy,Schindler:2006it} respectively. We take advantage of the re-derivation of these results up to $\mathcal{O}(p^3)$ to illustrate the application of the EOMS scheme in Lorentz covariant BChPT. For the nucleon mass, we need to consider the modifications of the nucleon propagator due to the term $c_1 \la \chi_+\ra \bar{\psi}\psi$ in $\mathcal{L}^{(2)}_{\pi N}$ and the self-energy diagram in Fig. \ref{selfenergydiagram}.
An explicit calculation using the covariant Lagrangian and dimensional regularization gives the following expression for $m_N$,
\begin{align}
 m_N&=m - 4 c_1 M^2- \frac{3g^2 m}{2 f^2}(2\bar{\lambda}(m^2+M^2)) +\frac{3 g^2 m M^2 }{32 \pi^2 f^2}  \\
  &-\frac{3 g^2 M^3}{64 \pi^2 f^2}\Big[ \frac{M}{m}\log\left(\frac{M^2}{m^2}\right) - 4 \sqrt{1-\frac{M^2}{4 m^2}}  \arccos\left( \frac{M}{2 m}\right) \Big], \label{Eq:mNChiralI}
\end{align} 
where
\begin{equation}
\bar{\lambda}= \frac{m^{d-4}}{16\pi^2}\Big\{\frac{1}{d-4}-\frac{1}{2}[\ln4\pi+\Gamma^\prime(1)+1]\Big\}, \label{Eq:div}
\end{equation}
and where we have chosen $\mu=m$ for the sake of simplicity. As it was anticipated in Sec.~\ref{Sec:eomsrenormalization}, the loop contribution is $\mathcal{O}(p^3)$ according to the power-counting formula~(\ref{n.def}) but it contains analytic and divergent pieces which count as $\mathcal{O}(p^0)$ and $\mathcal{O}(p^2)$. The EOMS method recovers the right hierarchy among the different contributions to $m_N$ by choosing a suitable (and systematic) renormalization scheme,   
\begin{align}
 m&\rightarrow m' + \frac{3 g^2 m^3}{2 f^2}(2\bar{\lambda}), \nonumber\\
 c_1&\rightarrow c_1' - \frac{3g^2 m}{8 f^2} (2\bar{\lambda})+\frac{3 g^2m }{128 \pi^2 f^2},  \label{Eq:RenmN}
\end{align}
leading to the final expression
\begin{align}
 m_N&=m' - 4 c_1' M^2 -\frac{3 g^2 M^3}{64 \pi^2 f^2}\Big[ \frac{M}{m}\log\left(\frac{M^2}{m^2}\right)-   4 \sqrt{1-\frac{M^2}{4 m^2}}  \arccos\left( \frac{M}{2 m}\right) \Big].\label{Eq:mNEOMS}
\end{align}
One can explicitly check that this formula fulfills the power-counting and recovers the $\mathcal{O}(p^3)$ HB result in a $1/m$ expansion,
\begin{equation}
m_N= m' - 4 c_1' M^2-\frac{3 g^2 M^3}{32 \pi f^2}+\mathcal{O}(\frac{M^4}{\Lambda_\chi^2m}).  
\end{equation}
On the other hand, the wave function renormalization can be obtained straightforwardly in covariant ChPT, 
\begin{align}\label{ZN}
 Z_N &= 1 -\frac{3 g^2}{2f^2} M^2\bar{\lambda} -\frac{3g^2 M^2}{32\pi^2 f^2 m^2(4m^2-M^2)}\left[2M(M^2-3m^2)\sqrt{4 m^2-M^2}\arccos\left( \frac{M}{2 m}\right)\right.\nonumber\\
&\left.+(M^2-4 m^2)\Big( (2 M^2- 3 m^2)\log\left(\frac{M}{m}\right) -2 m^2 \Big)  \right]. 
\end{align}

\begin{figure}[ht]
\begin{center}
\epsfig{file=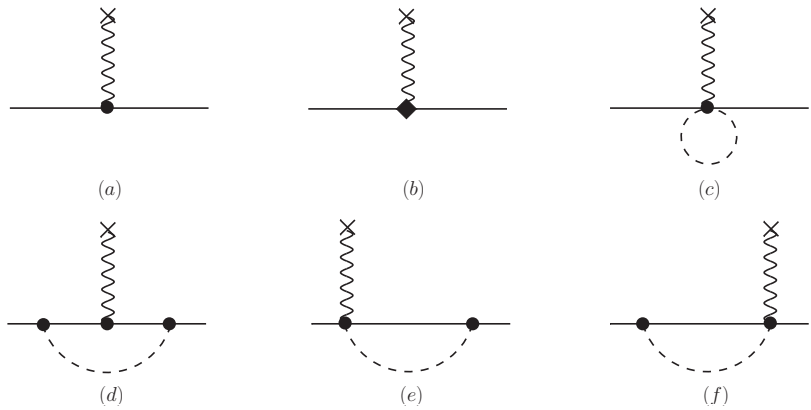,width=0.65\textwidth,angle=0}
\caption[pilf]{
\protect Diagrams that contribute to the nucleon axial form factor up to $\Opt$. 
The wavy line with the cross at the end corresponds to an external axial-vector source. The diamond coupling introduces the LEC $d_{16}$.  \label{axialformfactordiagram}}
\end{center}
\end{figure}

The set of diagrams in Fig~\ref{axialformfactordiagram} contribute to the chiral expansion of axial coupling $g_A$ up to $\mathcal{O}(p^3)$. A direct calculation in dimensional regularization gives, 
\begin{eqnarray}
g_A&=&g+4M^2\,d_{16}-\frac{2g}{f^2} \left(\left(g^2-2\right) m^2+\left(g^2-1\right) M^2\right)\bar{\lambda}-\frac{g^3 m^2}{16 f^2\pi ^2}\nonumber\\
&-&\frac{g M^2}{16 \pi ^2f^2 m^2} \left[\frac{\left(3 g^2+2\right) M^3-8 \left(g^2+1\right) m^2 M}{\sqrt{4 m^2-M^2}} \arccos\left(\frac{M}{2 m}\right)\right.\nonumber\\
&+&\left.   \left(\left(3 g^2+2\right) m^2+\left(\left(4 g^2+2\right) m^2-\left(3 g^2+2\right) M^2\right) \log
   \left(\frac{M}{m}\right)\right)\right],\label{Eq:gAfullR}
\end{eqnarray}
where we have used again $\mu=m$ to fix the renormalization scale. In the first line, we have explicitly shown the analytic and divergent pieces which can be canceled by redefining the bare LECs $g$ and $d_{16}$ in the EOMS scheme,
\begin{eqnarray}
g^{\prime}&=&g-\frac{2g\left(g^2-2\right)m^2}{f^2} \bar{\lambda}-\frac{g^3 m^2}{16 f^2\pi ^2},\nonumber\\
d_{16}^{\,\prime}&=&d_{16}-\frac{g \left(g^2-1\right)}{2f^2}\bar{\lambda}.\label{Eq:EOMSgandd16}
\end{eqnarray}
This leads to a renormalized expression of $g_A$ verifying the power-counting formula~(\ref{n.def}), as it can be explicitly seen by recovering the HB result, modulo an analytic $\sim M^2$ piece~\cite{Ando:2006xy,Schindler:2006it}, in the non-relativistic limit,
\begin{equation}
g_A=g^\prime+4M^2\,d^{\,\prime}_{16}-\frac{g^\prime M^2}{16 \pi ^2} \left(3 g^{\prime\, 2}+\left(4 g^{\prime\,2}+2\right) \log \left(\frac{M}{m}\right)+2\right)+\mathcal{O}(\frac{M^3}{\Lambda_\chi^2m}).
\end{equation}

\subsection{The $\pi N$ scattering amplitude}
\label{Sec:ScatterinAmp2}

\begin{figure}
\begin{center}
\epsfig{file=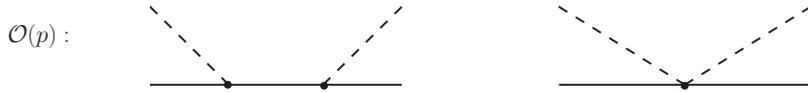,width=.65\textwidth,angle=0}
\caption[pilf]{Diagrams that contribute at LO, $\mathcal{O}(p)$, to the $\pi N$ scattering amplitude. The crossed $u$-channel has also to be considered. These diagrams enter at $\mathcal{O}(p^3)$ with wave-function renormalization contributions and corrections coming from the fact we use physical values $f_\pi$, $g_A$ and $m_N$ (see text for details). \protect  \label{piNdiagramsp}}
\end{center}
\end{figure}

\begin{figure}
\begin{center}
\epsfig{file=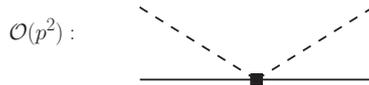,width=.3\textwidth,angle=0}
\caption[pilf]{Diagrams that contribute at NLO to the $\pi N$ scattering amplitude. The vertices denoted by a box introduce the $\mathcal{O}(p^2)$ LECs $c_{1-4}$. \protect  \label{piNdiagramsp2}}
\end{center}
\end{figure}

\begin{figure}
\begin{center}
\epsfig{file=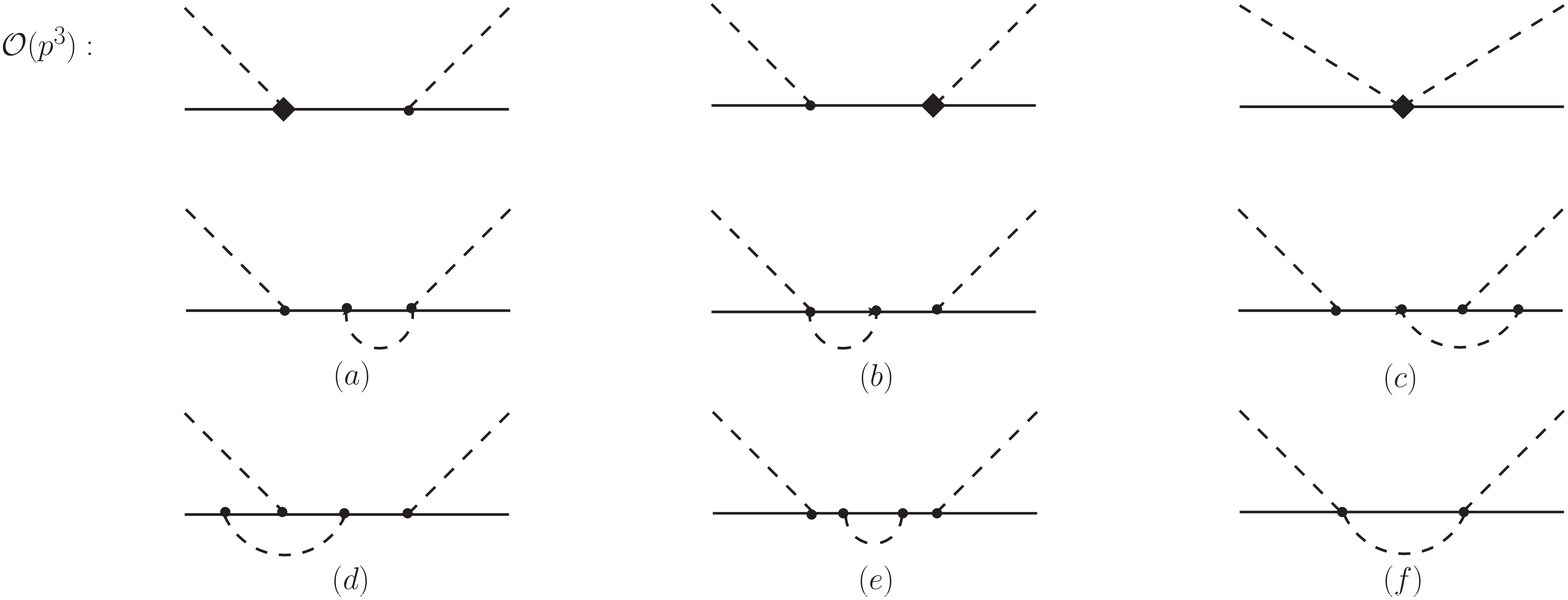,width=0.8\textwidth,angle=0}
\epsfig{file=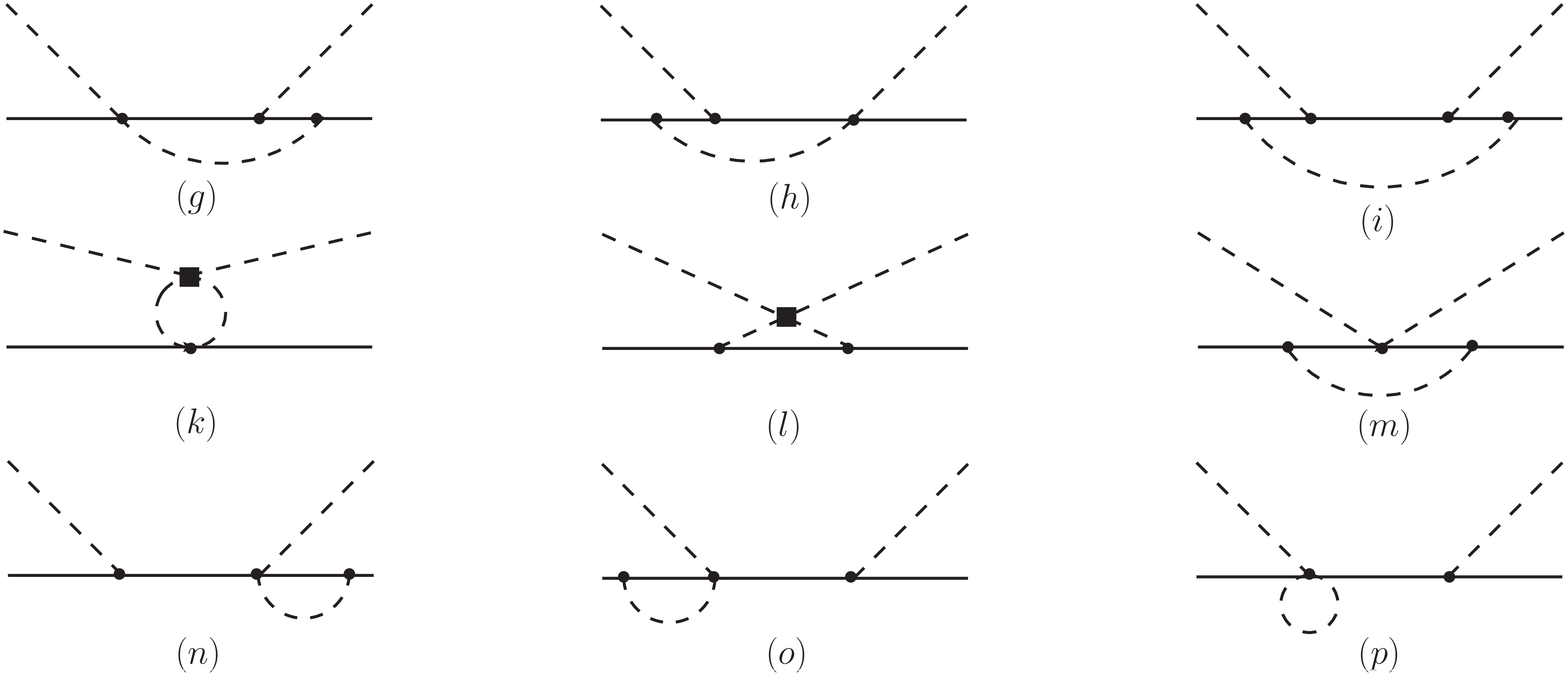,width=0.8\textwidth,angle=0}
\epsfig{file=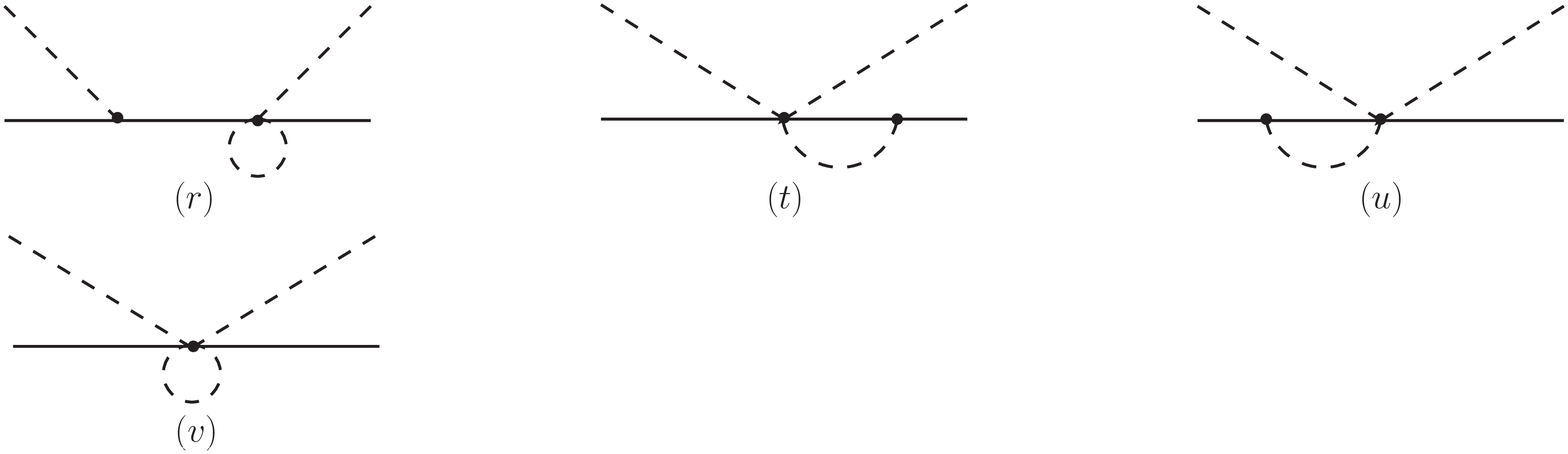,width=0.8\textwidth,angle=0}
\caption[pilf]{Diagrams that contribute at NNLO to the $\pi N$ scattering amplitude. The vertices in the loop diagrams are $\mathcal{O}(p)$ couplings, whereas in the tree level diagrams the diamonds introduce the $\mathcal{O}(p^3)$ LECs $d's$. The diagrams in the crossed $u$-channel has also to be considered with the appropriate topologies. \protect  \label{piNdiagramsp3}}
\end{center}
\end{figure}

According to the power counting formula~(\ref{n.def}), the $\pi N$ scattering amplitude includes the diagrams of Figs.~\ref{piNdiagramsp},~\ref{piNdiagramsp2} and~\ref{piNdiagramsp3} at orders $\mathcal{O}(p)$, $\mathcal{O}(p^2)$ and $\mathcal{O}(p^3)$, respectively. The contributions of the tree-level diagrams to the scattering amplitudes $D^{\pm}(s,t)$ and $B^{\pm}(s,t)$ are shown in the ~\ref{Sec:treelevelcalculations}. The calculation of the loop diagrams is done by means of the Passarino-Veltman decomposition of tensor integrals in terms of 1-, 2-, 3- and 4-points scalar integrals. The results in dimensional regularization, and prior to renormalization, are presented in the ~\ref{Sec:scalarintegrals}. This decomposition of the loop-results provides a simple method to obtain the ultraviolet divergences and PCBTs of any of the loops that contribute to the amplitude in terms of those of the few scalar integrals (see \ref{Sec:PCBTs}). This facilitates the renormalization of the 
final scattering amplitude and the application of the EOMS scheme to preserve, manifestly, the power counting.
       
The renormalization of the scattering amplitude is subtle. Only when $all$ the loop diagrams, including the pion wave function renormalization
\begin{equation}
Z_\pi=1-6\bar{\lambda}\frac{M^2}{f^2}-\frac{M^2}{f^2}\left( 2 \ell_4^r + \frac{1}{16\pi^2}\log\left(\frac{M^2}{\mu^2}\right)\right) + \mathcal{O}(M_\pi^4)\label{Eq:WFRpion}
\end{equation}
and the nucleon wave renormalization of the $\mathcal{O}(p)$ diagrams in Fig.~\ref{piNdiagramsp}, are properly included, one can set a redefinition of the available LECs that casts the amplitude into a finite function fulfilling the power counting. The necessary re-definitions of the bare LECs in the EOMS scheme can be found in the Appendix~\ref{Sec:eomsrenormalization}. A remarkable consequence of chiral symmetry and the consistency of this scheme (as in any other $d$-regularization scheme), is that the redefinition of the bare LEC $c_1$ required to make the $\pi N$ scattering amplitude finite and verifying the power counting is the same as the one demanded by the calculation of $m_N$, in Eq.~(\ref{Eq:RenmN}), and the pion-nucleon sigma term (see Sec.~\ref{Sec:sigmaterm}). An important check of the renormalization and, thus, of our calculation has been to confirm that our results are independent of the renormalization scale.\footnote{It is important to point out that some divergences at $\mathcal{O}(p^4)$ are generated by the 
relativistic structure in the amplitude $D^+$. This requires the inclusion of suitable $\mathcal{O}(p^4)$ LECs, although we do not consider them explicitly. We first check that the residual scale dependence is negligible in a very wide range of values of $\mu$ and then we remove by hand these divergences together with their associated scale dependence.} This feature of the conventional dimensional-regularization schemes contrasts with the IR scheme, which introduces a spurious (higher-order) dependence on $\mu$~\cite{becher}. 

Our calculation is finally given in terms of the physical quantities $f_\pi$, $m_N$ and $g_A$, instead of their chiral-limit values which are not very well known. This procedure implies some reshuffling of ${\cal O}(p^3)$ pieces. In particular, new contributions have to be considered arising from the $\mathcal{O}(p)$ diagrams in Fig.~\ref{piNdiagramsp} when expressing the renormalized $m$, $g$ and $f$ in terms of the physical values $m_N$, $g_A$ and $f_\pi$. For the former two, we use the EOMS expressions derived in Eqs.~(\ref{Eq:mNEOMS}) and~(\ref{Eq:gAfullR},~\ref{Eq:EOMSgandd16}) respectively. For $f_\pi$ we use the conventional expression        
\begin{equation}
f_\pi=f\left\{ 1+\frac{M^2}{f^2}\left[\ell_4^r- \frac{1}{16\pi^2}\log\left(\frac{M^2}{\mu^2}\right)\right]\right\} + \mathcal{O}(M_\pi^4).\label{Eq:fpi} 
\end{equation}
It is important to stress that we strictly keep the ${\cal O}(p^3)$ contributions stemming from these re-definitions (e.g. in $g^2$ and $m^2$ entering the Born terms), so we avoid introducing any high-order renormalization scale dependence. As a result, all the dependence on the LECs $\ell_4^r$ and $d^{\,\prime}_{16}$ disappear from the chiral amplitude. On the other hand, the physical values can be directly used for the $\Opd$ and $\Opt$ scattering amplitude because chiral corrections are of higher order. 

For the numerical evaluation of the scattering amplitude, we programmed the scalar loop integrals and checked thoroughly the numerical results provided by LoopTools~\cite{looptools}. All the final numerical results and fits to PWAs are done with these subroutines. We use the numerical values $f_\pi=92.4$~MeV, $M_\pi=139$~MeV, $m_N=939$~MeV and $g_A=1.267$.

\begin{figure}
\begin{center}
\epsfig{file=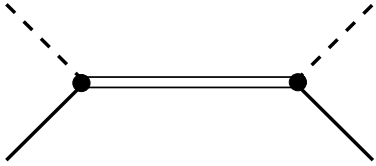,width=0.3\textwidth,angle=0}
\caption[pilf]{\protect Explicit contributions of the $\Delta(1232)$ resonance to the scattering amplitude up to $\mathcal{O}(p^3)$ in the $\delta$-counting. The crossed diagram is also included.  \label{piNdiagramsDelta}}
\end{center}
\end{figure}

Explicit contributions of the $\Delta(1232)$ appear up to $\mathcal{O}(p^3)$ only through the Born-term diagram, Fig.~\ref{piNdiagramsDelta}, and its crossed topology. At $\mathcal{O}(p^{3/2})$, one needs to include the terms given by the monomials in Eq.~(\ref{Eq:CnsBTLag}). At $\mathcal{O}(p^{5/2})$, Lagrangians in Eq.~(\ref{Eq:CnsBTLag2}) have also to be considered.  However, it has been recently shown that the latter couplings are redundant in the non-relativistic limit as their contributions can be absorbed in a redefinition of the coupling $h_A$ and the LECs $c_{1-4}$~\cite{long}. Nevertheless, we evaluate explicitly their contribution in order to numerically check the reach of these conclusions in a Lorentz-covariant framework. The results of these diagrams can be found in the ~\ref{Sec:treelevelcalculations}. Finally notice that loop diagrams involving $\Delta$ lines start at $\mathcal{O}(p^{7/2})$ and so they are beyond the accuracy considered in this paper.

\section{Description of partial wave phase shifts}
\label{Sec:phaseshifts}

In order to fix the LECs that appear in the Lagrangian and extract the physical information contained in the $\pi N$ scattering amplitude, we consider the $\pi N$ phase shifts provided by three different PWAs: 
The PWA of the Karlsruhe-Helsinki group~\cite{KA85} (KA85), the current solution of the George Washington University group~\cite{WI08} (WI08), and the low energy phase shift analysis of the Matsinos' group~\cite{EM06} (EM06). 

The aim of partial wave analysis is to determine the hadronic phase shifts\footnote{In order to obtain the hadronic amplitude is necessary to take into account the Coulomb and electromagnetic corrections, which is are usually treated following~\cite{tromborg}, although EM06 implements these corrections in a different fashion (see Refs.~\cite{EM06,EM02}).} from differential cross sections and polarization data. They use the available database of $\pi^+ p$ and $\pi^- p$ scattering together with the SCX reaction $\pi^- p\rightarrow \pi^0 n$. Assuming isospin invariance, all these reactions can be described by four invariant amplitudes, e.g. $A^\pm(s,t)$ and $B^\pm(s,t)$. One needs further theoretical constraints, in addition to unitarity (except near threshold), to provide a unique representation of the amplitude from the data~\cite{hoehler}. Such constraints are provided by fixed-$t$ analyticity that, together with isospin invariance, are strong enough to resolve the ambiguities of the 
PWAs. On the other hand, there are still uncertainties resulting from experimental errors and discrepancies among different data sets (which are frequently more important). 

In what concerns the PWAs used here, both KA85 and WI08 assume fixed-$t$ analyticity. While KA85 uses an old data set~\cite{KA85data}, WI08 employs the set in~\cite{WI08data}, which contains data collected in modern experiments. A very different approach is followed by EM06 which consists of a low-energy ($\sqrt{s}\lesssim 1.16$~GeV) phase shift analysis that employs hadronic potentials corrected electromagnetically~\cite{EM06}. In contrast with the methodology followed in KA85 and WI08, the Matsinos' analysis exclusively considers data in the elastic region.\footnote{It is worth mentioning that the Matsinos' group recently updated their PWA~\cite{EM12-1}. The new results are very similar to the ones obtained in Ref.~\cite{EM06}, which are those employed in this work.} Both EM06 and WI08 have in common the inclusion of the new data collected along the last 20 years in meson factories. Studying these three different solutions provides a handle on systematic discrepancies that could arise from the different methodologies or the data 
sets employed in each of these analyses.

After this brief introduction on the PWAs considered in this paper we proceed with the explanation of our fitting methodology. Our fits use Eq.~(\ref{des-pert}) for the calculation of the phase shifts, given that our calculation is perturbative. We also use the following $\chi^2$
\begin{align*}
 \chi^2=\sum_i \frac{(\delta_i-\delta_i^{th})^2}{\text{err}(\delta_i)^2},
\end{align*}
where $\delta_{th}$ corresponds to the theoretical phase shift, while $\delta$ and $\text{err}(\delta)$ denote the phase shifts provided by the PWAs and their errors, respectively. Since KA85 and WI08 do not give error for their phase shifts, we take for $\text{err}(\delta)$ a sum in quadrature of a systematic error ($e_s$) and a relative error ($e_r$), as was done in Ref.~\cite{nuestroIR}. For the systematic error we take the value $e_s=0.1$~degree, which is a value typically smaller than the value of the phase shifts. The introduction of a systematic error is advisable because avoids giving an excessive weight to the threshold region, where there is no experimental data. On the other hand, we take $e_r=2\%$ to take into account both the isospin breaking effects and the theoretical error coming from higher order corrections of $\Opc$, which are suppressed respect to the leading order by a factor $(p/\Lambda_\chi)^3$. It is important to stress that a reasonable variation of these values does not affect the conclusions 
presented in this work. The specific choice we take is quite conservative if we compare with the errors assigned in the EM06 analysis. We consider two different approaches, the one without explicit $\Delta$ degrees of freedom (that we will denote as $\slashed{\Delta}$-ChPT) and the one where we include the $\Delta$ as specified above (denoted as $\Delta$-ChPT).

\newpage

\subsection{$\slashed{\Delta}$-ChPT}
\label{Sec:deltalessfits}

\begin{figure}
\begin{center}
\epsfig{file=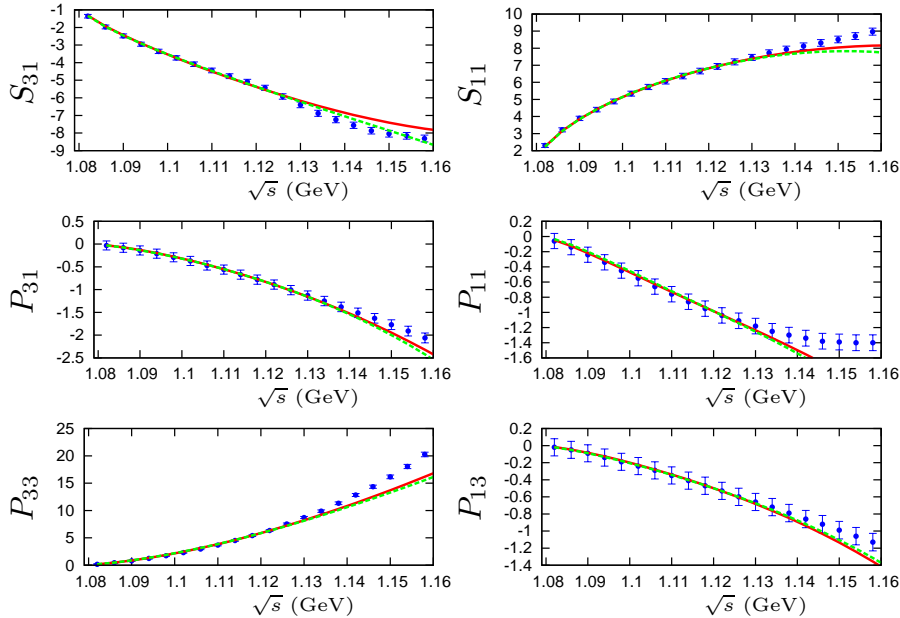,width=.71\textwidth,angle=0}
\caption[pilf]{\protect \small Fits to KA85~\cite{KA85} with $\slashed{\Delta}$-ChPT. The (red) solid lines correspond to the EOMS result and the (green) dashed ones to IR. Both fits are performed up to $\sqrt{s}_{max}=1.13$~GeV. \label{KA85-pert-strategyI}}
\end{center}
\end{figure}

\begin{figure}
\begin{center}
\epsfig{file=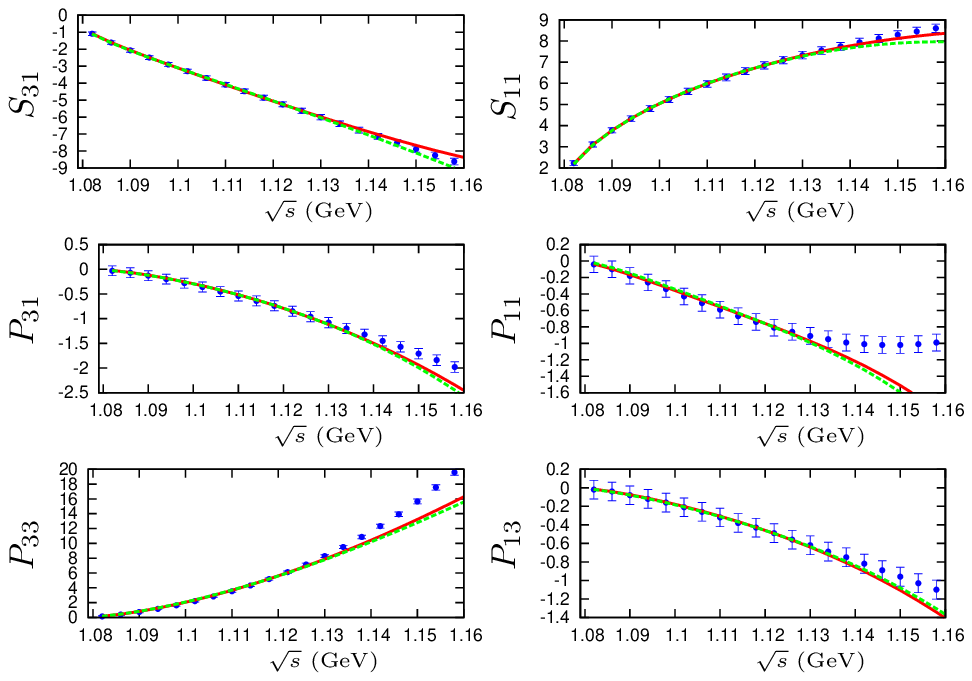,width=.71\textwidth,angle=0}
\caption[pilf]{\protect \small Fits to WI08~\cite{WI08} with $\slashed{\Delta}$-ChPT. The (red) solid lines correspond to the EOMS result and the (green) dashed ones to IR. Both fits are performed up to $\sqrt{s}_{max}=1.13$~GeV. \label{WI08-pert-strategyI}}
\end{center}
\end{figure}

\begin{figure}
\begin{center}
\epsfig{file=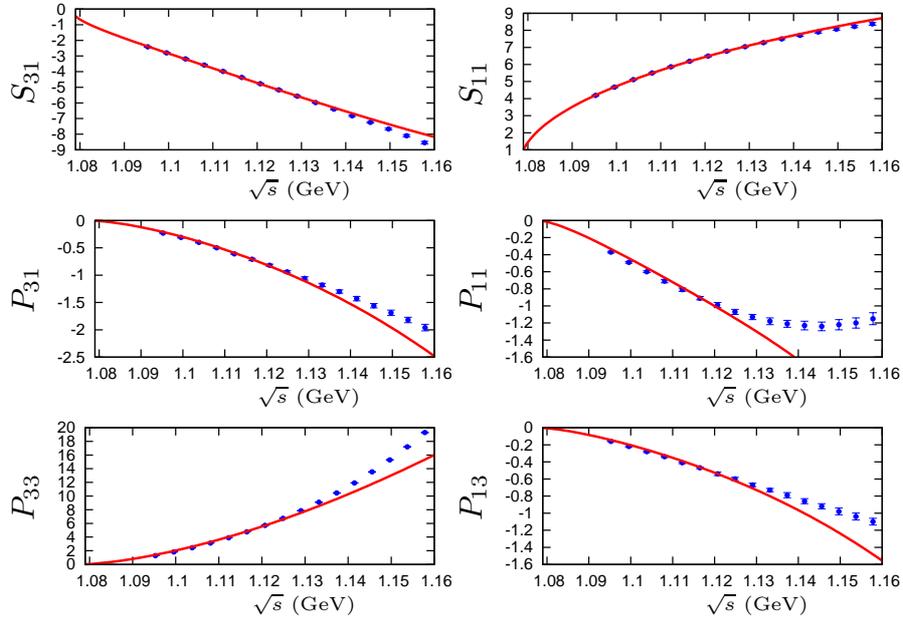,width=.71\textwidth,angle=0}
\caption[pilf]{\protect \small Fits to EM06~\cite{EM06} with $\slashed{\Delta}$-ChPT. Fits are performed up to $\sqrt{s}_{max}=1.13$~GeV. \label{EM06-pert-strategyI}}
\end{center}
\end{figure}

First, we compare $\slashed{\Delta}$ results that stem from our calculation in the EOMS scheme with those obtained in HB~\cite{fettes3} and IR~\cite{nuestroIR} at $\mathcal{O}(p^3)$. We fit the phase shifts up to energies $\sqrt{s}_{max}=1.13$~GeV taking only the pion and the nucleon as the relevant degrees of freedom. In Figs. \ref{KA85-pert-strategyI}, \ref{WI08-pert-strategyI} and \ref{EM06-pert-strategyI} we see qualitatively a good agreement between the EOMS and IR results for all the partial waves. On the other hand, in case of the fits to the EM06 solution (Fig. \ref{EM06-pert-strategyI}), there is not published IR result to compare with.

In Table~\ref{LECs-strategyI} we show the values of the LECs resulting from the fits using the EOMS scheme (columns $2 \textendash 4$) together with the ones obtained using the IR prescription~\cite{nuestroIR} (columns $5 \textendash 6$) and within the HBChPT formalism (last column). The fits are performed including the PWAs phase shifts up to $\sqrt{s}_{max}=1.13$ GeV and the error quoted for the LECs in the EOMS results is determined, as in Ref.~\cite{nuestroIR}, by adding in quadrature the {\it statistical} uncertainty resulting from this fit to the spread of values obtained by varying the maximum energy $\sqrt{s}_{max}=(1.11,1.12,1.13)$~GeV in the fit. Notice that the fit to the EM06 solution has a very poor quality in terms of the $\chi^2$, and we only list the results at $\sqrt{s}_{max}=1.13$ GeV with the corresponding statistical uncertainty. This difficulty can be traced back to the problems of $\slashed{\Delta}$-ChPT to describe the $P_{33}$ partial wave (see 
Sec. \ref{Sec:convergenceofthechiralseries1}). The large value of the $\chi^2_{\rm{d.o.f}}$ in this case affects considerably the theoretical prediction of the rest of the partial waves, in particular the $P_{11}$. 

From the comparison of the values listed in Table~\ref{LECs-strategyI}, we can see that those determined in the EOMS scheme and using the different PWAs are compatible within errors. On the other hand, the values of the LECs extracted from a given PWA and using either the EOMS or IR are also consistent with each other and with those reported in HB~\cite{fettes3}. Regarding the quality of the fits, in terms of the  $\chi^2_{\rm d.o.f.}$, let us stress that the EOMS and IR fits follow the same strategy, whereas the HBChPT result is taken from Ref.~\cite{fettes3},  which uses a different error assignment. 

\begin{table}\small
 \begin{center}
\begin{tabular}{|r|r|r|r|r|r|r|}
\hline
\small{LEC}       & KA85	        &    WI08         & EM06		                         & KA85-IR			    & WI08-IR					  & HBChPT			       \\
                  &  $\slashed{\Delta}$-ChPT & $\slashed{\Delta}$-ChPT  & $\slashed{\Delta}$-ChPT            &  \cite{nuestroIR}                  &  \cite{nuestroIR}  	                  &   \cite{fettes3}  \\
\hline							  			    													
$c_1$              &  $-1.26(14)$   &  $-1.50(7) $  & $-1.47(2)$	                            &  	-1.08(15)    &  -1.32 (14) & $(-1.71,-1.07)$		\\   
$c_2$              &  $ 4.08(19)$  &  $3.74(26) $   & $3.63(2) $	                             & 	4.7(6)	    & 4.3(6)	  & $(3.0,3.5)$ 		\\
$c_3$              &  $-6.74(38)$  &  $-6.63(31) $  & $-6.42(1)$	                             & 	-7.0(7)    & -6.9(6)  & $(-6.3,-5.8)$	\\
$c_4$              &  $3.74(16)$    &  $3.68(14) $   & $3.56(1)$	                       &3.72(32) &  3.66(31)		 & $(3.4,3.6)$ 	      \\
\hline		  					  			    														
$d_1+d_2$          &  $3.3(7)$    &  $3.7(6)$   &  $ 3.64(8)$	                         & 4.7(1.0)	    & 5.1(8)			  & $(3.2,4.1)$ 	  \\
$d_3$              & $-2.7(6) $   &  $-2.6(6)$  & $-2.21(8) $	                        &-3.9(9)	    & -3.8(6)		 & $(-4.3,-2.6)$      \\
$d_5$              & $0.50(35) $    &  $-0.07(16)$  & $ -0.56(4)$	                        &-0.3(5)	   & -0.83(24)				  & $(-1.1,0.4)$		     \\
$d_{14}-d_{15}$    & $-6.1(1.2)$    &  $-6.8(1.1)$  & $ -6.49(2)$	                         &  -5.4(1.3)     & -6.2(1.1)			  & $(-5.1,-4.3)$			       \\
$d_{18}$           & $-3.0(1.6)$    &  $-0.50(1.8)$  & $ -1.07(22)$	                          & -3.5(2.0)     &-1.1(1.5) 	  & $(-1.6,-0.5)$			    \\ 
\hline
$\chi^2_{\rm d.o.f.}$  &       $0.38$         &     $0.23$          &	  $25.08$                           & 0.45				    & 0.34				  & $(0.83-1.34)$      \\
\hline
\end{tabular}
\caption[pilf]{\small Comparison between LECs in the different approaches of $\slashed{\Delta}$-ChPT up to $\Opt$. 
The results of the present paper are given in the columns $2 \textendash 4$. The IR results~\cite{nuestroIR} are shown in columns $5 \textendash 6$, and the last column corresponds to the ones obtained in HBChPT~\cite{fettes3}.
The errors shown for the EOMS case are obtained as it is explained in the text.
  \label{LECs-strategyI}}
\end{center}
\end{table}

\subsubsection{Convergence of the chiral series}
\label{Sec:convergenceofthechiralseries1}

\begin{figure}
\begin{center}
\epsfig{file=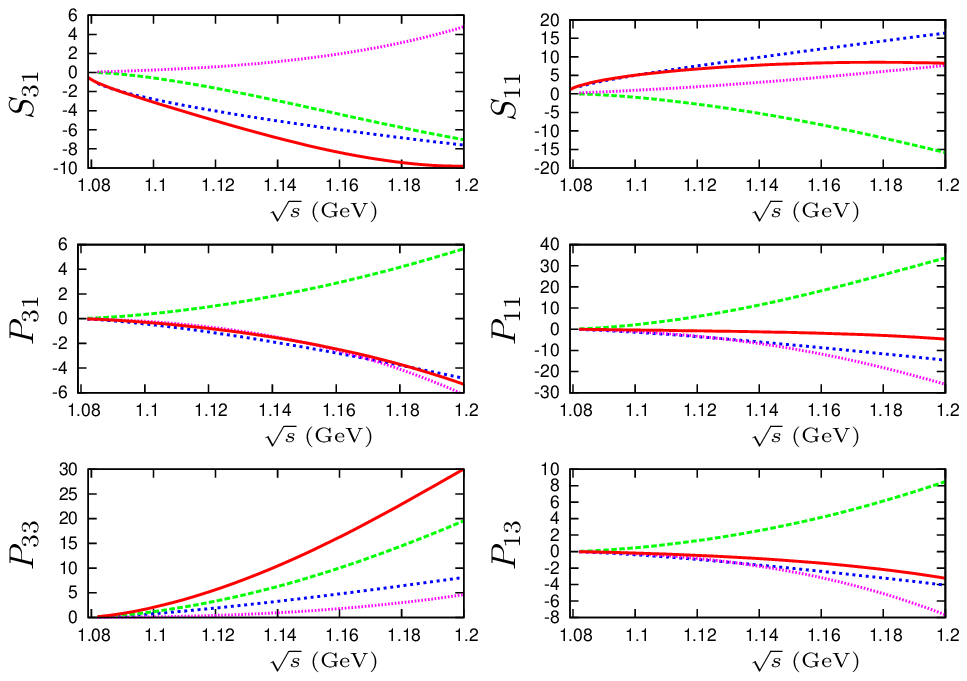,width=.7\textwidth,angle=0}
\caption[pilf]{\protect \small Convergence of the chiral series for the fit to WI08 in $\slashed{\Delta}$-ChPT. The short-dashed (blue), dashed (green), dotted (pink) and solid (red) lines correspond to the contributions of the $\Op$, $\Opd$, $\Opt$ and the total sum, respectively.   \label{convergence-WI08-fit3}}
\end{center}
\end{figure}

It is interesting to study the chiral expansion of the scattering amplitude calculated up to ${\cal O}(p^3)$ in the EOMS scheme by looking at the contribution of each order to the different PWA phase shifts. In Fig.~\ref{convergence-WI08-fit3}, we plot the respective contributions to the total result (red line) for the $\slashed{\Delta}$-ChPT fits to WI08. (Similar plots can be obtained for the KA85 and EM06 solutions.) This can be directly compared with Fig. $7$ of Ref.~\cite{fettes3} (HB) and Fig. $3$ of Ref.~\cite{nuestroIR} (IR) since both are also $\Deltaless$ fits to PWAs in the isospin limit. As it is discussed in these references, we see that there exists a cancellation between $\mathcal{O}(p^2)$ and $\mathcal{O}(p^3)$ contributions in almost all the partial waves. Furthermore, the size of the $\mathcal{O}(p^3)$ contributions can be very large and comparable to those given by the lower-order terms even at very low energies above threshold. Thus, the applicability of $\slashed{\Delta}$-ChPT to describe the PWA phase shifts at $\mathcal{O}(p^3)$ is questionable. Nevertheless, the HBChPT study of Ref.~\cite{fettes4} at ${\cal O}(p^4)$ obtains that the corrections to the ${\cal O}(p^3)$ result are more modest than those from the ${\cal O}(p^2)$ to the ${\cal O}(p^3)$ calculation, suggesting convergence.   

\subsection{$\Delta$-ChPT}
\label{Sec:Deltafits}

\begin{figure}
\begin{center}
\epsfig{file=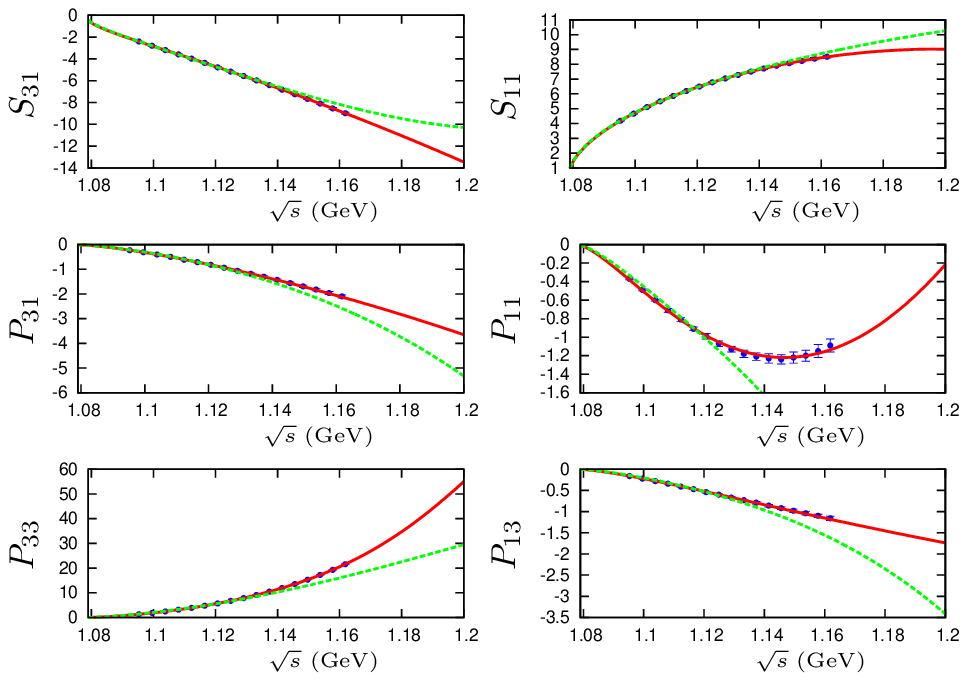,width=.71\textwidth,angle=0}
\caption[pilf]{\protect \small Fits to the EM06~PWA\cite{EM06} with $\Delta$-ChPT (red solid lines) compared with the $\slashed{\Delta}$-ChPT result (green dashed lines). The $\Delta$-ChPT fits are performed up to $\sqrt{s}_{max}=1.16$~GeV. \label{EM06-pert-strategyII}}
\end{center}
\end{figure}

As it has been argued in Sec.~\ref{Sec:DeltaIntrosection}, integrating out the $\Delta(1232)$-resonance from the chiral effective field theory is not well justified for the description of elastic $\pi N$ scattering above threshold. In this section, we study this in more detail by applying $\Delta$-ChPT within the EOMS scheme to the description of the PWA phase shifts. In the following, we present the results neglecting the $\mathcal{O}(p^{5/2})$ contributions of the $\pi N \Delta$ couplings $d_3^\Delta$ and $d_4^\Delta$ (see Sec.~\ref{Sec:DeltaIntrosection}). A discussion of the effects of these couplings is presented below. In Figs.~\ref{KA85-pert-strategyII}, \ref{WI08-pert-strategyII} and \ref{EM06-pert-strategyII} we show the results of the fits of $\Delta$-ChPT to the KA85, WI08 and EM06 PWAs, respectively, following the same procedure for the fits as done in the $\Deltaless$ theory. For the sake of comparison, we also plot the results previously obtained in $\slashed{\Delta}$-ChPT. In Table~\ref{LECs-strategyII}, we list the 
resulting values of the LECs. The central values stem from taking an average of the values obtained varying $\sqrt{s}_{max}$ from 1.14 to 1.20 GeV (up to its maximum 1.16 GeV in case of the EM06 solution), in intervals of 10~MeV.  For the $\mathcal{O}(p^3)$ LECs $d$'s, we take averages weighted by the corresponding statistical uncertainty to reflect the fact that these parameters are accurately determined only in the fits at the higher energies (all the values from the fits at lower-energies are perfectly consistent within errors). The errors are obtained adding in quadrature the statistical error at $\sqrt{s}_{max}=1.20$~GeV and the one resulting from the spread of central values at the different $\sqrt{s}_{max}$, which, in general, gives the larger contribution to the error.

\begin{figure}
\begin{center}
\epsfig{file=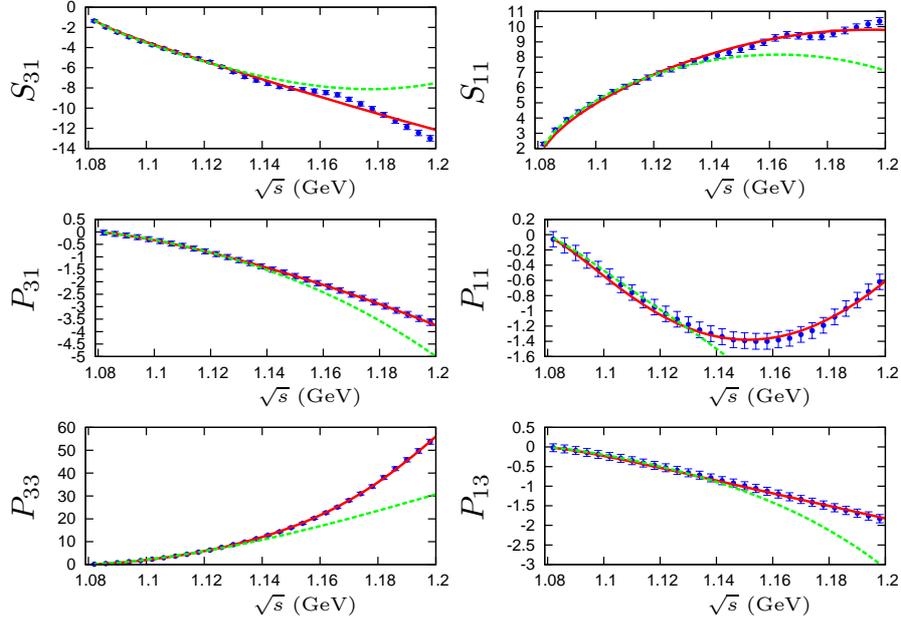,width=.71\textwidth,angle=0}
\caption[pilf]{\protect \small Fits to the KA85~PWA\cite{KA85} with $\Delta$-ChPT (red solid lines) compared with the $\slashed{\Delta}$-ChPT result (green dashed lines). The $\Delta$-ChPT fits are performed up to $\sqrt{s}_{max}=1.20$~GeV. \label{KA85-pert-strategyII}}
\end{center}
\end{figure}

\begin{figure}
\begin{center}
\epsfig{file=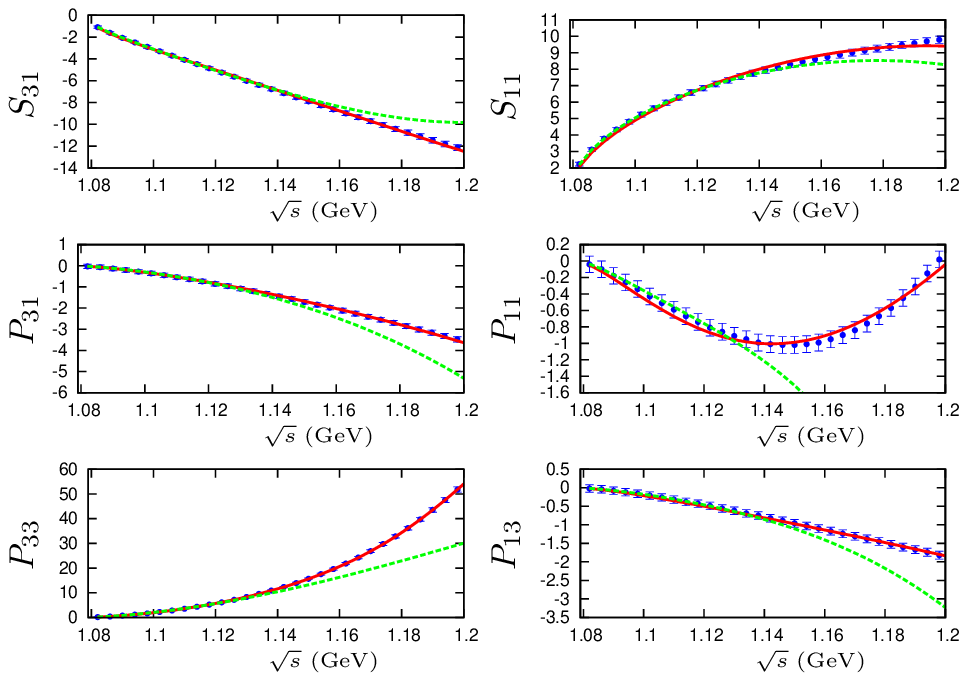,width=.71\textwidth,angle=0}
\caption[pilf]{\protect \small Fits to the WI08~PWA\cite{WI08} with $\Delta$-ChPT (red solid lines) compared with the $\slashed{\Delta}$-ChPT result (green dashed lines). The $\Delta$-ChPT fits are performed up to $\sqrt{s}_{max}=1.20$~GeV. \label{WI08-pert-strategyII}}
\end{center}
\end{figure}

As it can be deduced from these figures and the table, the inclusion of the $\Delta(1232)$ resonance has an important effect in the description of elastic $\pi N$ scattering, given that now we are able to describe accurately the $S$- and $P$-wave phase shifts up to energies of $\sqrt{s}=1.20$~GeV and with $\chi^2_{\rm d.o.f.}<1$. For the fits to the KA85 and WI08 PWAs, the $\chi^2_{\rm d.o.f.}$ remains well below $1$, whereas for the EM06 the improvement is more drastic since it reduces drastically the $\chi^2_{\rm d.o.f.}$, from $\approx 25$ ($\slashed{\Delta}$-ChPT) to $\approx 0.1$ ($\Delta$-ChPT). This illustrates the relevance of the $\Delta(1232)$ resonance even at very low energies and close to the threshold region.

\begin{table}
 \begin{center}
\begin{tabular}{|r|r|r|r|}
\hline
\small{LEC}       &   KA85	       &   WI08	   & EM06	                         \\
                  &	$\Delta$-ChPT    &	$\Delta$-ChPT	   &	$\Delta$-ChPT	 \\
\hline		   	   																	      
$c_1$             & $-0.80(6)$    &  $-1.00(4)$  &$-1.00(1)$  \\   
$c_2$             & $1.12(13)$    &  $1.01(4)$    &$0.58(3)$   \\
$c_3$             & $-2.96(15)$    & $-3.04(2)$   &$-2.51(4)$   \\
$c_4$             & $2.00(7)$     &  $2.02(1)$  &$1.77(2)$  \\
\hline		   					   													      
$d_1+d_2$         & $-0.15(21)$    & $0.15(20)$    & $-0.36(6)$   \\
$d_3$             & $-0.21(26)$    & $-0.23(27)$   & $0.28(4)$	  \\
$d_5$             & $0.82(14)$    &  $0.47(7)$    & $0.20(3)$	\\
$d_{14}-d_{15}$   & $-0.11(44)$    & $-0.5(5)$   & $0.35(9)$	 \\
 $d_{18}$          & $-1.53(27)$    & $-0.2(8)$   & $-0.53(12)$    \\ 
\hline  
$h_A$             &  $3.02(4)$   &  $2.87(4)$  & $2.99(2)$     \\
\hline
$\chi^2_{\rm d.o.f.}$  &       $0.77$       &     $0.24$          & $0.11$   \\
\hline
\end{tabular}
{\caption[pilf]{\protect \small Values of the LECs in $\Delta$-ChPT. The errors are obtained adding in quadrature the statistical uncertainties at $\sqrt{s}_{max}=1.20$ GeV and the spread of values produced using different values of $\sqrt{s}_{max}$, from $1.14$ to $1.20$~GeV, in intervals of 10~MeV. The $\chi^2_{\rm d.o.f.}$ is obtained at the maximum energies considered, $\sqrt{s}_{max}=1.20$ GeV.
\label{LECs-strategyII}}}
\end{center}
\end{table}

The effect of the $\Delta$ in the LECs is clearly shown in Table \ref{LECs-strategyII}. In general, one notices an important reduction, in absolute value, of these parameters. As we will see below, these new values are more $natural$ than the former ones as they lead to a good convergence of the chiral amplitude. It is interesting to compare the contribution of the $\Delta(1232)$ to the different LECs obtained explicitly from our fits with the one calculated using RSH~\cite{aspects}. An example of this comparison is given in Table \ref{saturation-estimation} for the WI08 fits. We observe that there is a good agreement between our results and the RSH approach, except for $c_1$, where we obtain that its value is shifted by an amount of $\approx 0.5$~GeV$^{-1}$. This can be interpreted as an indication of the fact that the LECs are stabilized once the tree-level $\Delta(1232)$ exchange contributions are taken into account (see also Sec. \ref{Sec:convergenceofthechiralseries2}).  

\begin{table}
 \begin{center}
\begin{tabular}{|c|ccc|c|}
\hline
& KA85&WI08&EM06&RSH\\
\hline
$\tilde{c}_1$&$-0.46$&$-0.50$&$-0.47$&$\sim0$\\
$\tilde{c}_2$&$2.96$& $2.73$&$3.06$&$1.9 \ldots 3.8$\\
$\tilde{c}_3$&$-3.78$&$-3.59$&$-3.91$&$-3.8\ldots -3.0$\\
$\tilde{c}_4$& $1.74$& $1.65$&$1.79$&$1.4 \ldots 2.0$\\
\hline
$\tilde{d}_1+\tilde{d}_2$&3.45&3.55&4.98&--\\
$\tilde{d}_3$&$-2.49$&$-2.37$&$-2.49$&--\\
$\tilde{d}_5$&$-0.32$&$-0.54$&$-0.76$&--\\
$\tilde{d}_{14}-\tilde{d}_{15}$&$-5.99$&$-6.30$&$-6.84$&--\\
$\tilde{d}_{18}$&$-1.47$&0.30&$-0.54$&-- \\
\hline   
\end{tabular}
{\caption[pilf]{\protect \small Estimation of the $\Delta(1232)$ contribution to the $\Opd$ and $\Opt$ LECs in units of GeV$^{-1}$ and GeV$^{-2}$, respectively, based on the results obtained with ($\Delta$-ChPT, Table \ref{LECs-strategyII}) and without ($\slashed{\Delta}$-ChPT, Table \ref{LECs-strategyI}) the inclusion of this resonance as an explicit degree of freedom.
In the last row we show the results of Refs.~\cite{aspects} and~\cite{beche2} employing RSH. \label{saturation-estimation}}}
\end{center}
\end{table}

On the other hand, the results obtained fitting the different PWAs are grossly consistent with each other. Interesting differences can be found though, and these translate into discrepancies in the $\pi N$ phenomenology derived from the various analyses. Note also the larger error obtained in some LECs for the KA85 analysis. This originates from the numerical instabilities in the $S$-waves of this solution, visible in the Fig.~\ref{KA85-pert-strategyII}, and which lead to an over-estimation of the uncertainties for some observables (e.g. scattering lengths) derived from this PWA. The relative stability on the values of the LECs in our $\Delta$-ChPT calculation contrasts very much with the results reported in $\Delta$-HBChPT~\cite{fettes_ep}, in which very large differences among different PWAs were reported, hindering a clear discussion of some related phenomenology, e.g. the pion-nucleon sigma term. For the value of the $\Delta(1232)$ axial coupling, $h_A$, we find 
that the WI08 solution gives a value that is perfectly compatible with the one directly extracted from the $\Delta(1232)$ Breit-Wigner width, $h_A=2.90(2)$. This is the width one should compare with because we are reproducing the phase shifts where one is sensitive to the Breit-Wigner shape of the resonance in the physical $s$-axis. For KA85, one obtains a coupling that is slightly larger, what could be related to the overestimation of the width resulting from this PWA~\cite{KA85}. The EM06 solution also leads to a larger $h_A$ than the one extracted from the width, although, we do not expect the EOMS analysis to describe accurately this quantity as it is focused in describing the phase shifts below the $\Delta$-resonance region. It follows that the WI08 solution is the only one that gives a $\Delta(1232)$ Breit-Wigner width compatible with the value quoted in the PDG. 

We have included the tree-level $\Delta$-contributions generated by the $\mathcal{O}(p^2)$ $\pi N \Delta$ couplings in Eq.~(\ref{Eq:CnsBTLag2}). As it was explained in Sec.~\ref{Sec:ScatteringAmp}, these couplings have been found to be redundant in the non-relativistic expansion since as they can be accounted for by a redefinition of the LECs $h_A$ and $c_{1-4}$~\cite{long}. We expect this reorganization of the chiral expansion to be effective also in the Lorentz covariant case because the leading contributions to the corresponding diagrams are their HB approximations. Nevertheless, we have checked this explicitly including the aforementioned pieces in the fits. We have found that: \textit{\textbf{(i)}} the two LECs are extremely correlated with each other and only one of them can be kept in order to obtain stable fits. (Similar conclusions were derived in Refs.~\cite{fettes3} and~\cite{Geng:2008bm}.)  \textit{\textbf{(ii)}} Large correlations are found between the remaining $d_i^\Delta$ and the LECs $h_A$ 
and $c_{1-4}$. Therefore, the inclusion of the new LEC does not change appreciably the quality of the fits. \textit{\textbf{(iii)}} The values obtained for the $d_i^\Delta$ and different $\sqrt{s}_{max}$ are stable and consistent with zero. In summary, we ratify the conclusions of Ref.~\cite{long}, what suggests that the chiral expansion can be organized also in covariant $\Delta$-ChPT so as to remove these higher-order $N\Delta$ couplings. Consequently, we do not include these contributions in the rest of the present work. 

Finally, we also studied the results of $\Delta$-ChPT within the IR scheme. Although the EOMS and IR representations of the amplitude give a completely equivalent description of the phase shifts near threshold in the $\Deltaless$ case, they give very different results when trying to describe them up to the $\Delta(1232)$ region in $\Delta$-ChPT. For instance, equivalent fits to those performed in the EOMS scheme at $\sqrt{s}_{max}=1.20$ GeV lead to a $\chi^2_{\rm d.o.f.}$ of 4.15 and 1.67 for the KA85 and WI08 PWAs, that are much larger than the ones obtained in EOMS, 0.77 and 0.24 respectively. Furthermore, the description of main observables related to the scattering amplitude, like the pion-nucleon sigma term, the GT discrepancy or the subthreshold coefficients, is not compatible with the results based on dispersive analyses which, on the other hand, are perfectly consistent with those obtained in the EOMS scheme (as we will see in the next sections). These difficulties arise from the IR-regularized loop contributions, which seem to develop some sensitivity to the $u$-channel unphysical cut at the energies reached in the $\Delta$-theory (see also the discussion in Sec.~\ref{Sec:unitarizedamplitudes}). For completeness and in order to justify these conclusions we present in~\ref{App:DeltaIR} a brief summary of main numerical results obtained in the $\Delta$-ChPT within the IR scheme. Given these problems of IRChPT to describe the higher energies considered in $\Delta$-ChPT we focus in the following on the representation of the $\pi N$ scattering amplitude obtained in the EOMS scheme.   

\subsubsection{Convergence of the chiral series}
\label{Sec:convergenceofthechiralseries2}

\begin{figure}
\begin{center}
\epsfig{file=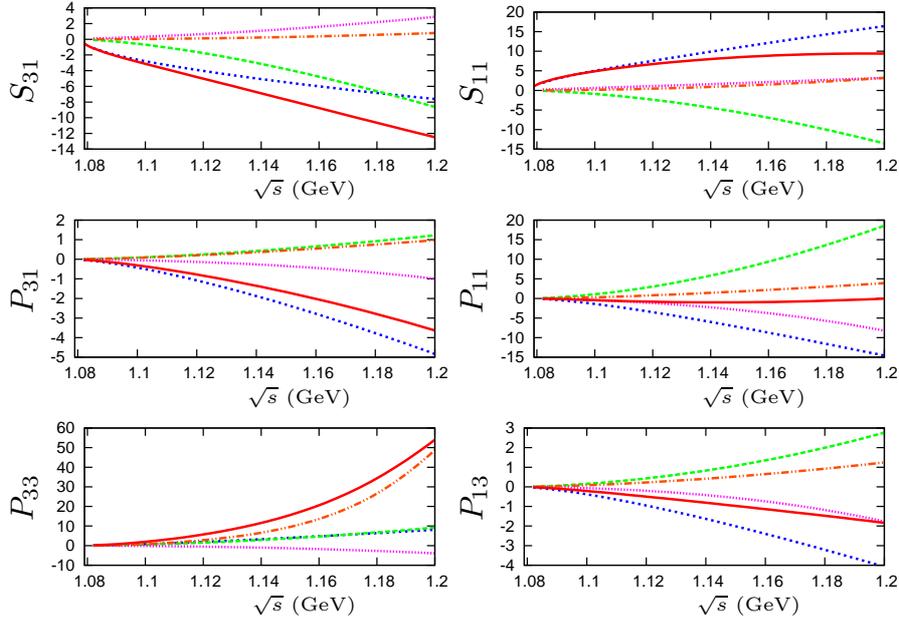,width=.71\textwidth,angle=0}
\caption[pilf]{\protect \small Convergence of the chiral series for the fit to WI08 in $\Delta$-ChPT. The short-dashed (blue), dashed (green), dotted (pink), dash-double-dotted (orange) and solid (red) lines correspond to the contributions of the $\Op$, $\Opd$, $\Opt$, $\Delta(1232)$ and the total sum, respectively. \label{convergence-WI08-fit14}}
\end{center}
\end{figure}

It has been repeatedly argued in this paper that the chiral representation of the $\pi N$ scattering amplitude in $\Delta$-ChPT presents better convergence than in the $\Delta$-less case. Although the dramatic improvement in the description of the different PWA phase shifts achieved in the former approach indicates that this is the case, a conclusion in this regard can be reached only after studying the contributions of the different orders to the amplitude. In Fig.~\ref{convergence-WI08-fit14}, we show this comparison for the  WI08 phase shift. Similar plots can be obtained for the KA85 and EM06 solutions. 

First of all, one confirms the expectation that the $\Delta(1232)$ is the main responsible for the rapid raise of the $P_{33}$ phase shifts. Besides, the $\Delta$-exchange gives a non-negligible contribution to the rest of the $P$-waves. Secondly, the contributions of the $\mathcal{O}(p)$ pieces are significantly more important than the $\mathcal{O}(p^2)$ ones in most part of the low-energy region and most of the partial waves (not for the $P_{11}$ and $P_{13}$ waves, for which these terms largely cancel each other). Nonetheless, the most important observation is that, in $\Delta$-ChPT, the $\mathcal{O}(p^3)$ contributions are completely subleading compared to the LO and NLO terms in all the low-energy region above threshold. In terms of LECs, this means that their values in $\Delta$-ChPT are smaller and more natural than in the $\Deltaless$-ChPT. Furthermore, in Sec.~\ref{Sec:subthresholdregion} we will see that the $\Delta$ is also an essential ingredient to connect, in CM energy $\sqrt{s}$, the 
subthreshold and the threshold regions. This whole picture in $\Delta$-ChPT is consistent with the chiral power counting and it corresponds to what is expected from a well-behaved chiral expansion of the $\pi N$ scattering amplitude. A remarkable consequence of these conclusions is that, while in $\slashed{\Delta}$-ChPT the poor convergence of the chiral expansion forces working in the subthreshold region~\cite{beche2,buettiker}, $\Delta$-ChPT can be applied to study the different phenomenology associated with the $\pi N$-scattering using, exclusively, the experimental information accessible in the physical region.  

\subsection{Unitarized amplitudes}
\label{Sec:unitarizedamplitudes}

Another important comparison between the covariant schemes, EOMS and IR, stems from the resulting scattering amplitudes after taking the perturbative ones as input  for nonperturbative $S$-matrix techniques.  
Proper unitarization approaches take care of the analytic properties associated with the right-hand cut and have proved to be very successful in the description of non-perturbative phenomena in ChPT~\cite{Oller:1998hw,NDdescription}.
In Ref.~\cite{nuestroIR}, the IR representation of the $\pi N$ scattering amplitude was unitarized using an approximate algebraic solution to the N/D method that is obtained by treating crossed channel dynamics~\cite{Oller:1998hw,NDdescription,pin}. Although the covariant amplitude achieved a good description of the phase shifts, the unphysical cut introduced by the IR-method spoiled the description for energies $\sqrt{s}\gtrsim 1.26$~GeV. Given that the EOMS scheme has the right analytical properties, it is interesting to see if we can improve the description of the data, and explore the potential of the unitarization techniques applied on a reliable BChPT kernel. 

We use the same unitarization method of~\cite{NDdescription,pin, plb} in order to compare with the IR approach Ref.~\cite{nuestroIR}. For the $P_{33}$ partial wave we include, in addition, a Castillejo-Dalitz-Dyson pole (CDD) \cite{pin} to take into account the contribution of the $\Delta(1232)$ resonance when unitarizing the amplitude of $\Deltaless$-ChPT.
In this method the unitarized amplitude, $T_{IJ\ell}$, is written as
\begin{align}
 T_{IJ\ell}(s)&=\frac{1}{\mathcal{T}_{IJ\ell}(s)^{-1}+g(s)}  &\text{for $I \neq 3/2$ or $J \neq 3/2$}, \nn \\
 T_{IJ\ell}(s)&=\left(\mathcal{T}_{IJ\ell}(s)^{-1}+\frac{\gamma}{s-s_P}+g(s)\right)^{-1}  &\text{ for $I = 3/2$ and $J = 3/2$}, \label{eq_master}
\end{align}
where $\mathcal{T}_{IJ\ell}$ is the interaction kernel, $\gamma$ and $s_P$ are the residue and pole position of the CDD pole, respectively. The function $g(s)$ corresponds to the unitary pion-nucleon loop,
\begin{align}
 g(s)&=g(s_0)-\frac{s-s_0}{\pi}\int^\infty_{s_{th}} ds' \frac{|\vp|}{8\pi \sqrt{s'}}\frac{1}{(s'-s)(s'-s_0)}\nn\\
     &=\frac{1}{(4\pi)^2}\left\{ a(\mu)+\log\left( \frac{m_N^2}{\mu^2} \right)-\frac{m_N^2-M_\pi^2+s}{2s}\logMpimN\right.\\ 
     &+\frac{|\vp|}{\sqrt{s}}\left[ \log(s-m_N^2+M_\pi^2+2\sqrt{s}|\vp|) + \log(s+m_N^2-M_\pi^2+2\sqrt{s}|\vp|) \right.\nn\\
     &\left. \left.- \log(-s+m_N^2-M_\pi^2+2\sqrt{s}|\vp|) - \log(-s-m_N^2+M_\pi^2+2\sqrt{s}|\vp|)\right]\right\}, \nn
\end{align}
where the subtraction constant is fixed by requiring that the $P_{11}$ unitarized partial wave keeps the nucleon pole at the same position as in the perturbative calculation, i. e.  physical nucleon mass. This condition translates onto the loop function as,

\begin{align}
 g(s=m_N^2)=0.
\end{align}

On the other hand, we follow Ref.~\cite{plb}, as it was done in Ref.~\cite{nuestroIR}, to extract the interaction kernel by matching the chiral amplitude obtained in $\Deltaless$-ChPT with the chiral expansion of Eq.~\eqref{eq_master}  order by order.
Namely, taking into account that $g(s)=\Op$, we use

\begin{align}\label{Tmatching}
T^{(1)}+T^{(2)}+T^{(3)}={\cal T}^{(1)}+{\cal T}^{(2)}+{\cal T}^{(3)}-g(s)\left({\cal T}^{(1)}\right)^2,~
\end{align}
where the superscript $(n)$ refers to the chiral order of the amplitudes.
Matching order by order we obtain the following relations between the chiral amplitude ($T(s)$) and the expansion of the interaction kernel $\mathcal{T}(s)$
\begin{align}
{\cal T}^{(1)}(s)&=T^{(1)}(s),~\nn\\
{\cal T}^{(2)}(s)&=T^{(2)}(s),~\nn\\
{\cal T}^{(3)}(s)&=T^{(3)}(s)+g(s)\left(T^{(1)}(s)\right)^2.~
\label{Tmatching2}
\end{align}

\begin{figure}
\begin{center}
 \epsfig{file=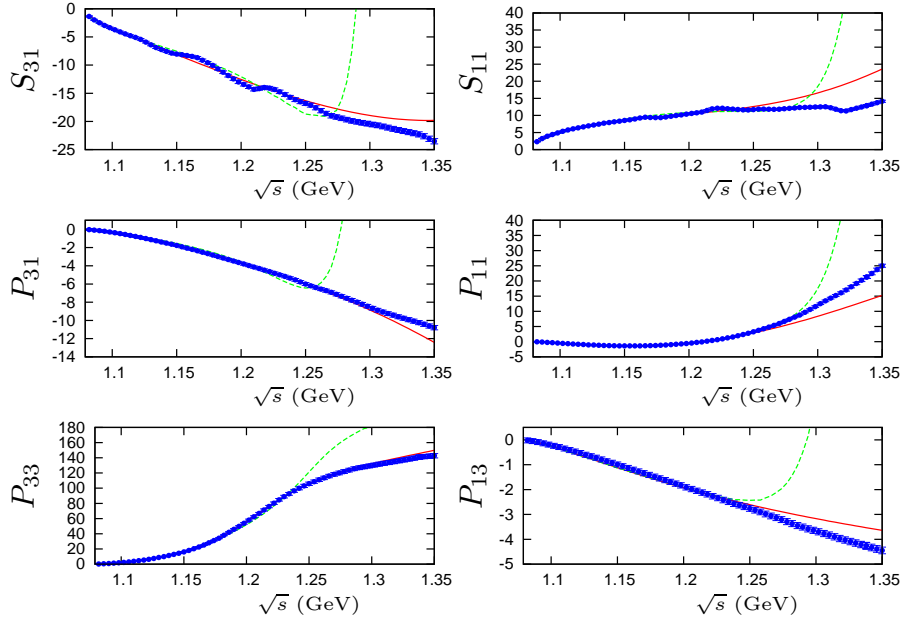,width=.71\textwidth,angle=0}
\caption[pilf]{\small Unitarized fits performed up to $\sqrt{s}_{max}=1.3$~GeV to the KA85 solution. Solid (red) line: EOMS. Dashed (green) line: IR~\cite{nuestroIR} \label{KA85-uni}} 
\end{center}
\end{figure}

\begin{figure}
\begin{center}
 \epsfig{file=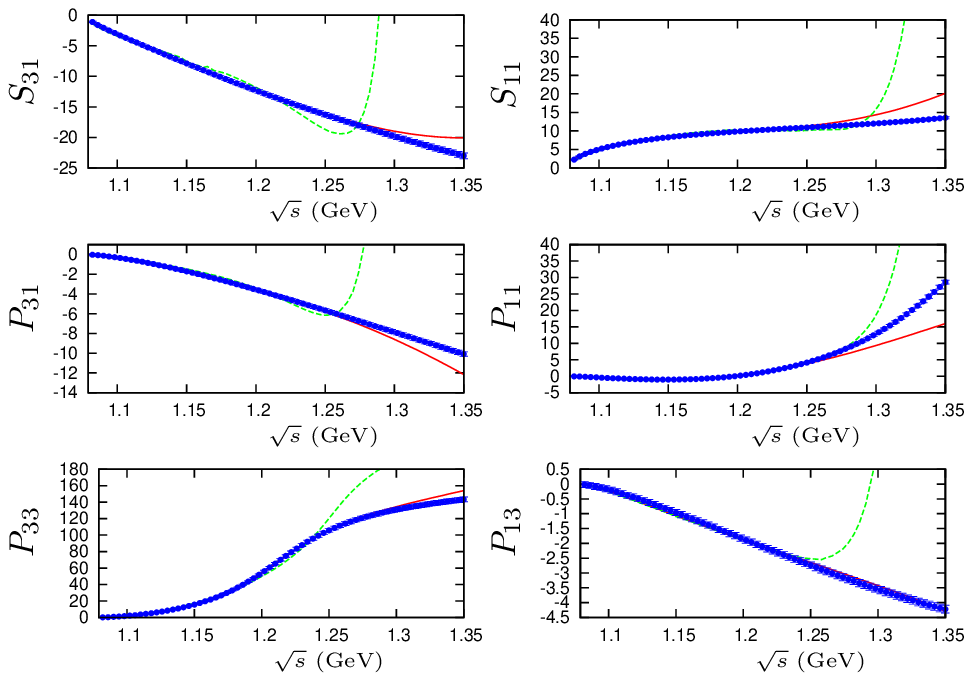,width=.71\textwidth,angle=0} 
\caption[pilf]{\small Unitarized fits performed up to $\sqrt{s}_{max}=1.3$~GeV to the WI08 solution. Solid (red) line: EOMS. Dashed (green) line: IR~\cite{nuestroIR} \label{WI08-uni}} 
\end{center}
\end{figure}

In Figs.~\ref{KA85-uni} and~\ref{WI08-uni} we  show the fits of the unitarized amplitudes to KA85 and WI08 PWAs (red solid line), and the previous result obtained within the IR scheme (green dashed line). We do not consider here the EM06 analysis because we consider data up to energies considerably higher than its upper limit. The EOMS fits are performed up to energies of $\sqrt{s}_{max}\sim1.3$~GeV and achieve a very good description of data up to energies of $\sqrt{s}\sim1.35$~GeV. The description of the WI08 phase shifts is better than for the KA85 solution, as it is reflected by a lower $\chi^2_{\rm d.o.f.}$ (see Table \ref{LECs-unitario}). Moreover, thanks to the CDD pole, the $P_{33}$ partial wave is described almost perfectly up to $\sqrt{s}=1.35$~GeV for both PWAs. Regarding the LECs, we see that fitting the unitarized amplitudes results in a set of values, for this range of energies, that is between the $\Deltaless$- and $\Delta$-ChPT perturbative results. This is not surprising since it is well known that the unitarization method used here respects the chiral order. Notice that by unitarizing the $\Deltaless$-ChPT amplitude the values for the $d_i$ LECs have a smaller size than those in Table \ref{LECs-strategyI}, except for $d_{18}$. On the other hand, the values for the different counterterms are very similar between the fits to KA85 and WI08 PWAs indicating that the procedure is stable.

From Figs. \ref{KA85-uni} and \ref{WI08-uni} it is also easy to see how the unphysical cut of IR (green dashed lines) affects the description of the phase shifts, giving rise to a sharp rise of the amplitude at around $\sqrt{s}\approx 1.26$~GeV. This problem is clearly absent in the results obtained using the EOMS kernel, which has the conventional analytic properties. Although this issue arises at energies well above the expected range of ChPT, it sets a clear warning for future applications of the IR scheme in combination with unitarization techniques. This problem could be quite severe in the $SU(3)$-flavor sector, where the effects of the unphysical IR cuts are already perceived in perturbative calculations~\cite{gorgorito}.

\begin{table}\small
 \begin{center}
\begin{tabular}{|r|r|r|r|r|}
\hline
\small{LEC}        & KA85	                   &    WI08       \\
                   &  $\slashed{\Delta}$-UChPT      & $\slashed{\Delta}$-UChPT  \\
\hline							  			    													
$c_1$              &        $-1.04(2)$            &   $-1.11(2)$    \\   
$c_2$              &        $2.48(3)$             & $ 2.54(3)$       \\
$c_3$              &        $-4.48(5)$            &  $-4.78(4)$ \\
$c_4$              &        $3.00(2) $             &  $3.04(2)$  \\
\hline		  					  			    														
$d_1+d_2$          &   $0.40(2)$                   &  $0.51(2)$  \\
$d_3$              &       $0.13(1)$              &  $0.07(1) $  \\
$d_5$              &         $0.53(2)$           &  $0.36(2) $   \\
$d_{14}-d_{15}$    &          $-0.33(4)$          &   $-0.46(3)$ \\
$d_{18}$           &          $-3.90(13)$         &  $-3.23(12)$  \\ 
\hline
$\gamma$           &      $0.0093(9)$          &       $0.0072(6)$  \\
$s_P$              &      $2.38(7)$              &     $2.21(5)$    \\
\hline
$\chi^2_{\rm d.o.f.}$  &       $2.71$                &     $1.25$         \\
\hline
\end{tabular}
\caption[pilf]{\small Value of the $\Opd$ (in GeV$^{-1}$) and $\Opt$ (in GeV$^{-2}$) LECs, together with the CDD parameters (in GeV$^{2}$), obtained in our best fits to the data of KA85~\cite{KA85} and WI08~\cite{WI08}.  
The label $\slashed{\Delta}$-UChPT means that we took the perturbative result without the $\Delta(1232)$ to obtain the interaction kernel according to the technique described in this section. \label{LECs-unitario}}
\end{center}
\end{table}


\section{$\pi N$ scattering phenomenology}
\label{Sec:piNpheno}

Once the LECs have been determined, ChPT is able to predict a full set of related observables. In this section we study the results obtained for the threshold parameters, the Goldberger-Treiman relation and the pion-nucleon sigma term, in this order. We focus on the results obtained in $\Delta$-ChPT although we also list those in $\Deltaless$-ChPT for the sake of comparing the two approaches. The results reported here for $\sigma_{\pi N}$ are already published elsewhere~\cite{nuestrosigmaterm}. 
 
\subsection{Threshold parameters}
\label{Sec:thresholdparameters}

The scattering lengths and volumes extracted from the different partial waves, together with the scalar-isoscalar ($a_{0+}^+$) and scalar-isovector ($a_{0+}^-$) scattering lengths, are shown in Table~\ref{tabla-thresholdparameters-fits}  for the $\slashed{\Delta}$-ChPT and $\Delta$-ChPT cases.\footnote{For the technical methods employed to determine these parameters from the amplitude see Ref.~\cite{mitesis}.} These results can be compared to the values reported by the PWAs ~\cite{KA85,WI08,EM06}, which are listed in Table~\ref{tabla-thresholdparameters-comparison}. One can see that there is a good agreement between the values extracted from the fits and the results of their respective PWAs. The only exception is the $P_{33}$ scattering volume in the $\slashed{\Delta}$-ChPT fits. This is due to the $\Delta(1232)$-resonance and to the inappropriate description of its effects in the $\Deltaless$-theory~\cite{nuestroIR}. In fact, we observe that the explicit inclusion of the $\Delta(1232)$ in our calculations 
improves the description of $a_{P_{33}}$. It is also remarkable that in $\Delta$-ChPT, one obtains an accurate description of the threshold region despite that the fits are performed up to energies significantly above threshold. In the following we will focus on the discussion of the results obtained in this case. 

In the last column of Table~\ref{tabla-thresholdparameters-comparison} we list the values of the scattering lengths obtained from pionic-atom data~\cite{baru}. We can see that the PWA of the George Washington group, WI08, is the one that presents the best agreement with these independent experimental results. Regarding the results quoted for the Matsinos' group~\cite{EM06}, it is important to point out that the values $a_{0+}^+=0.22(12) 10^{-2} M_\pi^{-1}$ and $a_{0+}^-=8.78(11) 10^{-2} M_\pi^{-1}$ are obtained by the same collaboration in studies of pionic hydrogen~\cite{EM07}. In this reference, only the scattering lengths for $\pi^- p \rightarrow \pi^- p$ and $\pi^- p \rightarrow \pi^0 n$ are provided and we calculate $a_{0+}^+$ and $a_{0+}^-$ using isospin relations. The comparison of the scalar-isoscalar scattering length $a_{0+}^+$ is specially interesting because this quantity is related to the not-very-well known scalar structure of the nucleon. In this sense, the result obtained from $\pi$-atoms data is perfectly compatible with our determinations of this quantity based on the modern WI08 and EM06 analyses. Besides, the error in the $a_{0+}^+$ value of the KA85 solution is overestimated due to the unphysical oscillations of its $S$-wave data. Keeping this in mind, we conclude that our determination based on this older PWA is only compatible with negative values of the $a_{0+}^+$.

\begin{table}
 \begin{center}
\begin{tabular}{|r|r|r|r|r|r|r|}
\hline
Partial &        KA85            	                        & WI08                                            & EM06                                     &    KA85                  & WI08	                      & EM06                                        \\
Wave   &       $\slashed{\Delta}$-ChPT   &	  $\slashed{\Delta}$-ChPT       & $\slashed{\Delta}$-ChPT  & $\Delta$-ChPT     &$\Delta$-ChPT            & $\Delta$-ChPT   \\
\hline
 $a_{0+}^+$     &  $ -0.8(8)$                      &  $ 0.4(8)$                                    & $0.6(4)$                                & -1.1(1.0)            	 &    -0.12(33)                  & 0.23(20)	                                                      \\
 $a_{0+}^-$     &  $ 9.2(10)$                      &  $ 8.4(10)$                                  & $ 7.7(4)$                              & 8.8(5)  	               	 &     8.33(44)                  &  7.70(8)	                                        \\ 
  $a_{S_{31}}$  & $-9.9(13)$                     & $ -8.0(12) $                                & $ -7.1(6) $                            & -10.0(1.1) 	          &   -8.5(6)                        & -7.47(22)	                                                   \\
 $a_{S_{11}}$   & $ 17.5(21)$                   & $ 17.2(21) $                               & $ 15.9(10) $                         & 16.6(1.5)                &    16.6(9)                      & 15.63(26)                                                  \\
  $a_{P_{31}}$   &  $-4.0(7)$                      &  $ -3.5(7)$                                  & $ -3.7(2)$                             & -4.15(35)	           &   -3.89(35)                   &  -4.10(9)	                                              \\
 $a_{P_{11}}$   &  $-7.7(18)$                    &  $ -6.0(18) $                               & $ -7.2(3) $                            & -8.4(5)	                     &   -7.5(1.0)                   & -8.43(18)	                                               \\
 $a_{P_{33}}$   &  $ 25.1(9)$                     &  $ 23.7(9) $                               & $ 23.6(2) $                            & 22.69(30)	            &    21.4(5)                     & 20.89(9)	                                              \\
 $a_{P_{13}}$   &  $ -2.7(7)$                       & $-2.3(6) $                                 & $ -2.7(3)$                              & -3.00(32)	             &  -2.84(31)                  & -3.09(8)                                                      \\
\hline
\end{tabular}
\caption[pilf]{\small Summary of the extracted values of the threshold parameters in $\Deltaless$-ChPT and $\Delta$-ChPT fits. The scattering lengths and volumes are shown in units of $10^{-2} M_\pi^{-1}$ and $10^{-2} M_\pi^{-3}$ respectively.} \label{tabla-thresholdparameters-fits}
\end{center}
\end{table}

\begin{table}
 \begin{center}
\begin{tabular}{|r|r|r|r|r|r|r|r|}
\hline
Partial                     &	  KA85          &     WI08              & EM06                       &    $\pi$-atoms\footnotemark      \\
Wave                      & 	\cite{KA85}  &	    \cite{WI08}  &    \cite{EM06}           &  \cite{baru}      \\
\hline
 $a_{0+}^+$           &$-0.8$	            &  $-0.10(12)$       & $0.22(12)$             &    $-0.1(1)$       \\
 $a_{0+}^-$            &  $9.2$	            &  $8.83(5)$           & $7.742(61) $        &    $8.71(10)$       \\ 
  $a_{S_{31}}$       &  $-10.0(4)$       &  $-8.4$                 & $-7.52(16)$           &  $-8.81(18)$         \\                 	
 $a_{S_{11}}$        &$17.5(3)$ 	   &  $17.1$                & $15.71(13)$         &    $17.5(3)$       \\
  $a_{P_{31}}$       & $-4.4(2)$  	   &  $-3.8$                 & $ -4.176(80)$      &    --                   \\
 $a_{P_{11}}$        & $-7.8(2)$	   &  $-5.8$                 & $ -7.99(16)$         &     --                     \\
 $a_{P_{33}}$        & $21.4(2)$  	   &  $19.4$                & $ 21.00(20)$        &    --                    \\
 $a_{P_{13}}$        &$-3.0(2)$  	   &  $-2.3$                 & $ -3.159(67)$       &     --                      \\
\hline
\end{tabular}
\caption[pilf]{\small Results for the threshold parameters obtained by the different PWAs (columns 2-4) and from $\pi$-atoms data in Ref.~\cite{baru} (last column) .
As before, the scattering lengths and volumes are shown in units of $10^{-2} M_\pi^{-1}$ and $10^{-2} M_\pi^{-3}$, respectively.} \label{tabla-thresholdparameters-comparison}
\end{center}
\end{table}

\footnotetext{Since the PWAs are based on $\pi^+p$ and $\pi^-p$ scattering data, the values given here are obtained from Table 6 of Ref.~\cite{baru}, where the $\pi^+p$ and $\pi^-p$ scattering threshold parameters are given in their isospin limit corrected by the isospin breaking corrections~\cite{Hoferichter:2009gn}.\label{pie2}}

\subsubsection{Chiral expansion of the threshold parameters}
\label{Sec:chiralexpansionofthescatteringlengths}

In 1966, Steven Weinberg used current algebra to predict the $\pi N$ scattering lengths \cite{WeinbergScattLengths}, and later, he re-derived them in the celebrated first application of non-linear chiral Lagrangians~\cite{Weinberg:1966fm}, 
\begin{align}
 a_{0+}^+&=\mathcal{O}(M_\pi^2), \\
 a_{0+}^-&=\frac{M_\pi m_N}{8\pi f_\pi^2 (M_\pi+m_N)}+\mathcal{O}(M_\pi^3).\label{Eq:WeinbergSLs}
\end{align}
In this section we investigate the corrections to these observables in ChPT \cite{BKMnueva} and we pay special attention to the convergence of their chiral expansions. The threshold parameters $a_{IJ\ell}$ can be decomposed order by order in the following fashion
\begin{align}
 a_{IJ\ell}=a_{IJ\ell}^{(1)}+a_{IJ\ell}^{(3/2)}+a_{IJ\ell}^{(2)}+a_{IJ\ell}^{(3)},\label{Eq:scattChExp}
\end{align}
where we use the physical values of $g_A$, $f_\pi$ and $m_N$ to calculate the leading-order terms. The Born-term, which counts as $\mathcal{O}(p)$, gives a contribution to $a_{0+}^+$ proportional to $\mathcal{O}(M_\pi^2)$. We include this contribution into the $\mathcal{O}(p^2)$ piece to avoid confusions with the Eqs.~(\ref{Eq:WeinbergSLs}). The numerical results of these expansions for $\Deltaless$- and $\Delta$-ChPT, and taking the results of the WI08 solution as an example, are shown in Table~\ref{chiralexpansionofthescatteringlengthsWI08fits}. In this Table we see that the convergence of the chiral expansion of the isovector scattering length is very fast in both approaches. In case of the isoscalar scattering length, the convergence seems to be slow in the $\Delta$-theory and completely broken in the $\Deltaless$ case. Besides that, the scattering volumes also present a problematic expansion in $\Deltaless$-ChPT, as $\mathcal{O}(p)$, $\mathcal{O}(p^2)$ and $\mathcal{O}(p^3)$ contributions are typically 
of the same size. This problem is alleviated considerably in the $\Delta$-ChPT where a clear hierarchy $\mathcal{O}(p)>\mathcal{O}(p^2)>\mathcal{O}(p^3)$ among the absolute values of the chiral corrections to each scattering volume is found. The $\Delta$ is important in these observables, in particular in the $P_{33}$ volume for which its contribution is as large as the LO one. For the rest of the partial waves, the $\Delta$ contributions are subleading effects, of the same order as the $\mathcal{O}(p^3)$ corrections. These results are consistent with the conclusions obtained from the inspection of the chiral expansions of the scattering amplitude discussed in Sec.~\ref{Sec:phaseshifts}.  

\begin{table}
\begin{center}
\begin{tabular}{r|r|r|r|r||r|r|r|r|r|}
\cline{2-10}
   &\multicolumn{4}{|c||}{WI08  $\slashed{\Delta}$-ChPT}     & \multicolumn{5}{|c|}{WI08  $\Delta$-ChPT}   \\
\cline{2-10}
               &   $\Op$ & $\Opd$ & $\Opt$ & Sum & $\Op$ & $\mathcal{O}(p^{3/2})$ &$\Opd$ & $\Opt$ & Sum \\
\hline
 \multicolumn{1}{|r|}{$a_{0+}^+$ }     & 0    & -0.48   & 0.91   & 0.43  & 0  & 0 & -1.04& 0.93 &-0.115  \\
 \multicolumn{1}{|r|}{$a_{0+}^-$ }     & 7.91     & 0 & 0.47   & 8.38  & 7.91   &0 & 0 & 0.42 &  8.33\\
\multicolumn{1}{|r|}{$a_{S_{31}}$ }    & -8.85    &0.46    & 0.44   & -7.95 & -8.85  & 0& -0.1& 0.51 & -8.44\\  
 \multicolumn{1}{|r|}{$a_{S_{11}}$ }   & 14.89    & 0.45   & 1.84   & 17.19 & 14.89  & 0 & -0.1& 1.77& 16.56\\
  \multicolumn{1}{|r|}{$a_{P_{31}}$ }  & -5.37    & 3.90   & -2.08  & -3.54 & -5.37  & 0.89 & 1.03 & -0.44 & -3.89\\ 
 \multicolumn{1}{|r|}{$a_{P_{11}}$ }   & -18.15   & 23.51  & -11.32 & -5.97 & -18.15 & 3.57& 11.85& -4.72& -7.45\\
 \multicolumn{1}{|r|}{$a_{P_{33}}$ }   &  9.78    & 13.11  &  0.84  & 23.73 &  9.78  & 7.67& 6.11& -2.18& 21.38\\
 \multicolumn{1}{|r|}{$a_{P_{13}}$ }   &  -4.89   &5.09    & -2.54  & -2.34 &  -4.89 & 0.95& 1.69&-0.59 & -2.84\\
\hline
\end{tabular}
\caption[pilf]{\small From left to right, numerical results for the chiral expansion of the threshold parameters, depending on whether the $\Delta$-resonance is included or not, and using the fits to the WI08 PWA shown in Tables~\ref{LECs-strategyII} and \ref{LECs-strategyI}, respectively.
They are shown in units of $10^{-2}$ $M_\pi^{-1}$ for the scattering lengths and $10^{-2}$ $M_\pi^{-3}$ for the scattering volumes.
The leading order coincides in both cases because we use the physical values of $g_A$, $f_\pi$ and $m_N$. \label{chiralexpansionofthescatteringlengthsWI08fits}}
\end{center}
\end{table}

\subsubsection{Lattice QCD results}
\label{Sec:LQCD}

In this section, we investigate semi-quantitatively the few results that have been reported by the LQCD community on the $\pi N$-scattering amplitudes. In particular, we study the results obtained using the quenched approximation by Fukugita {\itshape et. al.}~\cite{Fukugita:1994ve} and also a recent calculation of the scalar $I=\frac{1}{2}$ phase shift near threshold for a pion mass of $M_{\pi}\sim 400$~MeV reported by the NPLQCD collaboration~\cite{Beane:2009kya}. The latter is specially interesting because the value given is large and has the opposite sign compared to the experimental results. A natural question that follows is whether the chiral extrapolation can explain the sign flip for this observable and, therefore, for the corresponding scattering length.  In the following, we will use our $\Deltaless$-ChPT results (LECs) to study the extrapolation to unphysical pion masses.  Note that in this case we can not use the $\Delta$-theory because the $\delta$-counting is based on a hierarchy $\delta>M_\pi$, that would be broken in the extrapolation. We take into account the running of $g_A$, $f_\pi$ and $m_N$ with $M_{\pi}$, using the expressions that has been given in Sec.~\ref{Sec:ScatteringAmp} (we assume $d_{16}=0$). 

To our knowledge, the work of Fukugita {\itshape et al.}~\cite{Fukugita:1994ve} is the only LQCD calculation of the $\pi N$ scalar scattering lengths ($a_{0+}^+$ and $a_{0+}^-$) reported so far. It is important to recall here that using heavier-than-physical pion masses, as those required to connect with this LQCD calculation, necessarily slows down the convergence of the chiral expansion. The expansion parameter for describing the results of Ref.~\cite{Fukugita:1994ve} can be estimated by taking into account the corresponding results for the $\rho$-resonance mass ($M_\rho$) and $f_\pi$, taking naively that the chiral expansion scale lies between $M_\rho$ and $4\pi f_\pi$. Table \ref{fukugitaresults} shows that the convergence of the chiral series at the pion masses employed in the LQCD calculation is, of course, poorer than at its physical value, where the expansion parameter lies between $0.12-0.20$.

\begin{table}
 \begin{center}
\begin{tabular}{|c|c|c||c|c|}
\hline
                            $M_\pi$~(MeV)       & $M_\rho$~(MeV)        & $f_\pi$~(MeV) & $\frac{M_\pi}{M_\rho}$ & $\frac{M_\pi}{4\pi f_\pi}$	      \\
\hline
                            $732(5)$    & $989(8)$      &  $144(1)$          &     0.74 &  0.40     \\
                            $527(6)$    & $876(11)$      &  $120(2)$        &   0.60   &  0.35 \\
 \hline		   	   
\end{tabular}
{\caption[pilf]{\protect \small Results of Ref.~\cite{Fukugita:1994ve} for $M_\rho$ (second column) and $f_\pi$ (third column), for different values of $M_\pi$ (first column). 
The fourth and fifth columns show an estimation for the maximum an minimum values of the expansion parameter in the chiral series for the corresponding pion mass.
 \label{fukugitaresults}}}
\end{center}
\end{table}

\begin{figure}
\begin{center}
\hspace{0.9cm}KA85 \hspace{4.3cm} WI08 \hspace{4.3cm}  EM06
\epsfig{file=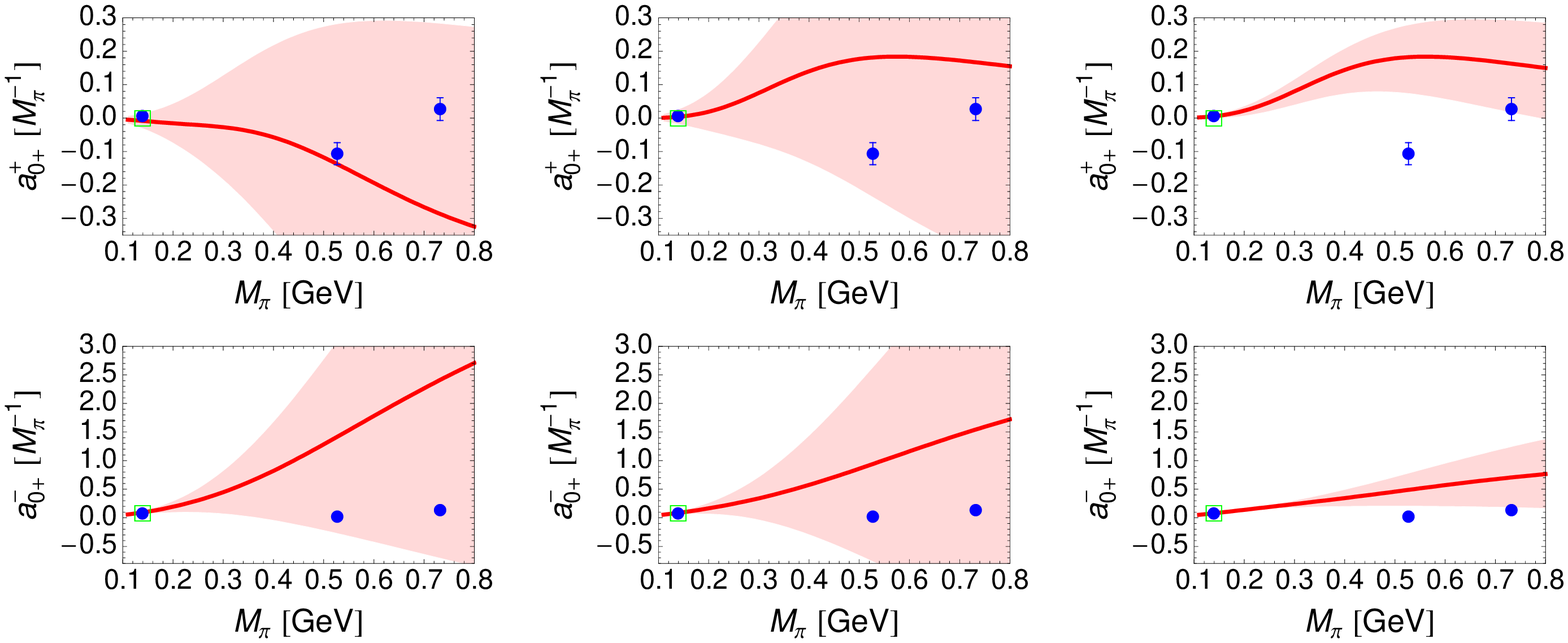,width=1.0\textwidth,angle=0}
\caption[pilf]{\small Results obtained from chiral extrapolations of the $\Deltaless$ case up to $M_\pi=800$~MeV. 
The errors taken for the LECs are taken from Table \ref{LECs-strategyI}. 
The points at the physical pion mass are taken from their corresponding PWAs, while the rest are taken from Ref.~\cite{Fukugita:1994ve}. 
On the other hand, the green square at the physical pion mass shows the Weinberg's predictions.\label{thresholdlattice}} 
\end{center}
\end{figure}

With all this in mind, we show in Figure~\ref{thresholdlattice} the extrapolation of the threshold parameters $a_{0+}^+$ and $a_{0+}^-$ from the physical point up to $M_\pi=800$~MeV for the $\Deltaless$-ChPT result. The uncertainties of the LECs considered only include the statistical ones. These propagate into the extrapolations giving rise to the bands shown in the figure. We see that the fits to KA85 and WI08 are compatible, within errors, with the LQCD results. On the other hand, the EM06 analysis seems to disagree with the LQCD points. Nevertheless, the important outcome of this exercise is to point to the fact that the extrapolation of the scattering lengths is very sensitive to certain combinations of LECs. Thus, LQCD calculations of the scattering parameters close to the physical point could be very important in the future to provide tight constraints onto the LECs relevant for $\pi N$ scattering and its related phenomenology.

\begin{figure}
\begin{center}
\hspace{0.9cm}KA85 \hspace{4.3cm} WI08 \hspace{4.3cm}  EM06
 \epsfig{file=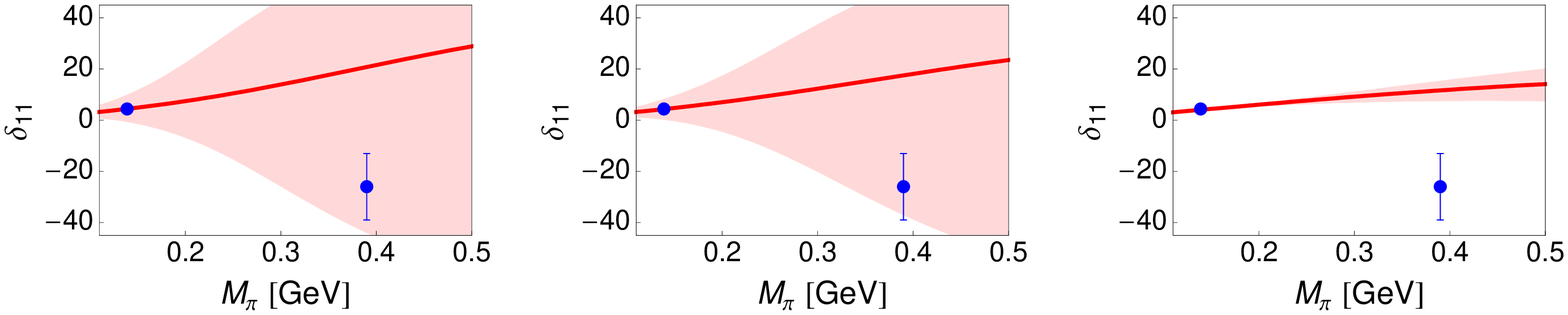,width=1.0\textwidth,angle=0}
\caption[pilf]{\small Extrapolation of the $S_{11}$ phase shift at energy $\delta E_{\pi N}=15.3$~MeV (see text for definitions) of the $\slashed{\Delta}$-ChPT results. 
The point at the physical pion mass corresponds to the value given by each PWA.\label{beanecomparision}} 
\end{center}
\end{figure}

In fact, such a calculation has been recently reported by the NPLQCD collaboration~\cite{Beane:2009kya}. In this work, the value of the $S_{11}$ phase shift at a CM energy $\delta E_{\pi N}=15.3\pm 1.8 \pm 3.2$~MeV, where $\delta E_{\pi N}$ is the energy respect to the $\pi N$ threshold, has been extracted. Curiously enough, at that energy and $M_\pi=390$~MeV, the NPLQCD collaboration obtains a value of $\delta_{11}=-26\pm 7 \pm 6$ degrees for this phase shift, which is large and has a different sign as compared with the experimental result. The interest is then to see if BChPT can explain a sign-flip in the chiral extrapolation of this observable. We proceed as in the previous section and we extrapolate, for the $S_{11}$ partial wave, the result of the fits to the different PWAs, obtained with the physical pion mass, up to $M_\pi=500$~MeV and for an energy above threshold of $\delta E_{\pi N}=15.3$~MeV. The results are shown in Fig.~\ref{beanecomparision} and, as we can see, ChPT could, in principle, explain a change of sign for 
the scattering parameters associated to the $S_{11}$ partial wave. As it also happened with the scattering lengths in Fig.~\ref{thresholdlattice} our results for the LECs fixed by fitting EM06 data are not compatible with the LQCD results. Nonetheless, it is worth recalling that the $\Deltaless$ fits to the EM06 solution have a very large $\chi^2_{\rm d.o.f.}$ and are not very reliable. A closer inspection to the chiral structure of the observable under study shows that the chiral extrapolation depends on the cancellation produced between the combination of LECs $c_2+c_3$, which is negative and $-2c_1$, which is positive. This is exactly the same cancellation occurring at $\mathcal{O}(p^2)$ in the iso-scalar scattering length $a_{0+}^+$. Therefore, a sign flip as the one described in Ref.~\cite{Beane:2009kya}, could provide pertinent information on the scalar structure of the nucleon.   

\subsection{The Goldberger-Treiman relation}
\label{Sec:GT}

The Goldberger-Treiman relation~\cite{Goldberger:1958tr} is one of the most important and earliest results of the ideas of chiral symmetry applied to the strong interactions. It unveils a connection between a purely hadronic quantity, the pion-nucleon coupling $g_{\pi N}$, and $g_A$, which describes the hadronic coupling of the axial part of the weak current to the nucleon, 
\begin{align}
 g_{\pi N}=\frac{g_A m_N}{f_\pi}(1+\Delta_{GT}). \label{Eq:DeltaGT}
\end{align}
This relation can be obtained directly using PCAC and the fact that the pseudoscalar current serves as an interpolating field for a pion. The breakdown of the relation, $\Delta_{GT}$, is of order $\mathcal{O}(M_\pi^2)$ due to the fact that these two couplings are evaluated at different kinematical points, with $g_{\pi N}$ at $t=M_\pi^2$ and $g_A$ at $t=0$. The smallness of the GT discrepancy inferred from this power-counting argument is ratified experimentally by studies based on $NN$ PWAs~\cite{rentcc} and pion-atoms data~\cite{baru}, leading to $\Delta_{GT}=1-3\%$.

In fact, the chiral expansion of $\Delta_{GT}$ does not contain non-analytic pieces up to $\mathcal{O}(M_\pi^4)$~\cite{fettes4}. A calculation of $g_{\pi N}$ in BChPT up to $\mathcal{O}(p^3)$ proceeds essentially as it is shown in Fig.~\ref{axialformfactordiagram}, where the wavy lines are replaced by the pseudoscalar current coupled through a pion pole to the nucleonic lines. The non-vanishing contributions to the GT-discrepancy at this order only arise from the operator accompanying the LEC $d_{18}$ and from the loop correction of the type \textbf{\textit{(d)}} in Fig.~\ref{axialformfactordiagram}, that we label as $\Delta^{(3)}_{EOMS}$. The Goldberger-Treiman deviation can be written as     
\begin{equation}
\Delta_{GT}=-\frac{2 M_\pi^2 d_{18}}{g_A}+\Delta^{(3)}_{\rm EOMS},\label{Eq:GTChPT}
\end{equation}
where 
\begin{equation}
\Delta^{(3)}_{\rm EOMS}=\frac{g_A^2\,M_\pi^2}{96\pi^2f_\pi^2}\left(1 +\frac{M_\pi^2}{m_N^2}\log\left(\frac{M_\pi}{m_N}\right)\right)+\mathcal{O}(M_\pi^4). \label{Eq:GTeoms}
\end{equation}
This result is exactly the same as in the relativistic calculation of Gasser {\it et al.}~\cite{gasser2}. The main contribution to $\Delta^{(3)}_{\rm EOMS}$, of about 0.4$\%$, comes from the $\mathcal{O}(M_\pi^2)$ piece, which in other schemes such as IR and HBChPT, is absorbed through a redefinition of $d_{18}$. The $\mathcal{O}(M_\pi^4)$ pieces from Eq.~\eqref{Eq:GTeoms},  according to a na\"ive power counting estimate, contribute only with a tiny bit $\sim0.01\%$, suggesting that the chiral expansion in $\Delta_{GT}$ converges at very fast pace. Therefore, $\Delta_{GT}$ is described, in very good approximation and with a negligible theoretical uncertainty, by Eq.~(\ref{Eq:GTChPT})~\cite{fettes4}. The pion-nucleon coupling $g_{\pi N}$, and the respective GT-discrepancy, can be extracted from the $\pi N$ scattering amplitude using numerical methods related to the determination of the residue of the amplitude at the nucleon pole~\cite{nuestroIR,mitesis}. We have checked that the application of these methods to the EOMS-renormalized amplitude naturally lead to the same values of $g_{\pi N}$ as those obtained by the direct application of Eq.~(\ref{Eq:GTChPT}). 

\begin{table}
 \begin{center}
\begin{tabular}{|c|c|c|c||c|c|c|}
\hline
                                     &   KA85  &  WI08       & EM06        & KA85 & WI08 & EM06 	     \\
                                    &  $\slashed{\Delta}$-ChPT & $\slashed{\Delta}$-ChPT &$\slashed{\Delta}$-ChPT    & $\Delta$-ChPT&$\Delta$-ChPT&$\Delta$-ChPT\\
\hline
 $\Delta_{GT}$                      & $10(4) \%$ &  $2(4)\%$ & $3.7(7)\%$ &$ 5.1(8)\%$& $ 1.0(2.5)\%$ & $ 2.0(4)\%$            \\
\hline		   	   
 $g_{\pi N}$ &$14.1(5)$ &$13.1(5)$ &$13.29(10)$ & 13.53(10) & 13.00(31) & 13.13(5) \\
\hline

\end{tabular}
{\caption[pilf]{\protect \small Results for $\Delta_{GT}$ and $g_{\pi N}$ from the different fits considered in this paper.\label{Table:GT-deviation1}}}
\end{center}
\end{table}
\begin{table}
\begin{center}
\begin{tabular}{|c|c|c|c|c|c|}
\hline
                                     	& KA85                            & WI08                              & EM06  & $NN$ scattering  & Pionic atoms    \\
                                       &  \cite{KA85}  &     \cite{WI08}  &      \cite{EM06}             &     \cite{rentcc} &      \cite{baru} \\                                                    
\hline
 $\Delta_{GT}$                       & $4.5(7)\%$& $2.1(1)\%$ &  $0.2(1.0)\%$                     &    $1\%$     &        $1.9(7)\%$           \\
\hline		   	   
 $g_{\pi N}$                          & $13.46(9)$& $ 13.15(1)$ & $12.90(12)$                  &    $\simeq 13.0$          &    $13.12(9)$     \\
\hline
\end{tabular}
{\caption[pilf]{\protect \small Results for $\Delta_{GT}$ and  $g_{\pi N}$ obtained from the PWAs, $NN$-scattering and pionic-atom data.\label{Table:GT-deviation2}}}
\end{center}
\end{table}

In Table~\ref{Table:GT-deviation1}, we display the results on $g_{\pi N}$ and $\Delta_{GT}$ obtained from the different fits in the EOMS scheme considered in this work. The uncertainties are propagated from the errors in $d_{18}$ shown in Tables~\ref{LECs-strategyI} and~\ref{LECs-strategyII}. In Table~\ref{Table:GT-deviation2}, we show also the values reported by the different PWAs and the ones that are obtained from independent experimental sources, $NN$-scattering~\cite{rentcc} and pionic-atom data~\cite{baru}. First of all, notice the larger errors in the $\Deltaless$-ChPT values, due to the troublesome convergence and, correspondingly, larger uncertainty in the determinations of the LECs in this approach. Also, it is worth pointing out the larger error in the $\Delta$-ChPT value obtained for WI08 results due to the sensitivity to $d_{18}$ in the fits to this PWA when varying $\sqrt{s}_{max}$. 

The comparison between both tables shows that the determinations from the fits of $\Delta$-ChPT to the PWA phase-shifts are quite consistent with the values reported by the respective collaborations. Interestingly enough, we see that not all the PWAs are consistent with the extractions of $g_{\pi N}$ obtained from alternative experimental sources. In particular, we see that the WI08 and EM06 results agree with those obtained from $NN$ and pion-atom data. On the other hand, the KA85 value is considerably larger than any of the other determinations, leading to a GT-discrepancy with a size that is currently considered implausible \cite{beche2}. Therefore, the analysis of $g_{\pi N}$ gives phenomenological support to the WI08 and EM06 solutions, in detriment to the KA85's one.

\subsection{The pion-nucleon sigma term}
\label{Sec:sigmaterm}

The definition of the pion-nucleon sigma term can be given in terms of the commutators~\cite{jameson},
\begin{align}\label{sigmatermdefinition1}
 \sigmaterm=\frac{1}{2m_N} \sum_{a=1}^3 \frac{1}{3}\langle N(p)|[Q_A^a,[Q_A^a,\mathcal{H}_{SB}]]|N(p)\rangle,
\end{align}
with $Q_A^a$ the axial charge and $\mathcal{H}_{SB}$ is the chiral-symmetry breaking part of the QCD Hamiltonian. This quantity is related to the explicit breaking of the chiral symmetry and, therefore, it should be small compared to $\Lambda_\chi$. From Eq.~\eqref{sigmatermdefinition1} it is straightforward to obtain 
\begin{align}\label{sigmatermdefinition2}
 \sigmaterm = \frac{\hat{m}}{2m_N}\langle N(p)|  (\bar{u}u+\bar{d}d) |N(p)\rangle,
\end{align}
where $\hat{m}=(m_u+m_d)/2$. The sigma term, in this form, can be identified with the nucleon scalar form factor $\bar{u}(p^\prime)\sigma (t)u(p) = \langle N(p')| \hat{m} (\bar{u}u+\bar{d}d) |N(p)\rangle$ evaluated at $t=0$. This matrix element can be derived also by means of the Hellmann-Feynman theorem from the quark-mass dependence of the nucleon mass, 
\begin{align} \label{sigmaterm-HF}
 \sigmaterm = M_\pi^2 \frac{\partial m_N}{\partial M_\pi^2}.
\end{align}
An explicit calculation of the scalar form factor or the nucleon mass in EOMS up to $\mathcal{O}(p^3)$ gives 
\begin{align}
 \sigmaterm=-4c^{\prime}_1 M_\pi^2-\frac{3 g_A^2 M_\pi^3}{16 \pi^2 f_\pi^2 m_N}  \left( \frac{3m_N^2-M_\pi^2}{\sqrt{4 m_N^2-M_\pi^2}}\arccos\frac{M_\pi}{2 m_N}+M_\pi \log \frac{M_\pi}{m_N} \right),\label{Eq:sigmapiNEOMS}
\end{align}
where $c_1^\prime$ is the LEC renormalized in the EOMS scheme (see \ref{Sec:lecsrenormalization}).

As it was discussed in the introduction, the pion nucleon sigma term is a quantity related to the structure of the nucleon which is important to understand the origin of the mass of the ordinary matter and the nature of the breaking of chiral symmetry in QCD. An accurate value of this matrix element is also required in to reduce the hadronic uncertainties that enter in the phenomenology of direct searches of dark matter. The main method to determine $\sigma_{\pi N}$ experimentally is by analytical continuation of the isoscalar scattering amplitude to the CD point~\cite{cheng-dashen-theorem} (see also Sec.~\ref{Sec:CDtheorem}). However, there is still no consense on the value of $\sigma_{\pi N}$ because it varies depending on the PWA taken as input~\cite{formfactorsigmaterm,GWsigmatermlarge,Olsson}.  In this respect,  Ref.~\cite{nuestrosigmaterm} made an important step forward by obtaining perfectly compatible results between ChPT and those reported from the dispersive analyses based on the same PWA. It should be stressed that both methods are well sound and 
model independent. The agreement reported in Ref.~\cite{nuestrosigmaterm} makes then clear that the problem to fix $\sigma_{\pi N}$ rests on the data basis employed and not on theory. 

In this section we briefly review the results presented in~\cite{nuestrosigmaterm}. The chiral Ward identity relating the scattering amplitude and $\sigma_{\pi N}$ at the CD point, can be accessed in a more elegant way using ChPT. Indeed, a value of $\sigma_{\pi N}$ can be predicted once the relevant LEC $c_1$ is properly determined from the scattering data. However, a reliable value for  $\sigma_{\pi N}$ would only follow from a representation of the scattering amplitude with a well behaved chiral expansion above threshold.


\begin{table}
 \begin{center}
\begin{tabular}{c|c|c|c|c|c|c|}
\cline{2-7}
                                  &      KA85                  &  WI08                   & EM06                     & KA85            & WI08         &  EM06                \\
                                 &      {\small $\slashed{\Delta}$-ChPT} &  {\small $\slashed{\Delta}$-ChPT} &  {\small $\slashed{\Delta}$-ChPT}  &  {\small  $\Delta$-ChPT}  &  {\small $\Delta$-ChPT}&  {\small $\Delta$-ChPT}   \\
\hline      
\multicolumn{1}{|r|}{$\sigmaterm$  (MeV)  } & 79(11)  & 97(7) & 95(3)    & 43(5) & 59(4) & 59(2)    \\
\hline
\end{tabular}
{\caption[pilf]{\protect Results for $\sigmaterm$ in MeV for the cases without and with the $\Delta(1232)$, $\slashed{\Delta}$-ChPT (Table \ref{LECs-strategyI}) and $\Delta$-ChPT (Table \ref{LECs-strategyII}), respectively.  \label{sigmatermstable}}}
\end{center}
\end{table}

\begin{table}
 \begin{center}
\begin{tabular}{c|c|c|c|}
\cline{2-4}
                                     &  KA85     \cite{formfactorsigmaterm}      & WI08 \cite{GWsigmatermlarge}       & EM06   \cite{Olsson}      \\
\hline      
\multicolumn{1}{|r|}{$\sigmaterm$ (MeV)}    & 45(8) & 64(7) & 56(9)                                \\
\hline
\end{tabular}
{\caption[pilf]{\protect Results for $\sigmaterm$, in MeV, extracted by the different PWAs.  \label{sigmatermsPWAstable}}}
\end{center}
\end{table}

In Table~\ref{sigmatermstable}, we list the results for $\sigma_{\pi N}$ obtained for the different PWAs. The results derived from  the $\slashed{\Delta}$-ChPT are shown for completeness. In Table~\ref{sigmatermsPWAstable} we also give the results obtained from dispersive analyses and using the CD theorem. As we can see, the results obtained in the $\Delta$-theory are quite accurate and perfectly consistent with the dispersive results. On the other hand, the values obtained for the $\Deltaless$-case tend to be larger. As it was discussed in Sec.~\ref{Sec:Deltafits}, this sizable effect of the $\Delta$ in the values of $c_1$ and $\sigma_{\pi N}$ is not expected on the grounds of the RSH. Finally, we reported in Ref.~\cite{nuestrosigmaterm} the value $\sigmaterm=59(7)$~MeV as it is extracted from the WI08 and EM06 PWAs. The error includes systematic uncertainties added in quadratures to the theoretical uncertainty that is estimated from the explicit calculation of higher-order diagrams. See Ref.~\cite{nuestrosigmaterm} for a detailed analysis of this determination and for the discussion of the consequences that a relatively large value of $\sigma_{\pi N}$ have in phenomenology. In relation with the strangeness content in the nucleon we address the reader to the recent reanalysis of the so-called ``strangeness puzzle'' presented in Ref.~\cite{nuestrosigmas}. There it is shown that the relatively large value reported here for $\sigma_{\pi N}$ is not at odds with a small strangeness content in the nucleon.

\section{Subthreshold region}
\label{Sec:subthresholdregion}

A proper description of the subthreshold region is very important in BChPT because the so-called soft point, $s=u=m_N^2$ and $t=0$, is the one about which the chiral expansion is performed (this point is defined in the chiral limit). Previous BChPT analyses have found difficulties to connect the information around the soft-point (as derived from dispersive studies) with the experimental data~\cite{fettes3,fettes4,beche2}. As a result, it has been concluded that BChPT at one loop is not accurate enough to relate the subthreshold and physical regions~\cite{beche2}. In this section we study the extrapolation of the chiral representation of the scattering amplitude into the subthreshold region in BChPT within the EOMS scheme. We also revisit important low-energy theorems established at specific points of the subthreshold region: The CD point~\cite{cheng-dashen-theorem} and the Adler point~\cite{adler}. 

The starting point is the so-called subthreshold expansion,
\begin{equation}
X^{\pm}(\nu,t)=x_{00}^{\pm}+x_{10}^{\pm}\nu^2+ x_{01}^{\pm}t+ x_{20}^{\pm}\nu^4+  x_{02}^{\pm}t^2+\ldots, \label{Eq:subthrExp} 
\end{equation}
with $X^\pm=\bar{D}^+,\bar{D}^-/\nu,\bar{B}^+/\nu,\,\bar{B}^-$, the Born-subtracted scattering amplitudes~\cite{hoehler}. 
Notice that $\nu=0$, $t=0$ corresponds to $s=u=m_N^2+M_\pi^2$, so that the expansion in Eq.~(\ref{Eq:subthrExp}) is done around the same point as the chiral expansion but for physical pion masses and with the coefficients $x_{ij}^{\pm}$ non-analytic functions of $M_\pi$. The usual procedure to analyze the subthreshold region of the $\pi N$ scattering amplitude is to determine the values of the leading coefficients in Eq.~(\ref{Eq:subthrExp}) using  dispersive analyses~\cite{hoehler}. The resulting values can then be used to fix the LECs of BChPT and to analyze the consequences of chiral symmetry around the soft point (low-energy theorems)~\cite{buettiker,beche2}.

\begin{table}
 \begin{center}
\begin{tabular}{|c|c|c|c||c|c|c|}
\hline
                                     &   KA85                          &  WI08                       & EM06                       & KA85                & WI08                   & EM06	               \\     
                                     &   $\slashed{\Delta}$-ChPT       &  $\slashed{\Delta}$-ChPT    &  $\slashed{\Delta}$-ChPT   &  $\Delta$-ChPT      &    $\Delta$-ChPT       &     $\Delta$-ChPT     \\      
\hline
 $d_{00}^+$ ($M_\pi^{-1}$)            &  $-2.02(41)$ & $-1.65(28)$ & $-1.56(5)$ & $-1.48(15)$ & $-1.20(13)$ &  $-0.98(4)$                                                                   \\    
 $d_{01}^+$ ($M_\pi^{-3}$)            &  $1.73(19)$  & $1.70(18)$  & $1.64(4)$ & $1.21(10)$   & $1.20(9)$ & $1.09(4)$                                                                       \\    
 $d_{10}^+$ ($M_\pi^{-3}$)            &  1.81(16)  & 1.60(18)   & 1.532(45)  & 0.99(14)   &  0.82(9)  & 0.631(42) \\                                                        
 $d_{02}^+$ ($M_\pi^{-5}$)            &  $0.021(6)$    & $0.021(6)$    & $0.021(6)$   &  0.004(6)   &    0.005(6) & 0.004(6) \\                                                        
 $b_{00}^+$ ($M_\pi^{-3}$)            & -6.5(2.4)  & -7.4(2.3)   & -7.01(1.1) & -5.1(1.7)    & -5.1(1.7)     &  -4.5(9) \\                                                        
 $d_{00}^-$ ($M_\pi^{-2}$)            &  1.81(24)  & 1.68(16)   & 1.495(28) &  1.63(9)  & 1.53(8)   & 1.379(8) \\                                                        
 $d_{01}^-$ ($M_\pi^{-4}$)            &  -0.17(6)  & -0.20(5)  & -0.199(7)& -0.112(25)  & -0.115(24)    & -0.0923(11)\\                                                        
 $d_{10}^-$ ($M_\pi^{-4}$)            & -0.35(10)  & -0.33(10) & -0.267(14) &  -0.18(5)  &   -0.16(5)    & -0.0892(41) \\                                                        
 $b_{00}^-$ ($M_\pi^{-2}$)            & 17(7)  & 17(7)   & 16.8(7) & 9.63(30)   & 9.755(42)    & 8.67(8) \\                                                        
\hline		   	   
\end{tabular}
{\caption[pilf]{Results for different subthreshold coefficients obtained from the LECs shown in Tables \ref{LECs-strategyI} and \ref{LECs-strategyII} obtained from fits to the PWA phase shifts in $\Deltaless$- and $\Delta$-ChPT, respectively.
\label{subthreshold-results}}}
\end{center}
\end{table}

\begin{table}
 \begin{center}
\begin{tabular}{|c|c|c|c||c|c|c||c|c|}
\hline
                                          &  KA85  & WI08    \\
                                          & \cite{KA85} &\cite{WI08} \\ 
\hline
 $d_{00}^+$ ($M_\pi^{-1}$)                &   $-1.46$    & $-1.30$         \\
 $d_{01}^+$ ($M_\pi^{-3}$)               &    $1.14$    &  $1.19$           \\ 
 $d_{01}^+$ ($M_\pi^{-3}$)               & 1.14(2)    & --                      \\
 $d_{02}^+$ ($M_\pi^{-5}$)               &    $0.036$    &  $0.037$           \\ 
 $b_{00}^+$ ($M_\pi^{-3}$)               & -3.54(6) &   --                   \\
 $d_{00}^-$ ($M_\pi^{-2}$)               & 1.53(2)  &  --                  \\
 $d_{01}^-$ ($M_\pi^{-4}$)               &  -0.134(5) & --               \\
 $d_{10}^-$ ($M_\pi^{-4}$)               & -0.167(5)  & --               \\
 $b_{00}^-$ ($M_\pi^{-2}$)               & 10.36(10)  & --               \\
\hline		   	   
\end{tabular}
{\caption{Results on subthreshold coefficients from the Karlsruhe and George Washington groups.\label{subthreshold-resultsPWA}}}
\end{center}
\end{table}

In this work we follow the inverse procedure as we fix the LECs in the physical region, where data actually exists, and we investigate the resulting description in the subthreshold region. In Table~\ref{subthreshold-results} we show the results for the leading subthreshold coefficients obtained from the fits to the PWA phase shifts in $\Deltaless$-ChPT and $\Delta$-ChPT. In Table~\ref{subthreshold-resultsPWA} we list the results for the same coefficients as they have been reported by the different PWAs~\cite{KA85,GWsigmatermlarge}. The first thing that is worth noticing from the comparison between the two tables is to confirm that $\Deltaless$-ChPT in the EOMS scheme fails to connect the physical and subthreshold regions at $\mathcal{O}(p^3)$. Indeed, the numerical values of the subthreshold coefficients are not consistent with the dispersive results, even though the ChPT values have a sizable uncertainty. This problem is not likely to be solved at $\mathcal{O}(p^4)$ in $\Deltaless$-ChPT in EOMS in the light of the results obtained in HBChPT~\cite{fettes4} and IR~\cite{beche2} at this order.   

On the other hand, $\Delta$-ChPT gives a description of the subthreshold region that is, in general, perfectly consistent with the dispersive results. The only disagreement concerns the coefficient $d_{02}^+$, which has its physical origin in the incapacity of BChPT to reproduce properly, at this order, the curvature induced by the two-pion threshold at $t=4M_\pi^2$~\cite{formfactorsigmaterm}. (This has a very important consequence on the determination of $\sigma_{\pi N}$ using the value of the scattering amplitude at the CD point, as we will see in the next section.) We conclude, then, that the explicit $\Delta$-exchange contribution is a fundamental ingredient to bridge the gap between the physical and subthreshold regions. This is an important result as it paves the road for studying all the phenomenology related to $\pi N$ scattering in a systematic manner within $\Delta$-ChPT, using directly scattering data and without any other dispersive input. This contrasts with the the conclusions derived in $\Delta$-HBChPT~\cite{fettes_ep}. On the other hand, further studies at higher orders in the chiral expansion 
and including the $\Delta(1232)$ degrees of freedom in a coherent way should corroborate this particular finding of the present work. Such a self-consistent framework to study the $\pi N$ scattering amplitude, based exclusively on BChPT, is  complementary to other model-independent approaches based on a pure dispersive treatment of the amplitude, e.g.~\cite{Ditsche:2012fv}.     

Finally, it is also interesting to compare the results obtained for the subthreshold coefficients from the different analyses and in $\Delta$-ChPT. The KA85 and WI08 results closely agree with each other, besides an important discrepancy in $d_{00}^+$ which is related to the different pion-nucleon sigma terms reported by the two solutions. However, comparing those with the novel results obtained for EM06 solution, we see that the latter gives a physical picture of the subthreshold region around the point ($\nu=0$, $t=0$) that is quite different to the former ones. In fact, the values of most of the subthreshold coefficients obtained from the fits to the EM06 phase shifts are not compatible with the ones extracted from the KA85 or the WI08 solutions. The fact that the two latter PWAs grossly agree gives support to their solution in the subthreshold region. This discussion could also take place at the level of the values of the LECs (as shown in Table~\ref{LECs-strategyII}), which the subthreshold coefficients ultimately depend on. Nevertheless, it is worth remarking that meaningful comparisons among different PWAs can only be done based on observable quantities. 

\subsection{The Adler consistency condition and the Cheng-Dashen theorem}
\label{Sec:CDtheorem}

The isoscalar scattering amplitude $D^+(\nu,t)$ is subject to a couple of important low-energy theorems. Its extrapolation onto  $t>0$ at $\nu=0$ is constrained by the Adler consistency condition~\cite{adler} at $t=M_\pi^2$ and relates the amplitude to the pion-nucleon sigma-term at the CD point, $t=2 M_\pi^2$~\cite{cheng-dashen-theorem}. The Adler's consistency condition states that
\begin{equation}
 D^+(\nu=0,t=M_\pi^2)\simeq \frac{g_{\pi N}^2}{m_N},\label{Eq:Adler}
\end{equation}
which is equivalent to the statement that the Born-subtracted isoscalar amplitude has a zero in the neighborhood of $t=M_\pi^2$, $\bar{D}^+(\nu=0,t=M_\pi^2)\simeq0$. Once the LECs have been fixed by fitting our theoretical amplitude to the PWAs data, it is interesting to check the consistency of our BChPT calculations with the Adler's condition. 

In Table~\ref{adlercondition} we display the results for $f_\pi^2 \bar{D}^+(\nu=0,t=M_\pi^2)$ and the different fits performed in this work. We also list the relative deviation with respect to the exact fulfillment of the Adler condition, Eq.~(\ref{Eq:Adler}). As shown in this table, there is no much difference between the results obtained for $\slashed{\Delta}$-ChPT and $\Delta$-ChPT, that is not surprising since the Adler condition is a direct consequence of PCAC and so it has to be satisfied in both cases. On the other hand, the WI08 and EM06 analyses fulfill the Adler condition better, although KA85 is also within the theoretical bounds established by Adler~\cite{adler}. Finally, these results can be compared with those given by the dispersive calculation, which can be obtained using Eq.~(\ref{Eq:subthrExp}) and the respective values of the subthreshold coefficients, leading to $-16$ MeV and $-4$ MeV for KA85 and WI08 solutions, in order. These values compare well with the ones obtained in both versions of BChPT.

\begin{table}
 \begin{center}
\begin{tabular}{|c|c|c|c||c|c|c|}
\hline
                                     &   KA85 &  WI08     & EM06       & KA85 & WI08  	       &  EM06  \\
                                     & $\slashed{\Delta}$-ChPT& $\slashed{\Delta}$-ChPT& $\slashed{\Delta}$-ChPT& $\Delta$-ChPT & $\Delta$-ChPT& $\Delta$-ChPT \\
\hline
  $f_\pi^2 \bar{D}^+(0,M_\pi^2)$   & $-16(18)$  & $4(11)$ & $6.0(1.7)$ & $-17(8)$ & $-4.4(4.5)$ &   $7.3(2.1)$                      \\
$\%$                               & $1.0(1.1)$  & $0.3(8)$ & $0.40(11)$ & $1.0(5)$ & $0.28(28)$ &    $0.50(14)$                  \\
\hline		   	   
\end{tabular}
{\caption[pilf]{\protect \small Check of the Adler condition. The second and third row show the deviation from this condition, in MeV, and the relative value of this deviation, respectively.\small  \label{adlercondition}}}
\end{center}
\end{table}

The second low-energy theorem is more important as it relates the isoscalar amplitude to the pion-nucleon sigma term at the CD point, 
\begin{equation}
\Sigma\equiv f_\pi^2 \bar{D}^+(\nu=0,t=2 M_\pi^2)=\sigma_{\pi N}+\Delta_\sigma+\Delta_{\rm R}. \label{Eq:CDTheorem} 
\end{equation}
In this equation, $\bar{D}^+(\nu,t)$ is the Born-subtracted isoscalar $\pi N$ scattering amplitude, $\Delta_\sigma=\sigma(2 M_\pi^2)-\sigma_{\pi N}$ and $\Delta_{\rm R}$ is a remainder originating from the translation of the exact relation at the soft-point (and thus with off-shell pions) to the CD point~\cite{cheng-dashen-theorem}. The different pieces appearing at both sides of this equation can be obtained in BChPT~\cite{gasser2,remanente}. The difference of the scalar form factor at $t=2M_\pi^2$ and $t=0$ in the EOMS scheme gives the same result as the one obtained by Gasser {\it et al.} in their seminal paper~\cite{gasser2}, $\Delta_\sigma\simeq4.7$ MeV. The remainder of the CD theorem, $\Delta_{\rm R}$, is an analytic piece of order $\mathcal{O}(M_\pi^4)$~\cite{gasser2,remanente}, which comes out to be numerically very small $\Delta_{\rm R}\lesssim1$ MeV~\cite{gasser2}. In Table~\ref{Sigmaresults}, we show the results for $\Sigma$ in the different schemes treated in this work which can be compared with those of $\sigma_{\pi N}$ in Table~\ref{sigmatermstable} in order to confirm the fulfillment of the CD theorem in our calculations.

\begin{table}
\begin{center}
\begin{tabular}{|c|c|c|c||c|c|c|}
\hline
                             &           KA85                          &  WI08                       & EM06                       & KA85                & WI08                   & EM06	               \\     
                            &            $\slashed{\Delta}$-ChPT       &  $\slashed{\Delta}$-ChPT    &  $\slashed{\Delta}$-ChPT   &  $\Delta$-ChPT      &    $\Delta$-ChPT       &     $\Delta$-ChPT     \\     
\hline
 $\Sigma$ (MeV)& $84(11)$  & $103(7)$    & $103(3)$  &  $48(5)$    & $64(4) $ & $64(2)$                                                                                              \\   
\hline		   	   
\end{tabular}
{\caption[pilf]{Results, in MeV, for the $\Sigma$-terms obtained in our analysis.~\label{Sigmaresults}}}
\end{center}
\end{table}

However, it is known that, at $\mathcal{O}(p^3)$, BChPT fails to catch the full strength of the two-pion threshold in the extrapolation of the scalar form factor of the nucleon to $t=2 M_\pi^2$~\cite{formfactorsigmaterm}.\footnote{This effect is enhanced by the particularly large scalar isoscalar $\pi\pi$ partial wave amplitude driving the scalar form factor of the pion \cite{Oller&Roca}.} Namely, dispersive calculations of the difference between these two kinematical points lead to the value $\Delta_\sigma\simeq15$ MeV~\cite{formfactorsigmaterm}. This issue could limit the applicability of the CD theorem at $\mathcal{O}(p^3)$ as the determination of $\sigma_{\pi N}$ from the $\pi N$ scattering amplitude might be afflicted by a systematic uncertainty of $\sim$10 MeV. On the other hand, comparing the $\Delta$-ChPT results in Table~\ref{Sigmaresults} for the KA85 and WI08 solutions with those given by the dispersive analyses, $\Sigma\simeq60$ MeV~\cite{formfactorsigmaterm} and $\Sigma=79(7)$~\cite{GWsigmatermlarge} MeV respectively, indicates that the same problem afflicts the $t$-dependence of the scattering amplitude on the left-hand side of Eq.~(\ref{Eq:CDTheorem}). 

Namely, the quantity $\Sigma$ can be rewritten as~\cite{formfactorsigmaterm}
\begin{equation}
\Sigma=\Sigma_{\rm d}+\Delta_{\rm D},\label{Eq:Sigmad} 
\end{equation}
where $\Sigma_{\rm d}=f_\pi^2(d_{00}^++2M_\pi^2d_{01}^+)$ and $\Delta_{\rm D}$ is the remainder given, in very good approximation ($\delta\Sigma\sim 1$ MeV~\cite{GWsigmatermlarge}), by the curvature term, $\Delta_{\rm D}=4f_\pi^2M_\pi^4 d_{02}^+$. Neglecting $\Delta_{\rm R}$, the CD theorem now takes the form
\begin{equation}
\sigma_{\pi N}=\Sigma_{\rm d}+\Delta_{\rm D}-\Delta_\sigma. \label{Eq:CDtheoremSigmad} 
\end{equation}
In the previous section, we found that the values obtained in $\Delta$-ChPT for the first two terms of this expansion ($d_{00}^+$ and $d_{01}^+$) agreed with the ones extracted from the dispersive analyses, so there is also agreement on the determination of $\Sigma_{\rm d}$. All the discrepancy between the values of $\Sigma$ extracted in $\Delta$-ChPT or the dispersive analyses thus originates from the discrepancy on the $d_{02}^+$ coefficient. As it was discussed above, this is related to the fact that the scattering amplitude at $\mathcal{O}(p^3)$ does not catch the full strength of the two-pion threshold, which translates into an underestimation of $\sim10$ MeV in the value of $\Delta_{\rm D}$ and, hence, of $\Sigma$, when extrapolating to the CD point. 

In conclusion, the crucial point in the determination of $\sigma_{\pi N}$ is not the value of $\Sigma$ but rather the one of $\Sigma_{\rm d}$, which is properly given by $\Delta$-ChPT, together with the value of the difference $\Delta_{\rm D}-\Delta_\sigma$. In this sense, the very same effect curving the $t$-dependence of the isoscalar scattering amplitude, enhances the slope of the scalar form factor at $t=0$ such that these contributions largely cancel in the extraction of the sigma term. Indeed, although $\Delta$-ChPT fails to give a reliable description of $\Delta_{\rm D}$ and $\Delta_\sigma$ individually, it gives a value for their difference, $\Delta_{\rm D}-\Delta_\sigma=-3.5(2.0)$ MeV, which is perfectly consistent with the dispersive result 
$\Delta_{\rm D}-\Delta_\sigma=-3(1)$ MeV~\cite{formfactorsigmaterm}. This analysis clarifies the reason why the pion-nucleon sigma term can be accurately pinned down in $\Delta$-ChPT at $\mathcal{O}(p^3)$ with the residual uncertainty produced by these effects well within the theoretical error estimated through the explicit calculation of $\mathcal{O}(p^{7/2})$ and $\mathcal{O}(p^4)$ diagrams~\cite{nuestrosigmaterm}.

\section{Summary and conclusions}
\label{Sec:Conclusions}

We have presented a novel analysis of the $\pi N$ scattering amplitude in Lorentz covariant B$\chi$PT within the EOMS scheme up to ${\cal O}(p^3)$ and considering the inclusion of the $\Delta(1232)$ explicitly in the $\delta$-counting. We first studied the phase shifts in partial waves provided by the Karlsruhe-Helsinki, George-Washington and Matsinos' groups, that we use as experimental data to fit our LECs. While the $\Deltaless$-ChPT approach has the same difficulties to describe the region above threshold as those found previously in the HB and IR schemes, the $\Delta$-ChPT perfectly describes the phase shifts up to energies below the $\Delta$-resonance region.  The improvement achieved in the latter case is clearly illustrated by the analysis of the EM06 solution, for which the $\Deltaless$ fit gives a very large $\chi^2_{\rm d.o.f.}$, that is put well below 1 once the $\Delta$ is explicitly included. Differently to a previous analysis up to the same accuracy in HB and SSE scheme, the 
values of our LECs are  stable against the PWAs phase-shifts used as input,  allowing for a  clear discussion of different $\pi N$ phenomenology.

Once the LECs are determined, we study thoroughly all the observables associated with the $\pi N$ scattering amplitude. In particular, we discuss the results and chiral expansion of the threshold coefficients, the Goldberger-Treiman relation, the pion-nucleon sigma term and the extrapolation onto the subthreshold region. Also, we investigated semi-quantitatively the extrapolation of the scarce results on scattering lengths and $S_{11}$ phase shift reported by the LQCD community. In general, we conclude that the $\Delta$-ChPT converges much better than the $\Deltaless$ approach and that, in the former case, one obtains a phenomenology perfectly consistent with the one reported by the PWAs. From the comparison among the results on observables that are obtained by an analysis of the different PWAs, we conclude that the WI08 solution is the most consistent with those extracted from alternative 
experimental sources ($NN$-scattering and pion-atom data). We remind here that the KA85 analysis gives rise to a value for $h_A$ that is not compatible with the value obtained from the $\Delta(1232)$ Breit-Wigner width (in agreement with the KA85 overestimation of this observable) and to a value for $g_{\pi N}$ that leads to a sizable violation of the GT relation, which is nowadays theoretically implausible.  As for our study of the EM06 PWA, we found a value for the isovector scattering length that is too small as compared with the accurate values obtained from pion-atoms data. Besides, the picture of the subthreshold region arising from this solution around the point ($\nu=0$, $t=0$) is quite different to the ones given by KA85 and WI08.     

The most important conclusion of our work is that the scattering amplitude in $\Delta$-ChPT converges well from the subthreshold region up to energies well above threshold. This shows that a systematic framework to analyze the $\pi N$ elastic scattering data without spoiling the structure of the amplitude in the subthreshold region is possible using the $\Delta$-ChPT approach developed in this paper. This is a remarkable result since it should allow to extract all the observables related to the $\pi N$ scattering amplitude directly from the differential cross sections at low energies and in a completely model-independent fashion. Also, a calculation up to a $\mathcal{O}(p^{7/2})$ in $\Delta$-ChPT is called for to confirm the good behavior of the chiral series in our approach and to improve the theoretical uncertainties of our determinations.

\section*{Acknowledgements}

This work is partially funded by the Spanish Government and FEDER funds under contract FIS2011-28853-C02-01 and the grants FPA2010-17806 and Fundaci\'on S\'eneca 11871/PI/09. We also thank the financial support from  the BMBF grant 06BN411, the EU-Research Infrastructure Integrating Activity ``Study of Strongly Interacting Matter" (HadronPhysics2, grant n. 227431) under the Seventh Framework Program of EU and the Consolider-Ingenio 2010 Programme CPAN (CSD2007-00042). JMA acknowledges support by the Deutsche
Forschungsgemeinschaft DFG through the Collaborative Research Center ÒThe Low-Energy Frontier of the Standard
ModelÓ (CRC 1044). JMC acknowledges the STFC [grant number ST/H004661/1] for support.

\newpage

\appendix
\numberwithin{equation}{section}

\section{Partial Wave Decomposition}
\label{Sec:partialwavedecomposition}

The free one-particle states are normalized according to the Lorentz-invariant normalization,
\begin{align}
\la \vp',\sigma';\gamma|\vp,\sigma;\gamma\ra=
  2 E_p (2\pi)^3\delta(\vp'-\vp) \delta_{\sigma\sigma'}\delta_{\gamma\gamma'}
\end{align}
where $E_p$ is the energy of the particle with three-momentum $\vp$, $\sigma$ the spin of the nucleon and $\gamma$ indicates any internal quantum number. A free two-particle state is normalized accordingly and it can be decomposed in states with  well defined total spin $S$ and total angular momentum $J$. In the CM frame one has,
\begin{align}
|\pi(-\vp;a)N(\vp,\sigma;\alpha)\ra&=\sqrt{4\pi}\sum_{\ell,m} (m \sigma \mu|\ell S J) Y_\ell^m(\hat{\mathbf{p}})^*|J \mu \ell;a \alpha \ra,
\label{waves}
\end{align}
with $\hat{\vp}$ the unit vector of the CM nucleon three-momentum $\vp$, $a$ and $\alpha$ the isospin third-components in the Cartesian basis of the pion and nucleon, respectively, $\ell$ the orbital angular momentum, $m$ its third component, $ \mu=m+\sigma$ the third-component of the total angular momentum and $S$ the total spin, with $S=1/2$ for $\pi N$ scattering. 

The Clebsch-Gordan coefficients are denoted by $(m_1 m_2 m_3|j_1 j_2 j_3)$, corresponding to the composition of the spins $j_1$ and $j_2$ (with third-components $m_1$ and $m_2$, in order) to give the third spin $j_3$, with third-component $m_3$. 
The state with well-defined total angular momentum, $|J \mu \ell;a \alpha\ra$, satisfies the normalization condition,
 \begin{align}
\la J' \mu' \ell';a' \alpha'|J \mu \ell;a \alpha\ra=\delta_{J J'}\delta_{\mu'\mu}\delta_{\ell \ell'}
\frac{4\pi \sqrt{s}}{|\vp|} \delta_{a'a}\delta_{\alpha'\alpha}.
\label{jdef.norma}
\end{align}
The partial wave expansion of the $\pi N$ scattering amplitude can be worked out straightforwardly from Eq.~\eqref{waves}. 
By definition, the initial baryon three-momentum $\vp$ gives the positive direction of the ${\mathbf{z}}$-axis. 
Inserting the series of Eq.~\eqref{waves} one has for the scattering amplitude,
\begin{align}
\la \pi(-\vp';a')N(\vp',\sigma';\alpha')|T|\pi(-\vp;a)N(\vp,\sigma;\alpha)\ra= \nonumber\\
4\pi\sum_{\ell,m,J}Y_\ell^0(\hat{\mathbf{z}})(m\sigma'\sigma|\ell\frac{1}{2}J)
(0\sigma\sigma|\ell\frac{1}{2}J) Y_{\ell}^m(\hat{\vp}') T_{J\ell}(s),~
\label{series_t}
\end{align}
where $T$ is the T-matrix operator and $T_{J\ell}$ is the partial wave amplitude with total angular momentum $J$ and orbital angular momentum $\ell$. Notice that in Eq.~\eqref{series_t} we   made use of the fact that $Y_\ell^m(\hat{\mathbf{z}})$ is non-zero only for $m=0$. Recall also that because of parity conservation partial wave amplitudes with different orbital angular momentum do not mix.
From Eq.~\eqref{series_t} it is straightforward to isolate $T_{J\ell}$ with the result,
\begin{align}
T_{J\ell}(a',\alpha';a,\alpha)&=\frac{1}{\sqrt{4\pi(2\ell+1)}(0\sigma\sigma|\ell\frac{1}{2}J)} \nonumber \\
\times\sum_{m,\sigma'} & \int d\hat{\vp}'\,\la \pi(-\vp';a')N(\vp',\sigma';\alpha')|T|\pi(-\vp;a)N(\vp,\sigma;\alpha)\ra  (m\sigma'\sigma|\ell\frac{1}{2}L) Y_\ell^m(\hat{\vp}')^*,
\label{tjl}
\end{align}   
in the previous expression the  resulting $T_{J\ell}$ is of course independent of choice of $\sigma$. 

The relation between the  Cartesian and charge bases is given by 
\begin{align}
|\pi^+\ra&=\frac{1}{\sqrt{2}}(|\pi^1\ra+i|\pi^2\ra)~,\nn\\
|\pi^-\ra&=\frac{1}{\sqrt{2}}(|\pi^1\ra-i|\pi^2\ra)~,\nn\\
|\pi^0\ra&=|\pi^3\ra~.
\label{pion_charged}
\end{align}
According to the previous definition of states $|\pi^+\ra=-|1,+1\ra$, $|\pi^-\ra=|1,-1\ra$ and $|\pi^0\ra=|\pi^3\ra=|1,0\ra$, where the states of the isospin basis are placed to the right of the equal sign. Notice the minus sign in the relationship for $|\pi^+\ra$. 
Then, the amplitudes with  well-defined isospin, $I=3/2$ or 1/2, are  denoted by $T_{IJ\ell}$ and can be obtained employing the appropriate linear combinations of $T_ {J\ell}(a',\alpha';a,\alpha)$, Eq.~\eqref{tjl}, in terms of standard  Clebsch-Gordan coefficients.

Due to the normalization of the states with well-defined total angular momentum, Eq.~\eqref{jdef.norma}, the partial waves resulting from Eq.~\eqref{tjl} with well defined isospin satisfy the unitarity relation,
\begin{align}
\hbox{Im}T_{IJ\ell}=\frac{|\vp|}{8\pi \sqrt{s}}|T_{IJ\ell}|^2.
\label{unita}
\end{align}
For $|\vp|>0$ and below the inelastic threshold due the one-pion production at $|\vp|\simeq 210$~MeV. 
Given the previous equation, the $S$-matrix element with well defined $I$, $J$ and $\ell$, denoted by  $S_{I J\ell}$, corresponds to
 \begin{align}
S_{I J\ell}=1+i\frac{|\vp|}{4\pi\sqrt{s}}T_{I J\ell},
\label{s.def}
\end{align}
satisfying $S_{I J \ell}S_{I J \ell}^*=1$ in the elastic physical region. In the same region we can then write
\begin{align}
S_{I J\ell}=e^{2 i\delta_{I J\ell}},
\label{s.def.2}
\end{align}
with $\delta_{I J\ell}$ the corresponding phase shifts. And, form Eqs.~\eqref{s.def} and \eqref{s.def.2} one has
\begin{align}
T_{IJ\ell}=\frac{8\pi \sqrt{s}}{|\vp|}\sin\delta_{IJ\ell} e^{i \delta_{IJ\ell}}.
\label{desfasajeunitario}
\end{align}
However, if the calculation is perturbative, the $S$-matrix does not fulfil unitarity exactly and one cannot use Eq.~\eqref{desfasajeunitario} to calculate the phase shifts.
Instead, is necessary to perform a perturbative expansion of the previous equation up to the order considered to find a relation between the perturbative amplitude to its corresponding phase shift.
Following this procedure, we find that up to $\Opc$ the different phase shifts can be obtained from the perturbative amplitudes by means of the equation:

\begin{align}\label{des-pert}
 \delta_{IJ\ell}=\frac{|\vp|}{8\pi\sqrt{s}}\text{Re}T_{IJ\ell}
\end{align}

\section{Tree Level Calculations}
\label{Sec:treelevelcalculations}

In this Appendix we show the results concerning the tree level calculation of $\pi N$ scattering amplitude.
The Born-terms, which are expressed in terms of the Mandelstam variables $s$ and $u$, include also their crossed version.

\subsection{$\Op$}

\begin{itemize}
 \item  Born-term:
\end{itemize}

\begin{align*}
A^+(s,t,u)&= \frac{g^2 (m_2+m_N)}{4 f^2} \left[\frac{s-m_N^2}{s-m_2^2} + \frac{u-m_N^2}{u-m_2^2} \right]\hspace{8cm} \\
B^+(s,t,u)&= -\frac{g^2 }{4 f^2 }\left[ \frac{(s+ 2 m_2 m_N +m_N^2)}{(s-m_2^2)} - \frac{(u+ 2 m_2 m_N +m_N^2)}{(u-m_2^2)}\right]\\
A^-(s,t,u)&= \frac{g^2 (m_2+m_N)}{4 f^2} \left[\frac{s-m_N^2}{s-m_2^2} - \frac{u-m_N^2}{u-m_2^2} \right] \\
B^-(s,t,u)&= -\frac{g^2 }{4 f^2 }\left[ \frac{(s+ 2 m_2 m_N +m_N^2)}{(s-m_2^2)} + \frac{(u+ 2 m_2 m_N +m_N^2)}{(u-m_2^2)}\right]
\end{align*}

Where $m_2 \equiv m-4c_1 M^2$ includes the $\Opd$ correction to the nucleon mass.

\begin{itemize}
 \item  Contact term:
\end{itemize}

\begin{align*}
A^+(s,t,u)&= B^+(s,t,u)=  A^-(s,t,u)= 0\hspace{10cm} \\
B^-(s,t,u)&= \frac{1 }{2 f^2 }
\end{align*}

\subsection{$\Opd$}

\begin{itemize}
 \item  Contact term:
\end{itemize}

\begin{align*}
A^+(s,t,u)&= \frac{1}{f_\pi^2}\left[ -4 c_1 M_\pi^2 + \frac{c_2 (s-u)^2}{8 m_N^2} + c_3 (2 M_\pi^2-t)\right]+\Opc \hspace{2cm}\\
B^+(s,t,u)&=0 \\ 
A^-(s,t,u)&= -\frac{c_4 (s-u)}{2 f_\pi^2}\\
B^-(s,t,u)&= \frac{2 c_4 m_N}{f_\pi^2}
\end{align*}

\subsection{$\Opt$}

\begin{itemize}
 \item  Born-term:
\end{itemize}

\begin{align*}
A^+(s,t,u)&= \frac{4 g (2 d_{16}-d_{18}) m_N M_\pi^2}{f_\pi^2} \hspace{10.5cm}\\
B^+(s,t,u)&= \frac{4 g (2 d_{16}-d_{18}) m_N^2 M_\pi^2 (s-u)}{f_\pi^2 (s-m_N^2) (u-m_N^2)} \\
A^-(s,t,u)&= 0  \\
B^-(s,t,u)&= \frac{2 g (2 d_{16}-d_{18}) M_\pi^2 (3m_N^4- s u - m_N^2 (s+u))}{f_\pi^2(s-m_N^2)(u-m_N^2)} 
\end{align*}

\begin{itemize}
 \item  Contact term:
\end{itemize}

\begin{align*}
A^+(s,t,u)&= -\frac{(d_{14}-d_{15})(s-u)^2}{4 m_N f_\pi^2 } + \Opc \hspace{6cm} \\ 
B^+(s,t,u)&=  \frac{(d_{14}-d_{15})(s-u)}{ f_\pi^2 } + \Opt\\
A^-(s,t,u)&=  \frac{s-u}{2 m_N f_\pi^2} \left[ 2(d_1+d_2+2 d_5)M_\pi^2-(d_1+d_2)t+2d_3 (s-u)^2 \right] +\mathcal{O}(q^5)\\
B^-(s,t,u)&= 0
\end{align*}

\begin{itemize}
 \item $\Delta$ Born-term $\mathcal{O}(p^{3/2})$:
\end{itemize}

An explicit calculation of the Born-term in the $s$-channel gives: 
\begin{eqnarray}
&&A^{\pm}=-\frac{h_A^2}{4 f_\pi^2 m_\Delta^2}C_I^{\pm}\frac{1}{s-m_\Delta^2}\left(m_N^5-2 \left(M_{\pi }^2+2 s\right) m_N^3-2 m_{\Delta} \left(M_{\pi }^2+s\right)
   m_N^2\right.\nonumber\\
&&\left.+\left(M_{\pi }^4-4 s M_{\pi }^2+3 s (s+t)\right) m_N+2 m_{\Delta} \left(M_{\pi
   }^2-s\right)^2+3 m_{\Delta} s t\right) \nonumber\\
&&B^{\pm}=-\frac{h_A^2}{4 f_\pi^2 m_\Delta^2}C_I^{\pm}\frac{1}{s-m_\Delta^2}\left(m_N^4-2 \left(M_{\pi }^2+3 s\right) m_N^2-2 m_{\Delta} \left(m_N^2-M_{\pi }^2+s\right) m_N\right.\nonumber\\
&&\left.+\left(M_{\pi }^2-s\right)^2+3 s t\right)
\end{eqnarray}
With $C_I^+=1/9$ and $C_I^-=-1/18$.  

\newpage 

\begin{itemize}
 \item $\Delta$ Born-term $\mathcal{O}(p^{5/2})$:
\end{itemize}

\begin{eqnarray}
&&A_2^{\pm}=\frac{h_A}{2 f_\pi^2 m_\Delta^2}C_I^{\pm} \frac{F_A(s,t)}{s-m_\Delta^2}\nonumber\\
&&B_2^{\pm}=\frac{h_A}{2 f_\pi^2 m_\Delta^2}C_I^{\pm} \frac{F_B(s,t)}{s-m_\Delta^2}
\end{eqnarray}
With:
\begin{eqnarray}
&&F_A(s,t)= \frac{1}{6 m_{\Delta}}\left(d_4^\Delta \left(m_N^2-M_{\pi }^2-s\right) \left(m_N^5-2 \left(M_{\pi }^2+2 s\right) m_N^3-2 m_{\Delta} \left(M_{\pi }^2+s\right)
   m_N^2\right.\right.\nonumber\\
&&\left.\left.+\left(M_{\pi }^4-4 s M_{\pi }^2+3 s (s+t)\right) m_N+2 m_{\Delta} \left(M_{\pi }^2-s\right)^2+3 m_{\Delta} s t\right)\nonumber\right.\\
&&\left.+2 d_3^\Delta
   m_{\Delta} \left(-m_N^6+2 \left(M_{\pi }^2+2 s\right) m_N^4-\left(M_{\pi }^4-2 s M_{\pi }^2+s (5 s+3 t)\right) m_N^2\right.\right.\nonumber\\
&&\left.\left.+m_{\Delta}
   \left(\left(m_N^2-s\right)^2-M_{\pi }^4\right) m_N+s \left(2 \left(M_{\pi }^2-s\right)^2+3 s t\right)\right)\right),\nonumber
\end{eqnarray}
\begin{eqnarray}
&&F_B(s,t)=\frac{1}{6
   m_{\Delta}}\left(d^\Delta_4 \left(m_N^2-M_{\pi }^2-s\right) \left(m_N^4-2 \left(M_{\pi }^2+3 s\right) m_N^2\right.\right.\nonumber\\
&&\left.\left.-2 m_{\Delta} \left(m_N^2-M_{\pi
   }^2+s\right) m_N+\left(M_{\pi }^2-s\right)^2+3 s t\right)+2 d_3^\Delta m_{\Delta} \left(m_{\Delta} \left(3 m_N^4\right.\right.\right.\nonumber\\
&&\left.\left.\left.-4 \left(M_{\pi
   }^2+s\right) m_N^2+\left(M_{\pi }^2-s\right)^2+3 s t\right)-m_N \left(m_N^4-2 \left(M_{\pi }^2+2 s\right) m_N^2\right.\right.\right.\nonumber\\
&&\left.\left.\left.+M_{\pi }^4-4 M_{\pi
   }^2 s+3 s (s+t)\right)\right)\right).\nonumber
\end{eqnarray}

\section{Loop Level Calculations}
\label{Sec:scalarintegrals}

In this section we list the scalar and tensor integrals needed for the one-loop calculations performed in this work. 
These integrals, calculated in dimensional regularization, are denoted by $\HH_{mn}$, where the subscripts $m$ and $n$ correspond to the number of mesonic and baryonic propagators, in order, that each integral has.

We will use the following variables
\begin{align*}
 \Sigma^\mu&=(P+q)^\mu=(P'+q')^\mu, \\
 \Delta^\mu&=(q'-q)^\mu=(P-P')^\mu,\\
 Q^\mu&= (P'+P)^\mu,
\end{align*}
where $P$ ($P'$) corresponds to the incoming (outgoing) nucleon and $q$ ($q'$) to the incoming (outgoing) pion, so $P^2=P'^2=m_N^2$ and $q^2=q'^2=M_\pi^2$. 

\subsection{Definitions}
\label{definitions}

\begin{itemize}
 \item  {\bfseries 1 meson, 0 nucleons:} 
\end{itemize}

\begin{align}
\HH_{10}&=\frac{1}{i}\int\! \dkd \frac{1}{M^2-k^2}\nonumber\\
\HH_{10}&= 2 \bar{\lambda} M^2 + \frac{M^2}{16 \pi^2} \logMmu \label{H10} 
\end{align}

\begin{itemize}
 \item {\bfseries 0 mesons, 1 nucleon:}
\end{itemize}

\begin{align}
\HH_{01}&=\frac{1}{i}\int\! \dkd \frac{1}{m^2-k^2} \nonumber \\
\HH_{01}&=2 \bar{\lambda} m^2 + \frac{m^2}{16 \pi^2} \logmmu \label{H01} 
\end{align}

\begin{itemize}
 \item {\bfseries 2 mesons, 0 nucleon:}
\end{itemize}

\begin{align*}
\left\{\HH_{20},\HH_{20}^\mu\right\}&=\frac{1}{i}\int\! \dkd \frac{\{1,k^\mu \}}{(M^2-k^2)(M^2-(k-\Delta)^2)}\\
\HH_{20}^\mu&= \frac{\Delta^\mu}{2} \HH_{20}(t) \\
\HH_{20}^{\mu \nu}&= (\Delta^\mu \Delta^\nu - g^{\mu \nu}\Delta^2)\HH_{20}^{(1)}(t)+\Delta^\mu \Delta^\nu \HH_{20}^{(2)}(t)
\end{align*}

\begin{itemize}
 \item {\bfseries 1 meson, 1 nucleon:}
\end{itemize}

\begin{align}
 \left\{\HH_{11},\HH_{11}^\mu\right\}& =\frac{1}{i}\int\! \dkd \frac{\{1,k^\mu\}}{(M^2-k^2)(m^2-(P-k)^2)} \nonumber\\
\HH_{11}(P^2)&=-2 \bar{\lambda} + \frac{1}{16 \pi^2}\left\{ 1 + \log \left(\frac{\mu^2}{m^2}\right) -\frac{P^2-m^2+M^2}{2 P^2} \logMm \right. \nonumber  \\
       &  + \frac{\sqrt{4 M^2 P^2 - (P^2-m^2+M^2)^2}}{P^2} \nn\\
       & \times\left[\arctan \left( \frac{m^2-M^2-P^2}{\sqrt{4 M^2 P^2 - (P^2-m^2+M^2)^2}} \right) \right. \nonumber\\ 
       &-\left.\left. \arctan \left( \frac{m^2-M^2+P^2}{\sqrt{4 M^2 P^2 - (P^2-m^2+M^2)^2}} \right) \right] \right\} \label{H11}\\
\HH_{11}^\mu&=P^\mu \HH_{11}^{(1)}(P^2) \nonumber
\end{align}

\begin{itemize}
 \item {\bfseries 0 mesons, 2 nucleons:}
\end{itemize}

\begin{align*}
\HH_{02}&=\frac{1}{i}\int\! \dkd \frac{1}{(m^2-k^2)(m^2-(k-\Delta)^2)}\\
\end{align*}

\begin{itemize}
 \item {\bfseries 2 mesons, 1 nucleon:}
\end{itemize}

\begin{align*}
&\left\{\HH_{21},\HH_{21}^\mu,\HH_{21}^{\mu\nu}\right\}=\frac{1}{i}\int\! \dkd \frac{\{1,k^\mu,k^{\mu\nu} \}}{(M^2-k^2)(M^2-(k-\Delta)^2)(m^2-(P-k)^2)}\\
&\HH_{21}^\mu=Q^\mu \HH_{21}^{(1)}(t)+\frac{1}{2}\Delta^\mu \HH_{21}(t) \\
&\HH_{21}^{\mu\nu}=g^{\mu\nu}\HH_{21}^{(2)}(t)+Q^\mu Q^\nu \HH_{21}^{(3)}(t)+ \Delta^\mu \Delta^\nu \HH_{21}^{(4)}(t) +\frac{1}{2}(\Delta^\mu Q^\nu + Q^\mu \Delta^\nu)\HH_{21}^{(1)}(t)
\end{align*}

\begin{itemize}
 \item {\bfseries 1 meson, 2 nucleon:}
\end{itemize}

\begin{align*}
\left\{\HH_{12},\HH_{12}^\mu,\HH_{12}^{\mu\nu}\right\}&=\frac{1}{i}\int\! \dkd \frac{\{1,k^\mu,k^{\mu\nu} \}}{(M^2-k^2)(m^2-(P_1-k)^2)(m^2-(P_2-k)^2)}\\
\end{align*}

For the topologies displayed in Fig.~\ref{piNdiagramsp3}, one of the momenta is always on-shell. 
Choosing this momentum to be $P_1$, we have for the diagram ($m$): $P_1=P$ and $P_2=P'$. This case defines the integral $\HH_A(t)$ as follows:
\begin{align*}
\HH_A(t)=\HH_{12}(m^2,t). 
\end{align*}
For this case, the tensor decomposition is defined as,
\begin{align*}
 \HH_A^\mu(t)&=Q^\mu \HH_A^{(1)}(t)\\
 \HH_A^{\mu\nu}(t)&=g^{\mu \nu}\HH_A^{(2)}(t)+  Q^\mu Q^\nu \HH_A^{(3)}(t) + \Delta^\mu \Delta^\nu \HH_A^{(4)}(t).
\end{align*}
For the diagrams ($c$), ($d$), ($g$) and ($h$) we have instead $P_1=P$ and $P_2=P+q$, with the integral $\HH_B(s)$,
\begin{align*}
\HH_B(s)=\HH_{12}(s,M^2) 
\end{align*}
In this case, the tensor decomposition is defined as,
\begin{align*}
\HH_B^\mu&=Q^\mu \HH_B^{(1)}(s)+\Delta^\mu \HH_B^{(2)}(s).
\end{align*}

\begin{itemize}
 \item {\bfseries 0 mesons, 3 nucleon:}
\end{itemize}

\begin{align*}
\HH_{03}&=\frac{1}{i}\int\! \dkd \frac{1}{(m^2-k^2)(m^2-(k-P_1)^2)(m^2-(k-P_2)^2}\\
\end{align*}
This integral can appear in two different configurations. 
In the first one $P_1=q$ and $P_2=q'$, which is labeled in Secs. \ref{PVcoefficients} and \ref{resultsloopdiagrams} as $\HH_{03}(t,M^2)$.
For this case its dependence on the $t$ variable comes from the combination $(P_1-P_2)^2$.
For the second configuration, however, the $t$ dependence comes from $P_2^2$, because in this second case $P_1=q$ and $P_2=q-q'$. 
This configuration is labeled in \ref{PVcoefficients} and \ref{resultsloopdiagrams} as $\HH_{03}(M^2,t)$.

\begin{itemize}
 \item {\bfseries 1 meson, 3 nucleon:}
\end{itemize}

\begin{align*}
\left\{\HH_{13},\HH_{13}^\mu\right\}&=\frac{1}{i}\int\! \dkd \frac{\{1,k^\mu\}}{(M^2-k^2)(m^2-(P-k)^2)(m^2-(\Sigma-k)^2)(m^2-(P'-k)^2)} \\
\HH_{13}^\mu(s,t)&=Q^\mu \HH_{13}^{(1)}(s,t) + (\Delta+2 q)^\mu \HH_{13}^{(2)}(s,t)
\end{align*}

\vspace{10cm} 

\subsection{Coefficients of the Passarino-Veltman \\ Decomposition}
\label{PVcoefficients}

\begin{align}
 \HH_{11}^{(1)}(s)&=\frac{1}{2 s}\left[ (s-m^2+M^2)\HH_{11}(s)+\HH_{10}-\HH_{01}\right] \label{H111}\\
 \HH_A^{(1)}(t)&=\frac{\HH_{11}(m^2)+M^2 \HH_A(t)-\HH_{02}(t)}{4 m^2-t} \nn\\ 
 \HH_A^{(2)}(t)&=\frac{2 M^2\HH_{11}(m^2)+2M^2(M^2-4m^2+t) \HH_A(t)-(2M^2-4m^2+t)\HH_{02}(t)}{2(2-d)(4 m^2-t)}\nn\\
 \HH_A^{(3)}(t)&=\frac{M^2((1-d)M^2+4 m^2-t)\HH_A(t)+(1-d)M^2 \HH_{11}(m^2)}{(2-d)(4m^2-t)^2}\nn\\
               &-\frac{(2(1-d)M^2+(3-d)(4m^2-t))\HH_{02}(t)}{2(2-d)(4m^2-t)^2}+\frac{M^2 \HH_{11}(m^2)+\HH_{10}-\HH_{01}}{4m^2(4m^2-t)}\nn\\
 \HH_B^{(1)}(s)&=\frac{1}{{2(M^4+(m^2-s)^2-2M^2(m^2+s))}}\Big[(s-m^2+M^2)((s-m^2-2M^2)\HH_B(s)\nn\\ 
               &-\HH_{11}(s))+(s-m^2-M^2)\HH_{11}(m^2)+2M^2\HH_{02}(M^2)\Big]\nn\\
 \HH_B^{(2)}(s)&=\frac{1}{2(M^4+(m^2-s)^2-2M^2(m^2+s))}\Big[(s-m^2)(s+3m^2-3M^2)\HH_B(s)\nn\\
               &+(M^2-m^2-3s)\HH_{11}(s)+(s+3m^2-M^2)\HH_{11}(m^2)\Big]\nn\\
               &+\frac{2(s-m^2)\HH_{02}(M^2)}{2(M^4+(m^2-s)^2-2M^2(m^2+s))}\nn\\
 \HH_{21}^{(1)}(t)&=\frac{(2M^2-t)\HH_{21}(t)-2\HH_{11}(m^2)+2\HH_{20}(t)}{2(4m^2-t)}\nn\\
 \HH_{21}^{(2)}(t)&=\frac{-2(M^4+m^2(t-4M^2))\HH_{21}(t)+2(M^2-2m^2)\HH_{11}(m^2)+(t-2M^2)\HH_{20}(t)}{2(d-2)(4m^2-t)}\nn\\
 \HH_{21}^{(3)}(t)&=\frac{(4(d-1)M^4-4M^2((d-2)t+4m^2)+t((d-2)t+4m^2))\HH_{20}(t)}{4(d-2)(t-4m^2)^2}\nn\\
                  &+\frac{1}{4(d-2)m^2(t-4m^2)^2}\Big[((d-2)(M^2+2m^2)t+4(3-2d)M^2 m^2 \nn\\ 
                  &+ 8m^4)\HH_{11}(m^2) +(d-2)(4m^2-t)(\HH_{01}-\HH_{10})\Big]-\frac{(d-1)(t-2M^2)\HH_{20}(t)}{2(d-2)(t-4m^2)^2}\nn
\end{align}
\vspace{3cm}

\begin{align}
\HH_{13}^{(1)}(s,t)&=\frac{1}{4(M^4-2M^2(s+m^2)+m^2(m^2-2s)+s(s+t))}\Big[(4M^2+2m^2\nn \\
                   &-2s-t)\HH_{13}(s,t)+2(m^2-s-M^2)\HH_B(s)+(2s+t-2m^2-2M^2)\HH_A(t)\Big] \nn\\
                   &+\frac{(4M^2-t)\HH_{03}(M^2,t)}{4(M^4-2M^2(s+m^2)+m^2(m^2-2s)+s(s+t))}\nn\\
\HH_{13}^{(2)}(s,t)&=\frac{1}{4(M^4-2M^2(m^2+s)+m^4-2m^2s+s(s+t))}\Big[\big(M^2(2s+t-2m^2) \nn\\
                   &+(m^2-s)(4m^2-t)-2M^4\big)\HH_{13}(s,t)+(2s+t-2m^2-2M^2)\HH_{03}(M^2,t)\Big]\nn\\
                   &+\frac{2(s+m^2-M^2)\HH_B(s)+(t-4m^2)\HH_A(t)}{4(M^4-2M^2(m^2+s)+m^4-2m^2s+s(s+t))}\nn
\end{align}

\subsection{Results for the loop diagrams}
\label{resultsloopdiagrams}

We list in this section the results concerning to the loop integrals.
For the diagrams $a+b$, $c+d$, $e$, $f$, $g+h$, $i$, $n+o$ and $p+r$ only the direct version is shown.
To construct the full contribution of the mentioned diagrams is necessary to add its crossed version according to the following rules~\cite{beche2}: 

\begin{align*}
 A^\pm_{TOTAL}(s,t,u)=A^\pm(s,t) \pm A^\pm(u,t) \\
 B^\pm_{TOTAL}(s,t,u)=B^\pm(s,t) \mp B^\pm(u,t) \\
\end{align*}

\begin{itemize}
 \item {\bfseries Loops a+b}
\begin{align*}
 A_{a+b}^{+}(s)&=\frac{g^2 m}{2 f^4}\left[\HH_{01}+\HH_{10}+(s-m^2-M^2)\HH_{11}(s)-(s+m^2)\HH_{11}^{(1)}(s)\right]\hspace{3.5cm}\\
 B_{a+b}^{+}(s)&=-\frac{g^2}{8 f^4}\left[ -\frac{4(3m^2+s)(\HH_{01}-M^2 \HH_{11}(s))}{m^2-s}+4 (m^2+s) \HH_{11}(s) \right] \\
 A_{a+b}^{-}(s)&=A_{a+b}^{+}(s) \\
 B_{a+b}^{-}(s)&=B_{a+b}^{+}(s)\\
\end{align*}
\end{itemize}

\begin{itemize}
 \item {\bfseries Loops c+d}
\begin{align*}
A_{c+d}^{+}(s)&=\frac{g^4 m}{8 f^4 (s-m^2)}\left[ 4 m^2 \HH_{01} - 2 (m^2+s)\HH_{10}\right. \\
              & + 2(2m^2(s-m^2)+M^2(m^2+s))\HH_{11}(m^2) \\
              &-2 (s^2-m^4 + M^2 (3m^2+s))\HH_{11}(s)-4m^2(s+m^2)\HH_{11}^{(1)}(m^2) \\
              & +(m^4+10 m^2 s+5 s^2)\HH_{11}^{(1)}(s) + 4m^2 ( m^2 (m^2-2s)+s^2) \HH_B(s)\\
              & \left.+8 m^2 (m^2-s)(2(s+m^2)-M^2) \right] \\
B_{c+d}^{+}(s)&=\frac{g^4 }{8 f^4 (m^2-s)}\left[ -(s+7m^2)\HH_{01} + 4m^2 \HH_{10} \right.\\
              & + 4m^2(3 m^2+s)\HH_{02}(M^2) + 4 m^2 M^2 \HH_{11}(m^2) + (M^2(s+3m^2)\\
              & +4m^2(s-m^2))\HH_{11}(s)-8 m^4 \HH_{11}^{(1)}(m^2) - (s^2+6 s m^2+m^4)\HH_{11}^{(1)}(s) \\
              & - 4m^2M^2 (s+3m^2)\HH_B(s)+ 4 m^2 (s-m^2)^2 \HH_B^{(1)}(s) \\
              &\left.+ 4 m^2 (m^2-s) (3m^2+s) \HH_B^{(2)}(s) \right] \\
A_{c+d}^{-}(s)&=A_{c+d}^{+}(s)\\
B_{c+d}^{-}(s)&=B_{c+d}^{+}(s)\\
\end{align*}
%
%
%
 \item {\bfseries Loop e}
\begin{align*}
A_{e}^{+}(s)&=\frac{3g^4 m}{16 f^4 (m^2-s)}\left[ 2(s+3m^2)\HH_{01} + 2(s-m^2)\HH_{10} \right.\hspace{6.5cm} \\
            & \left.+  2((s-m^2)^2-M^2(3m^2+s))\HH_{11}(s) - (s-m^2)^2 \HH_{11}^{(1)}(s)\right]\\
B_{e}^{+}(s)&= \frac{3 g^2}{16 f^4(m^2-s)^2}\left[ (9m^4 + 6 m^2 s+s^2)\HH_{01}+4m^2(s-m^2)\HH_{10} \right.\\
            &\left. +(4 m^2(s-m^2)^2-M^2(3m^2+s)^2) \HH_{11}(s)- (m^2-s)^3\HH_{11}^{(1)}(s)\right]\\
A_{e}^{-}(s)&= A_{e}^{+}(s)\\
B_{e}^{-}(s)&=B_{e}^{+}(s)\\
\end{align*}
%
%
%
%
 \item {\bfseries Loop f}
\begin{align*}
A_{f}^{+}(s)&=\frac{m(s-m^2)\HH_{11}^{(1)}(s)}{2f^4}\hspace{11.5cm}\\
B_{f}^{+}(s)&=\frac{(4\HH_{01}-\HH_{10}-4M^2\HH_{11}(s)+4\HH_{11}^{(1)}(s))}{8 f^4} \\
A_{f}^{-}(s)&=  \frac{A_{f}^{+}(s)}{2}\\
B_{f}^{-}(s)&=  \frac{B_{f}^{+}(s)}{2}\\
\end{align*}
%
%
 \item {\bfseries Loops g+h}
\begin{align*}
A_{g+h}^{+}(s)&=\frac{g^2 m (s-m^2)}{2 f^4}\left[ -2 \HH_{11}(s) + \HH_{11}^{(1)}(s) +8m^2\HH_{B}^{(1)}(s) \right]\hspace{5.5cm}\\
B_{g+h}^{+}(s)&=\frac{g^2}{4f^4}\left[-2 \HH_{01}+\HH_{10}+8 m^2 \HH_{02}(M^2)+2M^2 \HH_{11}(s)-2m^2 \HH_{11}^{(1)}(m^2) \right.\\
              & \left.- 2 (m^2+s) \HH_{11}^{(1)}(s)- 8 M^2 m^2 \HH_{B}(s)-8m^2(m^2-s)(\HH_{B}^{(1)}(s)-\HH_{B}^{(2)}(s))\right]\\
A_{g+h}^{-}(s)&=0\\
B_{g+h}^{-}(s)&=0\\
\end{align*}
%
%
%

 \item {\bfseries Loop i}
\begin{align*}
A_{i}^{+}(s,t)&= -\frac{3 g^4 m}{16 f^4}\left[ 8m^2 (\HH_{02}(M^2)-\HH_{02}(t)) +2(4m^2-M^2)\HH_{11}(m^2) \right.\\
              & +(m^2-s)\HH_{11}^{(1)}(s) +2(M^2-(s+3m^2))\HH_{11}(s) +32m^4(m^2-s)\HH_{13}(s,t)  \\
              &  +8m^2M^2 \HH_{A}(t)  -32m^4 \HH_{A}^{(1)}(t)+8m^2(2s+t-2M^2-2m^2)\HH_{A}^{(3)}(t) \\ 
              &  -8m^2(M^2+m^2-s)\HH_{B}(s)+ 8m^2 (M^2+3m^2+s)\HH_{B}^{(1)}(s)\\
              & \left.+8m^2(M^2+m^2-s)\HH_{B}^{(2)}(s)\right]\\
B_{i}^{+}(s,t)&= -\frac{3 g^4}{16 f^4}\left[ -\HH_{01}+\HH_{10}+4m^2(2\HH_{02}(M^2)+\HH_{02}(t))+16 m^4 \HH_{03}(t,M^2) \right.\\
               & +(M^2-4m^2)\HH_{11}(s)-4m^2 \HH_{11}^{(1)}(m^2)+(m^2-s)\HH_{11}^{(1)}(s)+16 m^4 M^2 \HH_{13}(s,t) \\
               & +32m^4(m^2-s)\HH_{13}^{(2)}(s,t)-4m^2M^2 \HH_{A}(t) +8m^2 \HH_{A}^{(2)}(t)-8m^2M^2 \HH_{B}(s) \\
               & \left.+ 8m^2(s+3m^2)\HH_{B}^{(1)}(s)+  8m^2(m^2-s)\HH_{B}^{(2)}(s)  \right]\\
A_{i}^{-}(s,t)&= -\frac{A_{i}^{+}(s,t)}{3}\\
B_{i}^{-}(s,t)&= -\frac{B_{i}^{+}(s,t)}{3}\\
\end{align*}
%
%
%
 \item {\bfseries Loop k}
%
\begin{align*}
A_{k}^{+}(t)&=0 \hspace{14.5cm}\\
B_{k}^{+}(t)&=0\\
A_{k}^{-}(t)&=0\\
B_{k}^{-}(t)&=\frac{t \HH_{20}(t)}{f^4}\\
\end{align*}
%
%
 \item {\bfseries Loop l}
%
\begin{align*}
A_{l}^{+}(t)&=\frac{g^2 m}{2 f^4}\left[ -2 \HH_{01} + 2M^2 \HH_{11}(m^2)+(M^2-2t)(4m^2 \HH_{21}^{(1)}(t)-\HH_{20}(t))\right] \hspace{3.3cm}\\
B_{l}^{+}(t)&=0\\
A_{l}^{-}(t)&= -\frac{4 g^2 m^3}{f^4}(s-u)\HH_{21}^{(3)}(t) \\
B_{l}^{-}(t)&= -\frac{g^2 }{f^4}\left[ 4m^2 \HH_{21}^{(2)}(t)+t \HH_{20}^{(1)}(t)\right]\\
\end{align*}
%
%
%
 \item {\bfseries Loop m}
%
\begin{align*}
A_{m}^{+}(s,t,u)&= 0\hspace{13.8cm}\\
B_{m}^{+}(s,t,u)&= 0\\
A_{m}^{-}(s,t,u)&= -\frac{g^2 m^3}{f^4}(s-u)\HH_{A}^{(3)}(t)\\
B_{m}^{-}(s,t,u)&= -\frac{g^2 }{8 f^4}\left[ \HH_{10}-4m^2 (\HH_{11}^{(1)}(m^2)-\HH_{02}(t)+M^2 \HH_{A}(t)-2\HH_{A}^{(2)}(t)) \right]\\
\end{align*}
%
%
%
 \item {\bfseries Loops n+o}
%
\begin{align*}
A_{n+o}^{+}(s,t)&=\frac{g^2m}{ f^4}\left[\HH_{01}-M^2 \HH_{11}(m^2) \right] \hspace{9.5cm} \\
B_{n+o}^{+}(s,t)&=\frac{g^2}{4 f^4}\frac{(7m^2+s)(\HH_{01}-M^2\HH_{11}(m^2))}{(m^2-s)} \\
A_{n+o}^{-}(s,t)&=A_{n+o}^{+}(s,t) \\
B_{n+o}^{-}(s,t)&=B_{n+o}^{+}(s,t) \\
\end{align*}
%
%
 \item {\bfseries Loops p+r}
%
\begin{align*}
A_{p+r}^{+}(s,t)&= \frac{g^2m}{2f^4}\HH_{10} \hspace{12.7cm}\\
B_{p+r}^{+}(s,t)&= \frac{g^2}{4f^4}\frac{(s+3m^2)\HH_{10}}{m^2-s} \\
A_{p+r}^{-}(s,t)&=A_{p+r}^{+}(s,t) \\
B_{p+r}^{-}(s,t)&=B_{p+r}^{+}(s,t) \\
\end{align*}
%
%
 \item {\bfseries Loops t+u}
%
\begin{align*}
A_{t+u}^{+}(s,t)&=\frac{g^2 m}{ f^4} (\HH_{01}-M^2 \HH_{11}(m^2))\hspace{9.5cm} \\
B_{t+u}^{+}(s,t)&=\frac{g^2}{ 2 f^4} (\HH_{01}-M^2 \HH_{11}(m^2)) \\
A_{t+u}^{-}(s,t)&=0 \\
B_{t+u}^{-}(s,t)&=0 \\
\end{align*}
%
%
 \item {\bfseries Loop v}
%
\begin{align*}
A_{v}^{+}(s,t)&= 0 \hspace{14cm}\\
B_{v}^{+}(s,t)&= 0\\
A_{v}^{-}(s,t)&= 0\\
B_{v}^{-}(s,t)&= \frac{5}{8f^4}\HH_{10}
\end{align*}
\end{itemize}

\section{Identifying the power counting breaking terms}
\label{Sec:PCBTs}

In this Appendix we explain the method we used to extract analytically the power counting breaking terms from the $\Opt$ loop amplitude ($T_{loops}$). First, as we did with the full amplitude, we decompose $T_{loops}$ in terms of its scalar integrals using the Passarino-Veltman decomposition. 
\begin{align*}
 T_{loops}=\sum_{mn} \mathcal{C}_{mn}\HH_{mn},
\end{align*}
where the scalar integrals $\HH_{mn}$ are defined in \ref{Sec:scalarintegrals} and $\mathcal{C}_{mn}$ refers to its coefficients that result in the Passarino-Veltman decomposition.
Second, we calculate the infrared regular part~\cite{becher} of these scalar integrals ($\RR_{mn}$), because it contains all the PCBT. Its calculation, for each of the scalar integrals, is straightforward because the chiral expansion of the regular part commutes with the integration in the Feynman parameters~\cite{G&J&K}. 
The chiral order of each $\mathcal{C}_{mn}$ tell us up to which order in the chiral expansion we need to obtain the regular part of $\HH_{mn}$. 
So, finally, we obtain,
\begin{align}\label{tloops}
 T_{loops}=\sum_{mn} \mathcal{C}_{mn}\RR_{mn}. 
\end{align}
We expand Eq.~\eqref{tloops} in a chiral series to end with a string of terms that can be splitted into a part that has chiral order lower than three (these are the PCBT in our case) and an infinite series that respect the power counting. 
The (finite) terms that break the power counting have the same analytical structure than the monomials in the original Lagrangian and can be cancelled via a LECs redefinition (see the next Appendix).

\section{Low-Energy Constants Renormalization}
\label{Sec:lecsrenormalization}

In this appendix we show how to redefine the $\Opd$ and $\Opt$ LECs in order to cancel the divergences and the PCBT.
In this way we have full relativistic scale-independent chiral amplitudes free from divergences that respect the chiral power counting. 

\subsection{$\Opd$ LECs}

The $\Opd$ LECs are redefined in order to cancel both divergent parts, as well as PCBT. 

\begin{align*}
 c_1&\rightarrow c_1^{EOMS} - 2\bar{\lambda} \frac{3g^2 m}{8 f^2} +\frac{3 g^2 m }{128 \pi^2 f^2}  (1-\logmmu) \\
 c_2&\rightarrow c_2^{EOMS} + 2\bar{\lambda} \frac{(g^2-1)^2 m}{2 f^2} + \frac{m}{32 \pi^2 f^2} [(g^2-1)^2\logmmu-(2+g^4)]\\
 c_3&\rightarrow c_3^{EOMS} + 2\bar{\lambda} \frac{(g^4-6g^2+1)m}{4 f^2}+\frac{m}{64 \pi^2 f^2}[(g^4-6g^2+1) \logmmu + 9g^4]\\
 c_4&\rightarrow c_4^{EOMS} + 2\bar{\lambda} \frac{(3g^4-2g^2-1)m}{4 f^2}+ \frac{m}{64 \pi^2 f^2} [(3g^4-2g^2-1)\logmmu -g^2(5+g^2) ]\\
\end{align*}

\subsection{$\Opt$ LECs}

In contrast to the $\Opd$ LECs, the $\Opt$ ones only cancel divergent parts (along with their scale-dependent logarithms) because the $\Opt$ analytical terms do not break the power counting in our $\Opt$ calculation. They are renormalized within the $\widetilde{MS}$ scheme (also known as $\overline{MS}-1$).
\begin{align*}
 d_1+d_2&\rightarrow (d_1+d_2)^{\widetilde{MS}} + 2\bar{\lambda} \frac{3g^4-4g^2+1}{48 f^2}+ \frac{3g^4-4g^2+1}{768 \pi^2 f^2} \logmmu \\
 d_5&\rightarrow d_5^{\widetilde{MS}} - 2\bar{\lambda} \frac{g^2+8}{48 f^2} - \frac{g^2+8}{ 768 \pi^2 f^2} \logmmu\\
 d_{14}-d_{15}&\rightarrow (d_{14}-d_{15})^{\widetilde{MS}} + 2\bar{\lambda}\frac{(g^2-1)^2}{4f^2} + \frac{(g^2-1)^2}{64 \pi^2 f^2} \logmmu \\
 d_{16} &\rightarrow d_{16}^{\widetilde{MS}} -2\bar{\lambda}\frac{g(g^2-1)}{4f^2}+\frac{g(g^2-1)}{64\pi^2f^2}\logmmu
\end{align*}

\newpage 

\section{Summary of results for $\Delta$-ChPT in the IR scheme}
\label{App:DeltaIR}

In Table~\ref{IR-Delta-LECs} we show the results at $\sqrt{s}_{max}=1.20$~GeV of the LECs and the $\chi^2_{\rm d.o.f.}$ for $\Delta$-ChPT within the IR scheme, which is to be compared with the EOMS results in Table~\ref{LECs-strategyII}. In Table~\ref{IR-Delta-pheno} we give the results obtained at this energy for a selection of observables. For the calculation of $\Delta_{GT}$ we have used the methods shown in Sec.~\ref{Sec:GT}, whereas for the rest of the observables we have made use of the formulas for $d_{00}^+$, $d_{01}^+$ and $\sigmaterm$ given for IR in Refs.~\cite{becher,beche2}.

\begin{table}[H]\small
 \begin{center}
\begin{tabular}{|r|r|r|}
\hline
\small{LEC}        & KA85	 $\Delta$-IR                  &    WI08   $\Delta$-IR    \\
\hline							  			    													
$c_1$              &        $-0.196(31)$            &   $-0.371(30)$    \\   
$c_2$              &        $1.88(10)$             & $ 1.97(9)$       \\
$c_3$              &        $-2.90(13)$            &  $-3.16(12)$ \\
$c_4$              &        $1.81(6) $             &  $1.96(6)$  \\
\hline		  					  			    														
$d_1+d_2$          &   $0.91(9)$                   &  $1.23(9)$  \\
$d_3$              &       $-1.26(6)$              &  $-1.36(6) $  \\
$d_5$              &         $0.168(42)$           &  $-0.124(40) $   \\
$d_{14}-d_{15}$    &          $1.33(16)$          &   $-0.97(15)$ \\
$d_{18}$           &          $-2.66(26)$         &  $-1.79(25)$  \\ 
\hline
$h_A$           &      $3.096(35)$          &       $2.956(34)$  \\
\hline
$\chi^2_{\rm d.o.f.}$  &       $ 4.15$                &     $1.67$         \\
\hline
\end{tabular}
\caption[pilf]{\small Result for the LECs obtained in a fit up to energies of $\sqrt{s}_{max}=1.20$~GeV in IR $\Delta$-ChPT. \label{IR-Delta-LECs}}
\end{center}
\end{table}

\begin{table}[H]
 \begin{center}
\begin{tabular}{c|c|c|}
\cline{2-3}
                                 & KA85   {\small  $\Delta$-IR}          & WI08   {\small  $\Delta$-IR}          \\
\hline                                 
\multicolumn{1}{|c|}{$\Delta_{GT}$}  &      8.1(8)\%      & 5.5(8)\% \\
\hline
\multicolumn{1}{|c|}{ $d_{00}^+$ ($M_\pi^{-1}$)}  & -0.9(12)  &       -0.81(11)                                                              \\    
\hline
\multicolumn{1}{|c|}{ $d_{01}^+$ ($M_\pi^{-3}$) }  & 0.338(41)  & 0.420(38)                                                                        \\    
\hline      
\multicolumn{1}{|c|}{$\sigmaterm$  (MeV)  }  & -5.8(2.4)    & 7.7(2.3)    \\
\hline
\end{tabular}
{\caption[pilf]{\protect Summary of the results obtained in $\Delta$-IR for some of the observables studied in this paper. \label{IR-Delta-pheno}}}
\end{center}
\end{table}

\end{document}